%% file: lmcs-final.tex
\documentclass{LMCS}
\usepackage{latexsym}
\usepackage{amssymb,xspace,latexsym,epic,eepic,graphicx}
\usepackage{epsfig,amsmath}
\usepackage{oldgerm}
\usepackage{color}
\usepackage{enumerate,hyperref}

\newcommand{\OMIT}[1]{}
\newcommand{\dupl}{Player II}
\newcommand{\spoiler}{Player I}
\newcommand{\M}{{\mu}} 
  
\newcommand{\w}{{\bar{w}}}

\newcommand{\e}{\varepsilon}
\newcommand{\dm}{\Diamond}

\newcommand{\desc}{\preceq_{\rm desc}}
\newcommand{\sib}{\preceq_{\rm sib}}
\newcommand{\lsib}{\prec_{\rm sib}}

\newcommand{\sProof}[1]{\vspace{2mm}{\noindent\em Proof.~}#1\qed} 
\newcommand{\aProof}[2]{\vspace{2mm}{\noindent\em Proof of
#1.~}#2\qed}
\newcommand{\A}{{\mathcal{A}}}
\newcommand{\B}{{\mathcal{B}}}
\newcommand{\C}{{\mathcal{C}}}
\newcommand{\R}{{\mathcal{R}}}

\newcommand{\WW}{{\mathcal{W}}}
\newcommand{\MM}{\mathfrak{M}}

\newcommand{\G}{{\mathbf G}}

\newcommand{\X}{{\mathbf X}}

\newcommand{\U}{{\mathbf U}}

\renewcommand{\S}{{\mathbf S}}

\newcommand{\next}{\text{\raisebox{1pt}{$\bigcirc$}}}
\renewcommand{\X}{\next}
\newcommand{\Xd}{\X_{\!\downarrow}}
\newcommand{\Xr}{\X_{\!\rightarrow}}
\newcommand{\Yd}{\X_{\!\uparrow}}
\newcommand{\Yr}{\X_{\!\leftarrow}}
\newcommand{\Ud}{\U_{\!\downarrow}}
\newcommand{\Sd}{\S_{\!\downarrow}}

\newcommand{\T}{\mathfrak{T}}

\newcommand{\boxtheorem}{\hfill $\Box$}

\newcommand{\FO}{{\rm FO}}
\renewcommand{\l}{\ell} 
\renewcommand{\phi}{\varphi}
\newcommand{\fth}{\hfill $\Box$}

\newcommand{\wt}{\iota_{\text{\rm w-t}}}

\newcommand{\crc}[1]{#1^\circ}

\theoremstyle{plain}
\newtheorem{theorem}{Theorem}[section]
\newtheorem{lemma}[theorem]{Lemma}
\newtheorem{corollary}[theorem]{Corollary}
\newtheorem{proposition}[theorem]{Proposition}
\newtheorem{claim}[theorem]{Claim}

\theoremstyle{definition}

\newtheorem{definition}{Definition}[section]

\newcommand{\ppath}{\sigma} 
\newcommand{\Ul}{\U}
\newcommand{\Sl}{\S}
\newcommand{\Uc}{\U^c}
\newcommand{\aUc}{\U^{c'}}
\newcommand{\Sc}{\S^c}
\newcommand{\Ur}{\U^a}
\newcommand{\Sr}{\S^a}
\newcommand{\Up}{\U^\ppath}
\newcommand{\Sp}{\S^\ppath}

\newcommand{\Us}{\Up_s}
\newcommand{\Ss}{\Sp_s}
\newcommand{\Uss}{\Up_{ss}}
\newcommand{\Sss}{\Sp_{ss}}

\newcommand{\qdp}[1]{\mathrm{qdp}(#1)}       

\newcommand{\odp}[1]{\mathrm{odp}(#1)}       

\newcommand{\ie}{i.\,e.}                     

\newcommand{\false}{\bot}                    

\newcommand{\true}{\top}                     

\newcommand{\ucaret}{\text{unary-}\caret}
\renewcommand{\ucaret}{\text{unary-}\nwtl}
\newcommand{\utl}{\ucaret}

\newcommand{\fotwo}{\FO^2}   

\newcommand{\always}{\Box} 
\newcommand{\eventually}{\text{\Large $\Diamond$}}

\newcommand{\puntil}{\Up}

\newcommand{\prev}{{\text{\Large $\mathbf{\circleddash}$}}}

\newcommand{\dmd}{
  {\qbezier(0,0)(0.5,0.5)(1,1) \qbezier(0,0)(-0.5,0.5)(-1,1)
    \qbezier(0,2)(0.5,1.5)(1,1) \qbezier(0,2)(-0.5,1.5)(-1,1)}}

\newcommand{\dmdminus}{
  {\unitlength1.5mm
    \begin{picture}(2.2,2)(-1,0)
      \dmd \put(-0.5,1){\line(1,0){1}}
    \end{picture}}}

\newcommand{\peventually}{\dmdminus}

\newcommand{\until}{\U}

\newcommand{\EXPTIME}{{\sc Exptime}}

\newcommand{\caret}{\text{CaRet}}

\newcommand{\retr}{\mathit{ret}}
\renewcommand{\int}{\intt}
\newcommand{\AP}{\mathit{AP}}
\def\Cl{{\it cl\/}}

\def\buchi{B\"{u}chi}

\newcommand{\mret}{{\tt{mret}}}
\newcommand{\mcall}{{\tt{mcall}}}
\newcommand{\pret}{\tt{pret}}
\newcommand{\pcall}{\tt{pcall}}

\def\cd{\delta_c}
\def\rd{\delta_r}

\def\id{\delta_i}

\newcommand{\dpuntil}{{\,\U^{\ppath\downarrow}}}
\newcommand{\upuntil}{{\,\U^{\ppath\uparrow}}}

\newcommand{\ltlv}{\text{\rm LTL}^\M}
\newcommand{\ltlvp}{\ltlv}
\newcommand{\nwtl}{\text{\rm NWTL}}
\newcommand{\nwtlf}{\nwtl^{\text{\rm future}}}
\newcommand{\nwtlp}{\nwtl^+}
\newcommand{\nwtls}{\nwtl^s}
\newcommand{\nwtlss}{\nwtl^{ss}}
\newcommand{\intt}{{\tt int}} 
 
\newcommand{\tltree}{\text{\rm TL}^{\text{\rm tree}}}

\newcommand{\rett}{{\tt ret}}
\newcommand{\call}{{\tt call}}
\renewcommand{\retr}{\rett}

\newcommand{\dmm}{\dm_{\M}}
\newcommand{\dmminus}{\dm^-}
\newcommand{\dmmminus}{\dm_{\M}^-}

\renewcommand{\dm}{\next}
\renewcommand{\dmminus}{\prev}
\renewcommand{\dmm}{\dm_\M}
\renewcommand{\dmmminus}{\dmminus_{\M}}

\newcommand{\sleft}{s^{\leftarrow}}
\newcommand{\sright}{s^{\rightarrow}}
\newcommand{\efeq}{\equiv}
\newcommand{\nn}{{\mathbb N}}

\newcommand{\LRA}{\Leftrightarrow}

\newcounter{example}
\renewcommand{\theexample}{\arabic{example}}

\def\doi{4 (4:11) 2008}
\lmcsheading%
{\doi}
{1--44}
{}
{}
{Jan.~17, 2008}
{Mar.~\phantom03, 2011}
{}   

\begin{document}

\title[First-Order and Temporal Logics for Nested Words]{First-Order and Temporal Logics for Nested Words}

\author[R.~Alur]{Rajeev Alur\rsuper a}
\address{{\lsuper a}Department of Computer and Information Science, University of Pennsylvania}
\email{alur@cis.upenn.edu} 

\author[M.~Arenas]{Marcelo Arenas\rsuper b}
\address{{\lsuper b}Department of Computer Science, Pontificia Universidad Cat\'olica de Chile}
\email{marenas@ing.puc.cl}

\author[P.~Barcel\'o]{Pablo Barcel\'o\rsuper c}
\address{{\lsuper c}Department of Computer Science, Universidad de Chile}
\email{pbarcelo@dcc.uchile.cl}

\author[K.~Etessami]{Kousha Etessami\rsuper d}
\address{{\lsuper d}School of Informatics, University of Edinburgh, Edinburgh}
\email{kousha@inf.ed.ac.uk}

\author[N.~Immerman]{Neil Immerman\rsuper e}
\address{{\lsuper e}Department of Computer Science,University of Massachusetts}
\email{immerman@cs.umass.edu}

\author[L.~Libkin]{Leonid Libkin\rsuper f}
\address{{\lsuper f}School of Informatics, University of Edinburgh, Edinburgh}
\email{libkin@inf.ed.ac.uk}


\begin{revision}
  This is a revised and corrected version of the article originally
  published on November 25, 2008.
\end{revision}

\begin{abstract}
Nested words are a structured model of execution paths in procedural
programs, reflecting their call and return nesting structure.  Finite
nested words also capture the structure of parse trees and other
tree-structured data, such as XML.

We provide new temporal logics for finite and infinite nested words,
which are natural extensions of LTL, and prove that these logics are
first-order expressively-complete.  One of them is based on adding a
``within'' modality, evaluating a formula on a subword, to a logic $\caret$
previously studied in the context of verifying properties of recursive
state machines (RSMs). The other logic, NWTL, 
is based on the notion of a summary
path that uses both the linear and nesting structures. For NWTL
we show that satisfiability is EXPTIME-complete, and that
model-checking can be done in time
polynomial in the size of the RSM model and exponential in the size of
the NWTL formula (and is also 
EXPTIME-complete).

Finally, we prove that first-order logic over nested words has the
three-variable property, and we present a temporal logic for nested
words which is complete for the two-variable fragment of first-order.
\end{abstract}

\maketitle
\vfill\eject

\section{Introduction} 
\noindent An execution of a procedural program can reveal
not just a linear sequence of program states encountered during the
execution, but also the correspondence between each point during the
execution at which a procedure is called and the point when we return
from that procedure call.  This leads naturally to the notion of a
finite or infinite nested word (see \cite{nested,VPL,AEM04}).  A
nested word is simply a finite or $\omega$-word supplied with an
additional binary matching relation which relates corresponding call
and return points (and of course satisfies ``well-bracketing''
properties).  Finite nested words offer an alternative way to view any
data which has both a sequential string structure as well as a
tree-like hierarchical structure.  Examples of such data are XML
documents and parse trees.

Pushdown systems (PDSs), Boolean Programs, and Recursive State
Machines (RSMs), are equivalent abstract models of procedural
programs, with finite data abstraction but unbounded call stack.
Software model checking technology is by now thoroughly developed for
checking $\omega$-regular properties of runs for these models, when
the runs are viewed as ordinary words (see 
\cite{BallRajamani,moped,RSM}).  Unfortunately, temporal logic and
$\omega$-regular properties over ordinary words are inadequate for
expressing a variety of properties of program executions that are
useful in interprocedural program analysis and software verification.
These include Hoare-like pre/post conditions on procedures, stack
inspection properties, and other useful program analysis properties
that go well beyond $\omega$-regular (see \cite{AEM04} for some
examples).  On the other hand, many such program analysis properties
can easily be expressed when runs are viewed as nested words.  
Runs of Boolean Programs and RSMs can naturally be viewed as
nested words once we add ``summary edges'' between matching calls and
returns, and we can thus hope to extend model checking technology for
procedural programs using richer temporal logics over nested words
which remain tractable for analysis.

These considerations motivated the definition 
of Visibly Pushdown Languages (VPLs) \cite{VPL} and
the call-return temporal logic \caret{} \cite{AEM04}.
\caret\ is a temporal logic over nested words\footnote{Although
the ``nested word'' terminology was not yet used in that paper.}
which extends LTL with new temporal operators that allow for
navigation through a nested word both via its ordinary sequential
structure, as well as its matching 
call-return summary structure.
The standard LTL model checking
algorithms for RSMs and PDSs can be extended to allow model
checking of \caret{}, with essentially the same complexity \cite{AEM04}.
VPLs \cite{VPL} are a richer class of languages
that capture MSO-definable properties of nested
words.  
Recently, results about VPLs have been recast in light
of nested words, and in particular in terms of Nested Word Automata
\cite{nested} which offer a machine acceptor 
for ($\omega$-)regular nested words, with all the
expected closure properties.

Over ordinary words, LTL has long been considered
the temporal logic of choice 
for program verification, not only
because its temporal operators offer the right abstraction for reasoning
about events over time, but
because it provides a good balance
between expressiveness (first-order complete), 
conciseness (can be exponentially more succinct compared to automata), 
and the complexity of  
model-checking (linear time
in the size of the finite transition system, 
and PSPACE in the size of the temporal formula).

This raises the question: 
{\em What is the right temporal logic 
for nested words?
}

The question obviously need not have a unique answer, particularly
since nested words can arise in various application domains: for
example, 
program verification, as we already discussed, or navigation and querying XML
documents under ``sequential'' representation (see, e.g., \cite{SV02}).
However, it is reasonable to hope that any good temporal logic for
nested words should possess the same basic qualities that make LTL a
good logic for ordinary words, namely: 
\begin{enumerate}[(1)]
\item  {\em first-order
expressive completeness:} LTL has the same expressive power as
first-order logic over words, and we would want the same over nested
words (of course, even more expressiveness, such as full MSO, would
be nice but natural temporal logics are subsumed by first order logic
and any further expressiveness typically comes at a cost, even over
words, of some other desirable properties); 
\item {\em reasonable
complexity for model checking and satisfiability;} and 
\item {\em nice
closure properties}: LTL is closed under boolean combinations
including negation without any blow-up, and we would want the same
for a logic over nested words. 

Finally (and perhaps least easy to
quantify), we want 
\item  {\em natural temporal operators with simple and
intuitive semantics}.  
\end{enumerate}

Unfortunately, the logic \caret{} appears to be deficient with respect
to some of these criteria: although it is easily first-order
expressible, it is believed to be incomplete but
proving incompleteness 
appears to be difficult.  
\caret{}  can express
program 
path properties (for example, every {\it lock\/} operation is
eventually followed by an {\em unlock\/} operation) and local path
properties (for example, if a procedure executes a {\em lock\/}
operation then the same procedure later executes an {\em unlock\/}
operation before returning), but it seems incapable of expressing
scope-bounded path properties (for example, every {\it lock\/}
operation in a procedure is eventually followed by an {\em unlock\/}
operation before the procedure returns).  Such scope-bounded path
properties are natural program requirements, and are expressible in
the first-order logic of nested words.  There is much related work in
the XML community on logics for trees (see, e.g., surveys
\cite{KSS03,Lib05,vianu-pods}), but they tend to have different kinds
of deficiencies for our purposes: they concentrate on the hierarchical
structure of the data and largely ignore its linear structure; also,
they are designed for finite trees.

We introduce and study new temporal logics over nested
words.  The main logic we consider, \emph{Nested Word Temporal Logic}
($\nwtl$) extends LTL with both a future and past variant of the
standard Until operator, which is interpreted over {\em summary paths}
rather than the ordinary linear sequence of positions.  A summary
path is the unique shortest directed path one can take
between a position in a run and some future position, if one is
allowed to use both successor edges and matching call-return
summary edges.  We show that $\nwtl$ possesses all the desirable
properties we want from a temporal logic on nested words.  In
particular, it is both first-order expressively complete and has good
model checking complexity.  Indeed we provide a tableaux construction
which translates an $\nwtl$ formula into a Nested Word Automaton,
enabling the standard automata theoretic approach to model checking of
Boolean Programs and RSMs with complexity that is polynomial in the
size the model and EXPTIME in the size of the formula (and indeed
EXPTIME-complete).

We then explore some alternative temporal logics, which extend
variants of \caret{} with variants of unary ``Within'' operators
proposed in \cite{AEM04}, and we show that these extensions are also
FO-complete.  However, we observe that the model checking and
satisfiability problems for these logics are 2EXPTIME-complete.  
These logics are -- provably -- more concise than $\nwtl$, but we pay
for conciseness with added complexity.

It follows from our proof of FO-completeness for $\nwtl$ that over
nested words, every first-order formula with one free variable can be
expressed using only 3 variables.  More generally, we show, using EF
games, that 3 variables suffice for expressing any first order formula
with two or fewer free variables, similarly to the case of words
\cite{IK} or finite trees \cite{marx-tods}.  Finally, we show that a
natural unary temporal logic over nested words is expressively
complete for first-order logic with 2 variables, echoing a similar
result known for unary temporal logic over ordinary words
\cite{EVW02}.

\subsection*{Related Work}  

VPLs and nested words were introduced
in \cite{VPL,nested}. The logic \caret\ was defined in  \cite{AEM04} 
with the goal of expressing
and checking some natural non-regular program specifications.  The
theory of VPLs and \caret\ has been recast in light of nested words in 
\cite{nested}.
Other aspects of nested words (automata
characterizations, games, model-checking) were further studied in
\cite{RSM,nested,AEM04,LMS04}. It was also observed that nested
words are closely related to a sequential, or ``event-based'' API for
XML known as SAX \cite{sax} (as opposed to a tree-based DOM API
\cite{dom}). SAX representation is very important in streaming
applications, and questions related to recognizing classes of nested
words by the usual word automata have been addressed in
\cite{SV02,Baranystacs}. 

While finite nested words can indeed be seen as XML documents under
the SAX representation, and while much effort has been spent over the
past decade on languages for tree-structured data (see,
e.g., \cite{KSS03,Lib05,vianu-pods} for surveys), adapting the logics
developed for tree-structured data is not as straightforward as it
might seem, even though from the complexity point of view,
translations between the DOM and the SAX representations are easy
\cite{luc-pods03}.  The main problem is that most such logics rely on
the tree-based representation and ignore the linear structure, making
the natural navigation through nested words rather unnatural under the
tree representation. Translations between DOM and SAX are easy for
first-order properties, but verifying navigational properties
expressed in first-order is necessarily non-elementary even for words
if one wants to keep the data complexity linear \cite{FG02}.  On the
other hand, logics for XML tend to have good model-checking properties
(at least in the finite case), typically matching the complexity of
LTL \cite{GK-jacm,neven00}.  We do employ such logics (e.g., those in
\cite{marx-pods04,marx-tods,Schl92}) in the proof of the expressive
completeness of $\nwtl$, first by using syntactic translations that
reconcile both types of navigation, and then by combining them with a
composition game argument that extends the result to the infinite
case, which is not considered in the XML setting. This, however,
involves a nontrivial amount of work. Furthermore, ``within''
operators do not have any natural analog on trees, and the proof for
them is done by a direct composition argument on nested words.

\subsection*{Organization}
Basic notations are given in Section \ref{notations-sec}. Section
\ref{tl-sec} defines temporal logics on nested words, and Section
\ref{expcompl-sec} presents expressive completeness results. We study
model-checking in Section \ref{mc-sec}, and in Section \ref{fv-sec} we
prove the 3-variable property and present a logic for the 2-variable
fragment.

\section{Notations} 
\label{notations-sec}

\subsection{Nested Words}

A {\em matching\/} on $\nn$ or an interval $[1,n]$ of $\nn$
consists of a binary relation $\M$ and two unary relations 
$\call$ and $\retr$, satisfying the following: (1) if
$\M(i,j)$ holds then $\call(i)$ and $\retr(j)$ and $i < j$; 
(2) if $\M(i,j)$ and $\M(i,j')$ hold
then $j = j'$ and if $\M(i,j)$ and $\M(i',j)$ hold then $i = i'$; 
(3) if $i\leq j$ and $\call(i)$ and $\retr(j)$ then
there exists $i\le k\le j$ such that either $\M(i,k)$ or $\M(k,j)$.

Let $\Sigma$ be a finite alphabet.  A {\em finite nested word} of
length $n$ over $\Sigma$ is a tuple $\w=(w,\M,\call,\retr)$, where
$w=a_1\ldots a_n\in \Sigma^*$, and $(\M,\call,\retr)$ is a matching on
$[1,n]$.  A {\em nested $\omega$-word} is a tuple
$\w=(w,\M,\call,\retr)$, where $w=a_1\ldots \in \Sigma^\omega$, and
$(\M,\call,\retr)$ is a matching on $\nn$.

We say that a position $i$ in a nested word 
$\w$ is a {\em call} position if $\call(i)$ holds;
a {\em return} position if $\retr(i)$ holds;
and an {\em internal} position if it is
neither a call nor a return.
If $\M(i,j)$ holds, we say that $i$ is the matching call of $j$, and $j$ is
the matching return of $i$, and write $c(j)=i$ and $r(i)=j$. 
Calls without matching returns are {\em pending\/} calls, and returns without
matching calls are {\em pending\/} returns (sometimes we will
alternatively refer to such calls and returns as {\em unmatched}).
A nested word is said to be {\em well-matched\/} if no calls or returns are 
pending. Note that for well-matched nested words, the unary predicates
$\call$ and $\retr$ are uniquely specified by the relation $\M$.

A nested word $\w=(w, \M,\call,\retr)$ is represented as a first-order structure
$$\langle\, U\,,(P_a)_{a \in \Sigma}\,,<\,,\M\,,\call\,,\retr\,\rangle,$$ where $U$ is
$\{1, \ldots, n\}$ if $w$ is a finite word of length $n$ and $\nn$ if
$\w$ is a nested $\omega$-word; $<$ is the usual ordering, $P_a$ is
the set of positions labeled $a$, and $(\M,\call,\retr)$ is the matching
relation. When we talk about first-order logic (\FO) over nested words,
we assume \FO\ over such structures
 (i.e. the vocabulary is that of words plus the matching relation).

For a nested word $\w$, and two elements $i,j$ of $\w$, we denote by
$\w[i,j]$ the substructure of $\w$ (i.e. a finite nested word) induced
by elements $\l$ such that $i \leq \l \leq j$. If $j < i$ we assume
that $\w[i,j]$ is the empty nested word.
For nested $\omega$-words $\w$, we let $\w[i,\infty]$ denote the substructure
induced by elements $l\ge i$.

When this is clear from the context, we do not 
distinguish references to positions in subwords $\w[i,j]$ and $\w$
itself, e.g., we shall often write $(\w[i,j],i)\models\phi$ to mean
that $\phi$ is true at the first position of $\w[i,j]$.  

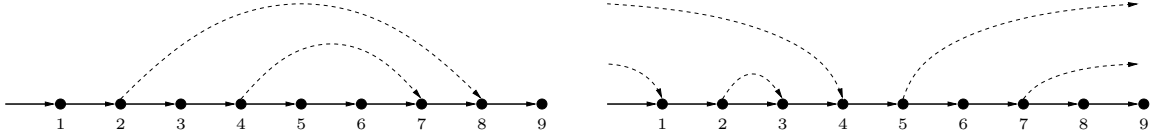
\begin{figure}
\begin{center}
\input{nw.pstex_t}
\caption{Sample nested words}
\label{fig:shape}
\end{center}
\end{figure}

Figure~\ref{fig:shape} shows two finite nested words (without the
labeling with alphabet symbols).  Nesting edges are drawn using dashed
lines.  For the first word, the relation $\M$ is $\{(2,8), (4,7)\}$,
the set $\call$ is $\{2,4\}$, and the set $\retr$ is $\{7,8\}$.
For the second word, the
relation $\M$ is $\{(2,3)\}$, the set $\call$ is $\{2,5,7\}$, and the
set $\retr$ is $\{1,3,4\}$.

Note that our definition allows a nesting edge from a position $i$ to
its linear successor, and in that case there will be two edges from
$i$ to $i+1$; this is the case for positions 2 and 3 of the second
sequence.  The second sequence has two pending calls and two pending
returns.  Pending calls are depicted by dashed outgoing edges and
pending returns are depicted by dashed incoming edges.  Note that
all pending return positions in a nested word appear before any of the
pending call positions (this is enforced by condition (3) of the
definition of matchings).

\subsection{Games and types}

The {\em quantifier rank} (or {\em quantifier depth}) of
an $\FO$ formula $\phi$ is the depth of quantifier nesting in $\phi$.
 The  {\em rank-$k$ type} of a structure $\MM$ over a relational vocabulary
is the set $\{\phi\mid \MM\models\phi \text{ and
 the quantifier rank of $\phi$ is $k$}\}$, 
where $\phi$ ranges over FO sentences over the vocabulary.
It is well-known that there are finitely many rank-$k$ types for all
$k$, and  
for each rank-$k$ type $\tau$ there is an $\FO$ sentence 
$\phi_\tau$ such that $\MM\models\phi_\tau$ iff
the rank-$k$ type of $\MM$ is $\tau$. 
Sometimes we 
associate types with formulas that define them.

Many proofs in this paper make use of {\em Ehrenfeucht-Fra\"iss\'e}
(EF) games, see for example \cite{I}. This game is played on two structures, $\MM$ and $\MM'$,
over the same vocabulary, by two players, {\em \spoiler} and {\em
\dupl}.  In round $i$ \spoiler\ selects a structure, say $\MM$, and an
element $c_i$ in the domain of $\MM$; \dupl\ responds by selecting an
element $e_i$ in the domain of $\MM'$.  \dupl\ {\em wins} in $k$
rounds, for $k \geq 0$, if $\{(c_i,e_i) \mid i \leq k\}$ defines a
partial isomorphism between $\MM$ and $\MM'$.  Also, if $\bar a$ is an
$m$-tuple in the domain of $\MM$ and $\bar b$ is an $m$-tuple in the
domain of $\MM'$, where $m \geq 0$, we write $(\MM,\bar a) \equiv_k
(\MM',\bar b)$ whenever \dupl\ wins in $k$ rounds no matter how
\spoiler\ plays, but starting from position $(\bar a,\bar b)$.

We write $\MM \equiv_k \MM'$ iff $\MM$ and $\MM'$ have the same
rank-$k$ type, that is for every $\FO$ sentence $\phi$ of quantifier
rank-$k$, $\MM \models \phi \Leftrightarrow \MM' \models \phi$. 
It is well-known that $\MM \equiv_k \MM'$ iff \dupl\ has a
winning strategy in the $k$-round Ehrenfeucht-Fra\"iss\`e game on
$\MM$ and $\MM'$. 

In the proof of Theorem \ref{3var th}, we shall also use {\em
  $k$-pebble games}. In such a game, \spoiler\ and \dupl\ have access
  to
 $k$ matching pebbles each, and each round consists of \spoiler\ 
  either removing, or placing, or replacing a pebble in one structure,
  and \dupl\ replicating the move in the other structure. The
  correspondence given by the matching pebbles should be a partial
  isomorphism. If \dupl\ can play while maintaining partial
  isomorphism for $m$ rounds, then the structures agree on all $\FO^k$
  sentences of quantifier rank up to $m$; if \dupl\ can play while
  maintaining partial
  isomorphism forever, then the structures agree on all $\FO^k$
  sentences.  ($\FO^k$ is first-order logic where at most $k$
  distinct variables may occur.)

\section{Temporal Logics over Nested Words}
\label{tl-sec}

\noindent We now describe our approach to temporal logics for nested words. It
is similar to the approach taken by the logic
$\caret$ \cite{AEM04}. Namely, we shall consider LTL-like
logics that define the next/previous and until/since operators for
various types of paths in nested words.

All the logics will be able to refer to propositional letters,
including the base unary relations $\call$ and $\retr$,
and will be closed under all Boolean  combinations. We shall write $\top$
for true and $\bot$ for false. For all the logics, we shall define the
notion of satisfaction with respect to a position in a nested word: we
write $(\w,i)\models\phi$ to denote that the formula $\phi$ is true in 
position $i$ of the word $\w$.

Since nested words are naturally represented as transition
systems with two binary relations -- the successor and the matching
relation -- in all our logics we introduce {\em next operators} $\dm$
and $\dmm$. The semantics of those is standard: $(\w,i)\models\dm
\phi$ iff $(\w,i+1)\models\phi$, 
$(\w,i)\models\dmm\phi$
iff $i$ is a call with a matching return $j$ (i.e., $\M(i,j)$ holds)
and $(\w,j)\models\phi$.
Likewise, we shall have {\em past} operators $\dmminus$ and $\dmmminus$:
that is,  $\dmminus \phi$ is true in position
$i>1$ iff $\phi$ is true in position $i-1$, and $\dmmminus\phi$ is true
in position $j$ 
if $j$ is a return position with matching call $i$ and $\phi$ is true at  $i$.

\subsection{Paths in Nested Words}

The {\em until/since operators} depend on what a path is.
In general, there are various notions of paths through a nested word.
We shall consider until/since operators for paths that are unambiguous: that is, for
every pair of positions $i$ and $j$ with $i < j$, there could be at
most one path between them. 
Then, with respect to any such given notion of a path, we
have the until and since operators with the usual semantics:
\begin{enumerate}[$\bullet$]
\item $(\w,i)\models\phi\U\psi$ iff there is a position $j\geq
  i$ and a path $i=i_0 < i_1 < \ldots < i_k=j$ between them such that
  $(\w,j)\models\psi$ and $(\w,i_p)\models\phi$ for every $0\leq
  p < k$.
\item $(\w,i)\models\phi\S\psi$ iff there is a position $j\leq
  i$ and a path $j=i_0 < i_1 < \ldots < i_k=i$ between them such that
  $(\w,j)\models\psi$ and $(\w,i_p)\models\phi$ for every $0<
  p \leq k$.
\end{enumerate}

The approach of $\caret$ was to introduce three types of paths, based on
the linear successor (called {\em linear paths}), the call-return
relation (called {\em abstract paths}), and  the innermost
call relation (called {\em call paths}).

To define those, 
we need the notions $\C(i)$ and $\R(i)$ for each position $i$
-- these are the innermost call within which the current action $i$ is
executed, and its corresponding return. 
Formally, $\C(i)$ is the greatest
matched call position $j < i$ whose 
matching return is after $i$ (if such
a call position exists), and $\R(i)$ is the least matched 
return position $\l > i$ whose matching call is before $i$.

\begin{definition}[Linear, call and abstract paths]
Given two
 positions $i < j$, a sequence $i=i_0 < i_1 < \ldots < i_k=j$
is 
\begin{enumerate}[$\bullet$]
\item a {\em linear path} if $i_{p+1}=i_p + 1$ for all $p< k$;
\item a {\em call path} if $i_p=\C(i_{p+1})$ for all $p < k$;
\item an {\em abstract path} if $$i_{p+1}=\begin{cases} 
r(i_p)\text{ if }i_p\text{ is a matched call}\\
i_p+1 \text{ if }i_p \text{ is not a call and } i_p+1 \text{ is not a
  matched return.}
\end{cases}$$
\end{enumerate}
We shall denote until/since operators corresponding to these paths by 
$\Ul/\Sl$ for linear paths, $\Uc/\Sc$ for call paths, and $\Ur/\Sr$
for abstract paths.
\end{definition}

Our logics will have some of the next/previous and until/since operators. 
Some examples are:
\begin{enumerate}[$\bullet$]
\item When we 
restrict ourselves to the purely linear fragment, our operators are
$\dm$ and $\dmminus$, and $\Ul$ and $\Sl$, i.e., precisely LTL 
(with past operators).
\item
The logic $\caret$ \cite{AEM04} has the following operators:
the next operators $\dm$ and $\dmm$;
the linear and abstract untils (i.e., $\Ul$ and $\Ur$), the call
since (i.e., $\Sc$) and a previous operator $\dmminus_c$,
defined by:  
$(\w, i) \models \dmminus_c \phi$ iff $\C(i)$ is defined and $(\w,
  \C(i)) \models \phi$.
\end{enumerate}

Another notion of a path combines both the linear and the nesting
structure. It is the shortest directed path between two positions 
$i$ and $j$. Unlike
an abstract path, it decides when to skip a 
call based on position $j$. Basically, a summary path from $i$ to $j$ moves along successor
edges until it finds a call position $k$. 
If $k$ has a matching return $\l$ such that $j$ appears after $\l$,
then the summary path skips the entire call from $k$ to $\l$ and
continues from $\l$; otherwise 
the path continues as a successor path.
Note that every abstract path is a summary path, but there are summary paths that are
not abstract paths.

\begin{definition}
A {\em summary path} between $i < j$ in a nested word $\w$ is a sequence
$i=i_0 < i_1 < \ldots < i_k=j$ such that for all $p < k$, 
$$i_{p+1} = 
\begin{cases}
r(i_p) \text{ if } i_p\text{ is a 
matched call and }j \geq r(i_p) \\
i_p+1 \text{ otherwise}\end{cases}$$
The corresponding until/since operators are denoted by $\Up$ and
$\Sp$. 
\end{definition}

We will also consider two special kinds of summary paths: {\em
summary-down\/} paths are allowed to use only {\em call} edges (from a
call position, $i$ to $i+1$ where $i+1$ is not a return), {\em nesting} edges (from
a call to its matching return), and {\em internal} edges (from some
$i$ to $i+1$ where $i$ is not a call and $i+1$ is not a return), and {\em
summary-up\/} paths are allowed to use only {\em return} edges (from a
position preceding a return to the return), nesting edges and internal
edges.  (In other words, summary-down paths are summary paths with no
return edges and summary-up paths are summary paths with no call edges.)

We will use $\dpuntil$ and $\upuntil$ to denote the
corresponding until operators.  
A general summary path is a concatenation of a summary-up path and
summary-down path: $\phi\puntil\psi$ is equivalent to
$\phi\upuntil(\phi\dpuntil\psi)$.

We will also study the expressiveness of various until modalities when
the logic is extended with the {\em within\/} operator, which allows
restriction to a subword.  If $\phi$ is a formula, then $\WW\phi$ is a
formula, and $(\w,i)\models\WW\phi$ iff $i$ is a call, and
$(\w[i,j],i)\models\phi$, where $j=r(i)$ if $i$ is a matched call,
$j=|\w|$ if $i$ is an unmatched call and $\w$ is finite, and $j =
\infty$ otherwise.  In other words, $\WW\phi$ evaluates $\phi$ on a
subword restricted to a single procedure.

To understand the various notions of paths in a nested word, let us
consider the left word shown in Figure~\ref{fig:shape} again.  An
abstract path 
uses internal and nesting edges; for
example, $\langle 1,2,8,9\rangle$ and $\langle 3,4,7\rangle$ are
abstract paths.  Summary-down paths, in addition, can use call edges;
for example, $\langle 1,2,3,4,7\rangle$ is a summary-down
(but not an abstract) path.  Summary-up paths 
can use internal and nesting edges, and can also go 
along return edges; for example, $\langle 3,4,7,8,9\rangle$ is a
summary-up path.  A summary path is a summary-up path followed by a
summary-down path; for example, $\langle 3,4,5,6,7\rangle$ 
in the right word of Figure ~\ref{fig:shape} is a
summary path (which also happens to be a linear path).  
Every two positions have a unique
summary path connecting them, and this is the ``shortest'' path in the
underlying graph between these positions.

\subsection{Specifying Requirements}
We now discuss how the various operators can be used for
specifying requirements for sequential structured programs.
In the classical linear-time semantics of programs,
an execution of a program is modeled as a word over program states.
In the nested-word semantics, this linear structure 
is augmented with nesting edges from
entries to exits of program blocks.
The main benefit is that using nesting edges one can 
skip procedure calls entirely, and continue to
trace a local path through the calling procedure.
A program is now viewed as a generator of nested words,
and requirements are written using temporal logics over nested words.

Suppose we want to express the requirement that, along a global
program execution, every write to a variable is followed by a read
before the variable is written again.  If $wr$ and $rd$ denote the
atomic propositions that capture write and read operations,
respectively, then the requirement is expressed by the until formula
over linear paths,

 \[ \always\ [\ wr\ \rightarrow (\neg\ wr)\ \until\
rd\ ]\] Here, $\always$ is defined in the usual manner from the linear
until: $\always\phi$ stands for $\neg(\top\,\until\neg\phi)$.  This
property is clearly already expressible in LTL and does not use nesting
edges at all.

Now let us review some of the properties expressible in the nested
call-return logic \caret{} of \cite{AEM04}, but
not expressible in LTL.  In the classical verification
formalisms such as Hoare logic, correctness of procedures is expressed
using pre and post conditions.  Partial correctness of a procedure $A$
specifies that if the pre-condition $p$ holds when the procedure $A$
is invoked, then if the procedure terminates, the post-condition $q$
is satisfied upon return. Total correctness, in addition, requires the
procedure to terminate.  Assume that all calls to the procedure $A$
are characterized by the proposition $p_A$. Then, the requirement
\[\always\ [\, (\call\ \wedge\ p\ \wedge\ p_A)\ \rightarrow \dmm\ q\
]\] expresses the total correctness, while \[ \always\ [\, (\dmm\top\
\wedge\ p\ \wedge\ p_A)\ \rightarrow \dmm\ q\ ]\] expresses the
partial correctness.  Both these specifications crucially rely upon
the abstract-next operator.

The abstract path starting at a position inside a procedure $A$ is
obtained by successive applications of internal and nesting edges,
 and skips over invocations of
other procedures called from $A$. 
Using the abstract versions of temporal operators, we 
 can specify properties of such
abstract paths. For example, suppose we want to specify that
if a procedure writes to a variable, then it (that is, the same 
invocation of the same procedure) will later read it and do so before writing to it again.
The requirement is expressed by the
until formula over abstract paths
\[ \always\ [\ wr\ \rightarrow \dm (\, \neg\ wr\,)\ \Ur\ rd\ ]\]

The call since-path starting at a position inside a procedure $A$ is
obtained by successively jumping to the innermost call positions, and
encodes the active stack at that position.  Stack inspection can
specify a variety of security properties. For instance, the
requirement that a procedure $A$ should be invoked only within the
context of a procedure $B$, with no intervening call to an overriding
module $C$, is expressed by the formula \[ \always\ [\ \call\ \wedge\
p_A\ \rightarrow\ (\neg p_C)\ \Sc\ p_B\ ].\]

Finally, we turn to {\em scope-bounded linear-path\/} properties.  For
a procedure, the corresponding scope-bounded linear path is the linear
path (that is, the path obtained by following linear edges) from its
call to it return.  That is, a scope-bounded path corresponding to a
procedure $P$ includes the executions of the procedures (transitively)
called by $P$, but terminates when the current invocation of $P$
returns.  Properties about scope-bounded paths are useful in asserting
contracts for modules.

Suppose we want to assert that a procedure $A$, and the procedures it
calls, do not write to a variable before it returns.  This is an
invariant of the scope-bounded path, and is captured by the formula:
\[\always\ [\ (\call\ \wedge\ p_A)\ \rightarrow\ \WW\ (\ \always\
\neg\ wr\ )\ ] \] 
Recall that the within operator $\WW$ restricts the evaluation of a
formula to a single procedure call. 
The same requirement can also be captured using summary
paths. It is even easier to state it using summary-down paths:
  \[\always\ [\ (\call\ \wedge\ p_A)\ \rightarrow\ \neg\ (\ \top\
\dpuntil\ wr\ )\ ]\]

Suppose we want to specify the requirement that if a procedure writes
to a variable then it is read along the scope-bounded path before
being written again.  We can use the within modality to express this
property: \[ \always\ [\ \call\ \rightarrow \ \WW\ \always\ (\ wr\
\rightarrow \dm (\neg\ wr)\ \until\ rd\ )\ ]\] This requirement can also
be alternatively specified using summary-down paths as follows:
\[\always\ [\ wr\ \rightarrow\ \dm (\ \neg\ wr\ \wedge\ (\retr\
\rightarrow\ \dmmminus\neg \top \dpuntil\ wr\ )\ )\ \dpuntil\ rd\ ]\]
The formula says that from every write operation, there is a read
operation along some summary-down path (and thus, within the same
scope) such that along the path, there is no write, and if the path
uses a summary edge, then the enclosed subword also does not contain a
write.

It is easy to see that the above requirements 
concerning scope-bounded paths are specifiable in the first-order
logic of nested words. It is conjectured that they
are not specifiable in \caret{}.

\section{Expressive Completeness}
\label{expcompl-sec}

\noindent In this section, we study logics that are expressively complete for
\FO, i.e. temporal logics that have exactly the same power as \FO\
formulas with one free variable over finite and infinite nested words.
In other words, for every formula $\phi$ of an expressively complete
temporal logic there is an \FO\ formula $\phi'(x)$ such that
$(\w,i)\models\phi$ iff $\w\models\phi'(i)$ for every nested word $\w$
and position $i$ in it, and conversely, for every \FO\ formula $\psi(x)$ 
there is a temporal formula $\psi'$ such that $\w\models\psi(i)$ iff
$(\w,i)\models\psi'$. 

Our starting point is a logic $\nwtl$ (nested-word temporal logic) based
on summary paths introduced in the previous section. We show that this
logic is expressively complete for FO, and of course remains expressively
complete with the addition of other first-order expressible operators
which may be  useful for
verification of properties of procedural programs. 
When we provide upper bounds on the complexity of model checking
for $\nwtl$, we shall in fact show that the upper bounds hold
with respect to an extension, $\nwtlp$,  which includes
a number of additional operators.

We then look at logics close to those in the verification literature,
i.e., with operators such as {\em call} and {\em abstract until} and {\em since}, and ask
what needs to be added to them to get expressive completeness. We
confirm a conjecture of \cite{AEM04} that a {\em within} operator is
sufficient.  Such an operator evaluates a formula on a nested
subword.
We then discuss the role of this within operator. We show that, if
added to $\nwtl$, it does not increase expressiveness, but makes the
logic exponentially more succinct.

\subsection{Expressive completeness and NWTL}

The logic $\nwtl$ ({\em nested words temporal logic}) has next and
previous operators, as well as until and since with respect to summary
paths. That is, its formulas are given by:
$$
\begin{array}{rcl} 
  \phi,\phi' & := &  \top\ \mid\ a \ \mid\ \call\ \mid\ \retr\ \mid \ \neg \phi \ \mid \ \phi \vee \phi' \
  \mid \\ 
&& \dm \phi \ \mid \  \dmm \phi \ \mid \ \dmminus \phi \ \mid\
  \dmmminus \phi \ \mid\\
&& \phi \Up \phi'  \ \mid \   \phi\Sp\phi'
\end{array}
$$
where $a$ ranges over $\Sigma$. We use abbreviations
$\intt$ for $\neg\call\wedge\neg\rett$ (true in an internal
position). Note that in the absence of pending calls and returns,
$\call$ and $\rett$ are definable as $\dmm\top$ and $\dmmminus\top$,
respectively.

\begin{theorem}
\label{nwtl-thm}
$\nwtl \ = \ \FO$ over both finite and infinite nested words.
\end{theorem}

\medskip
\noindent
{\em Proof}.
 We start with the easy direction $\nwtl\subseteq \FO$.

\begin{lemma}
\label{nwtl-to-fo-lemma}
For every $\nwtl$ formula $\phi$, there exists an \FO\ formula
$\alpha_\phi(x)$ that uses at most three variables $x,y,z$ such that
for every nested word $\w$ (finite or infinite), and every position,
$i$ in $\w$, we have
$(\w,i)\models\phi$ iff $\w\models\alpha_\phi(i)$.
\end{lemma}

\aProof{Lemma \ref{nwtl-to-fo-lemma}}{The proof is by induction
on the formulas and very simple for all the cases except $\Up$ and
$\Sp$: for example,
$$\alpha_{\dmm\phi}(x) \ =\ \exists y\ \big(\M(x,y) \wedge \exists
x\ (x=y \wedge \alpha_\phi(x))\big).$$ 

For translating $\Up$, we need a few auxiliary formulas. Our first
goal is to define a formula $\gamma_r(x,z)$ saying that $x$ is
$\R(z)$, i.e. the return of the innermost call within which $z$ is
executed. For that, we start with 
$\delta(y,z) \ = \ z<y\wedge\retr(y)\wedge \exists x\
(\M(x,y) \wedge x < z )$ saying that $y$ is a return that
is preceded by $z$ and whose matching call precedes $z$, 
that is, $y$ is a
candidate for $\R(z)$. Then the formula $\gamma_r(x,z)$ is given by
$$\exists y\ (y = x \wedge \delta(y,z)) \wedge 
\forall y\ (\delta(y,z) \to\ y \geq x).$$
Likewise, we define $\gamma_c(y,z)$ stating that that $y$ equals
$\C(z)$, that is, the innermost call within which $z$ is executed. 
Now define 
$$\chi_1(y,z) \ = \ \exists x\ \big(\gamma_r(x,z) \wedge x
\leq y\big), \ \ \ \  \chi_2(x,z) \ = \ \exists y\ \big(\gamma_c(y,z)
\wedge y \geq x\big)$$ 
and $\chi(x,y,z)$ as $\chi_1(y,z) \wedge \chi_2(x,z)$. Then this
formula says that the summary path from $x$ to $y$ does not pass
through $z$, assuming $x < z < y$. With this,
$\alpha_{\phi\Up\psi}(x)$ is given by
\begin{multline*}
\alpha_\psi(x) \vee \exists y\ 
\bigg(y > x \wedge \alpha_\varphi(x) \wedge \exists x\ (x=y
\wedge \alpha_\psi(x)) \ \wedge\\
\forall z\ \big((x < z < y \wedge \neg\chi(x,y,z)) \to
\exists x\ (x=z \wedge \alpha_\phi(x))\big)\bigg)
\end{multline*}
The proof for $\phi\Sp\psi$ is similar. This concludes
 the proof of
the lemma.}

In the proof of the other direction, $\FO\subseteq \nwtl$, we shall
use a tree representation of nested words. 
The translation is 
essentially the same as in \cite{nested}. For each nested word $\w$ we
have a binary tree $T_\w$ (i.e., its nodes are elements of $\{0,1\}^*$)
and a function $\wt: \w \to T_\w$ that maps each
position of $\w$ to a node of $T_\w$ as follows:
\begin{enumerate}[$\bullet$]
\item the first position of $\w$ is mapped into the root of $T_\w$;
\item if $s=\wt(i)$ then:
\begin{enumerate}[(1)]
\item if $i$ is an internal, or an unmatched call, or a matched call
  whose return is the last position of $\w$, or an unmatched
  return,
  and $i$ is not the last position of
$\w$, then $s$ has only child $s\cdot 0$ and $\wt(i+1)=s\cdot 0$;
\item if $i$ is a matched call whose return is not the last position
  in $\w$, then $s$ has both children $s\cdot 0$ and
  $s \cdot 1$ and $\wt(r(i)+1)=s\cdot 0$, and 
  $\wt(i+1)=s\cdot 1$.
\item
 if $i$ is a matched return, then $s$ has no children.
\end{enumerate}
\end{enumerate}
The $\Sigma$-labels of $i$ and $\wt(i)$ are the same. 
If $i$ was a pending call, we label $\wt(i)$ with $\pcall$, and if $i$
was a pending return, we label $\wt(i)$ with $\pret$. 

\begin{figure*}
\begin{center}

\setlength{\unitlength}{0.001in}
\begingroup\makeatletter\ifx\SetFigFont\undefined%
\gdef\SetFigFont#1#2#3#4#5{%
  \reset@font\fontsize{#1}{#2pt}%
  \fontfamily{#3}\fontseries{#4}\fontshape{#5}%
  \selectfont}%
\fi\endgroup%
{\renewcommand{\dashlinestretch}{30}
\begin{picture}(4397,2527)(0,1010)
\put(3196,1357){\makebox(0,0)[lb]{{\SetFigFont{10}{12.0}{\rmdefault}{\mddefault}{\updefault}5}}}
\put(721.000,1638.250){\arc{487.500}{3.5364}{5.8884}}
\put(46,1725){\blacken\ellipse{40}{40}}
\put(46,1725){\ellipse{40}{40}}
\put(496,1725){\blacken\ellipse{40}{40}}
\put(496,1725){\ellipse{40}{40}}
\put(946,1725){\blacken\ellipse{40}{40}}
\put(946,1725){\ellipse{40}{40}}
\put(1396,1725){\blacken\ellipse{40}{40}}
\put(1396,1725){\ellipse{40}{40}}
\put(1846,1725){\blacken\ellipse{40}{40}}
\put(1846,1725){\ellipse{40}{40}}
\put(2296,1725){\blacken\ellipse{40}{40}}
\put(2296,1725){\ellipse{40}{40}}
\put(3646,2330){\blacken\ellipse{40}{40}}
\put(3646,2330){\ellipse{40}{40}}
\put(3346,2025){\blacken\ellipse{40}{40}}
\put(3346,2025){\ellipse{40}{40}}
\put(3346,1425){\blacken\ellipse{40}{40}}
\put(3346,1425){\ellipse{40}{40}}
\put(3646,1725){\blacken\ellipse{40}{40}}
\put(3646,1725){\ellipse{40}{40}}
\put(3946,2025){\blacken\ellipse{40}{40}}
\put(3946,2025){\ellipse{40}{40}}
\put(4246,1725){\blacken\ellipse{40}{40}}
\put(4246,1725){\ellipse{40}{40}}
\path(46,1732)(496,1732)
\blacken\path(376.000,1702.000)(496.000,1732.000)(376.000,1762.000)(376.000,1702.000)
\path(496,1732)(946,1732)
\blacken\path(826.000,1702.000)(946.000,1732.000)(826.000,1762.000)(826.000,1702.000)
\path(946,1732)(1396,1732)
\blacken\path(1276.000,1702.000)(1396.000,1732.000)(1276.000,1762.000)(1276.000,1702.000)
\path(1396,1732)(1846,1732)
\blacken\path(1726.000,1702.000)(1846.000,1732.000)(1726.000,1762.000)(1726.000,1702.000)
\path(1846,1732)(2296,1732)
\blacken\path(2176.000,1702.000)(2296.000,1732.000)(2176.000,1762.000)(2176.000,1702.000)
\path(3646,2332)(3346,2032)
\blacken\path(3409.640,2138.066)(3346.000,2032.000)(3452.066,2095.640)(3409.640,2138.066)
\path(3646,2332)(3946,2032)
\blacken\path(3839.934,2095.640)(3946.000,2032.000)(3882.360,2138.066)(3839.934,2095.640)
\path(3946,2032)(4246,1732)
\blacken\path(4139.934,1795.640)(4246.000,1732.000)(4182.360,1838.066)(4139.934,1795.640)
\path(3946,2032)(3646,1732)
\blacken\path(3709.640,1838.066)(3646.000,1732.000)(3752.066,1795.640)(3709.640,1838.066)
\path(3646,1732)(3346,1432)
\blacken\path(3409.640,1538.066)(3346.000,1432.000)(3452.066,1495.640)(3409.640,1538.066)
\put(26,1499){\makebox(0,0)[lb]{{\SetFigFont{10}{12.0}{\rmdefault}{\mddefault}{\updefault}1}}}
\put(486,1499){\makebox(0,0)[lb]{{\SetFigFont{10}{12.0}{\rmdefault}{\mddefault}{\updefault}2}}}
\put(946,1499){\makebox(0,0)[lb]{{\SetFigFont{10}{12.0}{\rmdefault}{\mddefault}{\updefault}3}}}
\put(1396,1499){\makebox(0,0)[lb]{{\SetFigFont{10}{12.0}{\rmdefault}{\mddefault}{\updefault}4}}}
\put(1846,1499){\makebox(0,0)[lb]{{\SetFigFont{10}{12.0}{\rmdefault}{\mddefault}{\updefault}5}}}
\put(2296,1499){\makebox(0,0)[lb]{{\SetFigFont{10}{12.0}{\rmdefault}{\mddefault}{\updefault}6}}}
\put(3571,2407){\makebox(0,0)[lb]{{\SetFigFont{10}{12.0}{\rmdefault}{\mddefault}{\updefault}1}}}
\put(3196,2032){\makebox(0,0)[lb]{{\SetFigFont{10}{12.0}{\rmdefault}{\mddefault}{\updefault}6}}}
\put(4021,2032){\makebox(0,0)[lb]{{\SetFigFont{10}{12.0}{\rmdefault}{\mddefault}{\updefault}2}}}
\put(4321,1732){\makebox(0,0)[lb]{{\SetFigFont{10}{12.0}{\rmdefault}{\mddefault}{\updefault}3}}}
\put(3496,1732){\makebox(0,0)[lb]{{\SetFigFont{10}{12.0}{\rmdefault}{\mddefault}{\updefault}4}}}
\put(946.000,1063.250){\arc{2242.522}{3.7806}{5.6441}}
\end{picture}
}

\caption{A nested word and its tree translation}
\label{transl-fig}
\end{center}
\end{figure*}
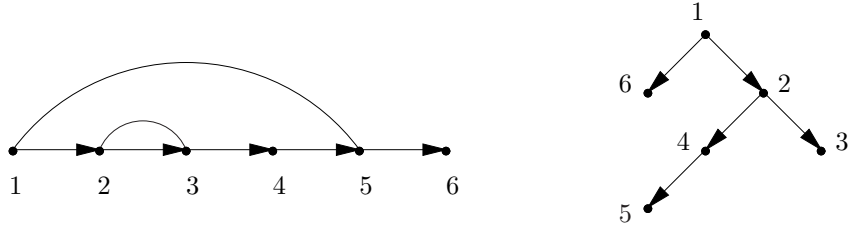

Note that $\wt$
is a bijection, and that labels $\pcall$ and $\pret$ may only occur on
the leftmost branch of $T_\w$. An example of a nested word and its
translation are given in Fig.~\ref{transl-fig}.

To relate paths in nested words and paths in their tree translations,
we introduce the notions of semi-strict and strict paths. Intuitively,
a semi-strict path in a nested word corresponds to a path on
its tree translation that, in addition to following tree edges, can
jump from a node with no children to its successor in the depth-first
traversal of the tree (where depth-first starts with the right subtree
and then moves to the left subtree). A strict path is just a
path that follows tree edges.  These are both slight modifications of
summary paths.

More precisely, 
a {\em semi-strict path}
between positions $i$ and $j$, with $i < j$, in a nested word $\w$, is
a sequence $i = i_0 < i_1 < \cdots < i_k = j$ such that
$$i_{p+1} = 
\begin{cases} 
r(i_p)+1 \text{ if } i_p\text{ is a 
matched call and }j > r(i_p) \\
i_p+1 \text{ otherwise.}
\end{cases}$$
That is, when skipping a call, instead of jumping to the matching return
position, a semi-strict path will jump to its successor. 

A {\em strict path} is a semi-strict path $i = i_0 <
i_1 < i_2 < \cdots < i_k = j$ in which no $i_p$ with $p < k$ is a
matched return position. In other words, a strict path stops if it
reaches a matched return position. In particular there may be positions $i <
j$ in a nested word such that no strict path exists between
them.  

For example, in Fig.~\ref{transl-fig}, $\langle 2,4,5,6 \rangle$ is a
semi-strict path. Although $\langle 2,4,5,6 \rangle$ is not a path in
the tree (we jump from 5 to 6), this is allowed under the definition
of semi-strict paths. Strict paths are exactly the paths on the
tree; for example, $\langle 1,2,4,5 \rangle$ is such a path.

The until/since operators for semi-strict paths and
strict paths will be denoted by $\Uss/\Sss$ and $\Us/\Ss$,
respectively. Versions of $\nwtl$ in which $\Up/\Sp$ are replaced by
$\Uss/\Sss$ ($\Us/\Ss$) will be denoted by $\nwtlss$ ($\nwtls$).

We will use $\mret$ for $\retr\wedge\prev_{\M}\top$, 
and $\mcall$ for $\call\wedge\next_{\M}\top$, to capture matching return and call
positions, respectively.

The proof is based on two lemmas.

\begin{lemma}
\label{nwtl-lemma-one}
$\nwtls \subseteq \nwtlss \subseteq \nwtl$.
\end{lemma}

\begin{lemma}
\label{nwtl-lemma-two}
$\FO\subseteq \nwtls$.
\end{lemma}

This of course implies the theorem: $\nwtl \subseteq \FO \subseteq
\nwtls \subseteq \nwtlss \subseteq \nwtl$. Note that as a corollary we
also obtain $\nwtls=\nwtlss=\FO$. 

\aProof{Lemma \ref{nwtl-lemma-one}}{For translating an  $\nwtls$
formula $\phi$ into an equivalent formula $\alpha_\phi$ of $\nwtlss$
we need to express $\psi \Us \theta$ with $\Uss$, which is simply
$(\alpha_\psi \wedge \neg\mret) \Uss \alpha_\theta$, and likewise for
the since operators. For translating each $\nwtlss$ formula $\phi$
into an equivalent $\nwtl$ formula $\beta_\phi$, again we need to consider
only the case of until/since operators. The formula $\psi \Uss \theta$ is
translated into
{\small
\begin{multline}
\label{trans-eq}
{
\beta_\theta \vee \bigg(\beta_\psi \wedge \bigg(\bigg((\beta_\psi \vee
\rett) \wedge (\neg \mcall \to \dm \beta_\psi) \wedge (\mcall \to (\dmm
\dm \beta_\psi \vee \dmm \dm \beta_\theta))\bigg) \Up} \\ 
{
\bigg((\beta_\psi \vee \rett) \wedge (\neg \mcall \to \dm \beta_\theta)
\wedge (\mcall \to (\dm \beta_\theta \vee \dmm \dm \beta_\theta \vee
\dm (\neg \rett \wedge \gamma)))\bigg)\bigg)\bigg),}
\end{multline}
where $\gamma$ is a formula defined as follows:
\begin{multline*}
{
\bigg((\beta_\psi \vee
\rett) \wedge (\neg \mcall \to \dm \beta_\psi) \wedge (\mcall \to (\dm
\beta_\psi \vee \dmm \dm \beta_\psi)) \wedge (\dm \rett \to
\call) \bigg)\ \Up}\\ 
{
\bigg((\beta_\psi \vee \rett) \wedge (\neg\mcall \to \dm \beta_\theta)
\wedge (\mcall \to (\dm \beta_\theta \vee \dmm  \dm  \beta_\theta))\bigg)}
\end{multline*}
}
The idea is that we split a semi-strict path into a semi-strict up
path (where call edges are excluded) followed by a semi-strict down
path (where return edges are excluded).  The first Until in
(\ref{trans-eq}) captures the semi-strict up path and the second Until
in $\gamma$ captures the semi-strict down path.
The translation for $\Sss$ is similar. 

The proof that the translation is correct is a rather detailed case
analysis which we have relegated to the appendix.}

\aProof{Lemma \ref{nwtl-lemma-two}}{We start with the finite case,
and then show how the inclusion extends to nested $\omega$-words.

As a tool we shall need a slight modification of a result from
\cite{Schl92,marx-pods04} providing an expressively complete temporal
logic for trees with at most binary branching.  We consider binary
trees whose domain $D$ is a prefix-closed subset of $\{0,1\}^*$, and
we impose a condition that if $s\cdot 1\in D$ then $s\cdot 0\in D$.
When we refer to \FO\ on trees, we assume they have two successor
relations $S_0, S_1$ and the descendant relation $\preceq$ (which is
just the prefix relation on strings) plus the labeling predicates,
which include two new labels $\pcall$ and $\pret$ (for pending calls
and returns). Each node can be labeled 
by either a letter from $\Sigma$, or by a letter from $\Sigma$ and
$\pcall$, or by a letter from $\Sigma$ and $\pret$ (i.e. labels
$\pcall$ and $\pret$ need not be disjoint from other labels).

We also consider the following logic $\tltree$:
$$
\begin{array}{rcl}
\phi & \mbox{:=} & a\ \mid\ \phi\vee \phi\ \mid\ \neg\phi\ \mid\\
&& \Xd\phi \ \mid\ \Yd\phi \ \mid \Xr\phi\ \mid\ \Yr\phi\ \mid\\
&& \phi\Ud\phi\ \mid\ \phi\Sd\phi
\end{array}
$$
where $a$ ranges over $\Sigma\cup\{\pcall,\pret\}$, with the following semantics:
\begin{enumerate}[$\bullet$]
\item $(T,s)\models \Xd\phi$ iff $(T,s\cdot i)\models \phi$ for some $i\in\{0,1\}$;
\item $(T,s\cdot i)\models \Yd\phi$ iff $(T,s)\models \phi$ (where
  $i$ is either $0$ or $1$);
\item $(T,s\cdot 0)\models \Xr\phi$ iff $(T,s \cdot 1)\models \phi$;
\item $(T,s\cdot 1)\models \Yr\phi$ iff $(T,s \cdot 0)\models \phi$;
\item $(T, s)\models \phi\Ud \psi$ iff there exists $s'$ such that $s
  \preceq s'$, $(T,s')\models \psi$, and $(T,s'')\models \phi$ for all
  $s''$ such that $s \preceq s'' \prec s'$;
\item $(T, s)\models \phi\Sd \psi$ iff there exists $s'$ such that $s'
  \preceq s$, $(T,s')\models \psi$, and $(T,s'')\models \phi$ for all
  $s''$ such that $s' \prec s'' \preceq s$.
\end{enumerate}

\begin{lemma}
\label{marx-lemma}
(see \cite{marx-pods04})\ For unary queries over finite binary trees,
$\tltree=\FO$.
\end{lemma}

This lemma is an immediate corollary of expressive completeness of
logic ${\mathcal X}_{\text{\it until}}$ from \cite{marx-pods04} on
ordered unranked trees, as for a fixed number of siblings, the until
and since operators can be expressed in terms of the next and previous
operators. The result of \cite{marx-pods04} applies to arbitrary
alphabets, and thus in particular to our labeling that may use
$\pcall$ and $\pret$.

The following is immediate by using the tree representation of nested
words and a straightforward
translation of formulae. 

\begin{claim}
\label{fo-to-fo-claim}
For every $\FO$ formula $\phi(x)$ over nested words there is an \FO\
formula $\phi'(x)$ over trees such that for every nested word $\w$ and
a position $i$ in it, we have $\w\models\phi(i)$ iff
$T_\w\models\phi'(\wt(i))$. 
\end{claim}

In fact the converse, that \FO\ over trees $\T_\w$ can be translated
into \FO\ over nested words, is true too, but we do not need it in this
proof.

Since $\FO=\tltree$ by Lemma \ref{marx-lemma}, all that remains to
prove is the following claim.

\begin{claim} 
\label{tltree-to-nwtl}
For every $\tltree$ formula $\phi$, there exists an $\nwtls$ formula
$\crc\phi$ such that for every nested word $\w$ and every position $i$
in it, we have 
$$(\w,i) \models \crc\phi \  \ \Leftrightarrow\ \ (T_\w,\wt(i))\models\phi.$$
\end{claim}

This is now done by induction, omitting the obvious cases of
propositional letters and Boolean connectives. 
We note that a path down the tree from $\wt(i)$ to $\wt(j)$
corresponds precisely to the strict path from $i$ to $j$ (that
is, if such a strict path is $i=i_0,i_1,\ldots,i_k=j$, then
$\wt(i_0),\wt(i_1),\ldots,\wt(i_k)$ is the path from $\wt(i)$ to
$\wt(j)$ in $T_\w$). Hence, the translations of until and since
operators are:
$$\crc{(\phi\Ud\psi)} = \crc\phi\Us\crc\psi, \ \ \ \ 
\crc{(\phi\Sd\psi)} = \crc\phi\Ss\crc\psi.$$
For translating next and previous operators, and pending calls/returns, define:
\begin{eqnarray*}
\mcall & \equiv & \dmm\top \ \text{(true in a matched call position);}\\
\mret & \equiv & \dmmminus\top \ \text{(true in a matched return position).}
\end{eqnarray*}
Then the rest of the translation is as follows:
\begin{eqnarray*}
\crc{\pcall} & \equiv & \call \wedge \neg\mcall \\
\crc{\pret} & \equiv & \rett \wedge \neg\mret \\
\crc{(\Xd\phi)} & \equiv & \neg\mret \wedge \Big(\dm\crc\phi \vee 
  (\call \wedge \dmm\dm\crc\phi)\Big)\\
\crc{(\Yd\phi)} & \equiv & \Big(\dmminus\rett \wedge
\dmminus\dmmminus\crc\phi\Big) \ \vee\  \Big(\dmminus \neg \mret
\wedge \dmminus\crc\phi\Big)\\ 
\crc{(\Xr\phi)} & \equiv & \dmminus\rett \wedge \dmminus\dmmminus\dm
\crc\phi\\ 
\crc{(\Yr\phi)} & \equiv & \dmminus\call \wedge
\dmminus\dmm\dm\crc\phi
\end{eqnarray*}
Now with the proof completed for finite nested words, we extend it to
the case of nested $\omega$-words. Note that Claim
\ref{fo-to-fo-claim} continues to hold, and Claim
\ref{tltree-to-nwtl} provides a syntactic translation that applies to
both finite and infinite nested words, and thus it suffices to  prove an
analog of Lemma \ref{marx-lemma} for trees of the form $T_\w$, where
$\w$ ranges over nested $\omega$-words. 

If $\w$ is a nested $\omega$-word, then $T_\w$ has exactly one
infinite branch, which consists precisely of all nodes of the form $\wt(i)$ where $i$ is an
{\em outer} position, i.e., not inside any (matched) call. We say that $i$ is
inside a call if there exists a call $j$ with a matching return $k$
such that $j < i \leq k$. If $i$ is an outer position, then we
shall call $\wt(i)$ an {\em outer} node in the tree $T_\w$ as well.

If $i$ is an outer position which is not a matched call, then $i+1$ is
also an outer position and
$\wt(i+1)$ is the left successor of $\wt(i)$. If $i$ is an outer
position and a call with $j > i$ being its matching return, then the
left successor of $\wt(i)$ on the infinite path is
$\wt(j+1)$. Furthermore, the subtree $t^\w(i)$, which has $\wt(i)$ as the
root, plus its right child, and all the descendants of the right child, is
finite and isomorphic to $T_{\w[i,j]}$ (note that $\w[i,j]$ has no
pending calls/returns). If $i$ is an outer position other than a
matched call, we let $t^\w(i)$ be a
single node tree labeled with $i$'s label in $\w$.

Let $\w$ now be a nested $\omega$-word. For each outer position $i$
we let $\tau_m^\w(i)$ be the rank-$m$ type of $t^\w(i)$. If $i$ is not
a matched call, such a type is completely described by $i$'s label
(which consists of a label in $\Sigma$ and potentially $\pcall$ or
$\pret$). 

If $j$ is not
an outer position, and $i$ is an outer position such that $i < j \leq
k$, where $k$ is the matching return of $i$, then $\tau_m^\w(j)$ is
the rank-$m$ type of $(T_\w[i,k],\wt(j))$ (i.e., the type of
$T_\w[i,k]$ with a distinguished node corresponding to $j$).

Next, for a nested $\omega$-word $\w$, let $s$ be a node in $T_\w$
such that $s=\wt(i)$. Let $i_1,i_2,\ldots$ enumerate all the outer
positions of $\w$, and assume that $i_p$ is such that $i_p \leq i <
i_{p+1}$ -- that is, $\wt(i)$ is a node in the subtree $t^\w(i_p)$. We
now define a finite word $\sleft_m(\w,s)$ of length $p-1$ such that
its positions $1,\ldots,p-1$ are labeled $\tau_m^\w(i_1), \ldots,
\tau_m^\w(i_{p-1})$, and an $\omega$-word $\sright_m(\w,s)$ such that
its positions $1,2,\ldots$ are labeled by $\tau_m^\w(i_{p+1}),
\tau_m^\w(i_{p+2}), \ldots$. Next we show:

\begin{claim}
\label{nwtl-comp-claim}
Let $\w, \w'$ be two nested $\omega$-words, and $s=\wt(i), s'=\wt(i')$
two nodes in $T_\w$ and $T_{\w'}$ such that:
\begin{enumerate}[\em(a)]
\item $\sleft_m(\w,s) \efeq_m \sleft_m(\w',s')$;
\item $\sright_m(\w,s) \efeq_m \sright_m(\w',s')$;
\item $\tau_m^\w(i) = \tau_m^{\w'}(i')$.
\end{enumerate}
Then $(T_\w,s) \efeq_m (T_{\w'},s')$.
\end{claim}

\sProof{A standard composition argument shows that \dupl\ wins. If $i_1,i_2,\ldots$
enumerate outer positions in $\w$ and $i_p \leq i < i_{p+1}$, then a
move by \spoiler, say, in $T_\w$, occurs either in $t^\w(j)$ with
$j < i$, or in $t^\w(i)$, or in $t^\w(j)$ with $j > i$. \dupl\
then selects $j'$ so that the response is in $t^{\w'}(j')$ according
to his winning strategy in games either (a) or (b) (if $j$ is in
$t^\w(i)$, then $j'$ is in $\tau_m^{\w'}(i')$), and then, since the
rank-$m$ types of $t^\w(j)$ and the chosen $t^{\w'}(j')$ are the same,
selects the actual response according to the winning strategy
$t^\w(j)\efeq_m t^{\w'}(j')$.}

Next we show how Claim \ref{nwtl-comp-claim} proves that $\FO$ is
expressible in $\tltree$ over infinite trees $T_\w$. 
First note that being an outer node is expressible: since $\Yr\top$ is
true in right children of matched calls, then 
$$\alpha_{outer}\ = \ \neg \big(\top \Sd (\Yr\top)\big)$$
is true if no node on the path to the root is inside a call, that is,
precisely in outer nodes. 

Next note that for each rank-$m$ type $\tau$ of a tree there is a $\tltree$
formula $\beta_\tau$ such that if $s=\wt(i)$ is an outer node of
$T_\w$, then $(T_\w,s)\models\beta_\tau$ iff the rank-$m$ type of
$t^\w(i)$ is $\tau$. If $i$ is not a matched call, then such a type is
uniquely determined by $i$'s label and perhaps $\pcall$ or $\pret$,
and thus is definable in $\tltree$. 

If $i$ is a matched call, the existence of such a formula $\beta_\tau$ 
follows from the fact that the rank-$m$ type of
$t^\w(i)$ is completely determined by the label of $i$ and the
rank-$m$ type $\tau'$ of the subtree $t^\w_0(i)$ of $t^\w(i)$ rooted
at the right child of $s$ (recall that the root only has a right
child, by the definition of $t^\w(i)$). Type $\tau'$ is expressible in
FO and, since $t^\w(i)$ is finite, by Lemma \ref{marx-lemma} it is
expressible by a $\tltree$ formula $\beta''_{\tau'}$. If we now
inductively take conjunction of every subformula in $\beta''_{\tau'}$
with $\neg\alpha_{outer}$, we obtain  
a formula $\beta'_{\tau'}$ such that 
$(T_\w,s\cdot 1)\models \beta'_{\tau'}$ iff 
$t^\w_0(i) \models \beta''_{\tau'}$ iff 
the rank-$m$ type of $t^\w_0(i)$ is
$\tau'$. 
Hence, $\beta_\tau$ is expressible in $\tltree$ as a Boolean
combination of propositional letters from $\Sigma$ and formulas $\Xd
\beta'_{\tau'}$. Note that in this case, $\beta_\tau$ does not use
$\pcall$ and $\pret$. 

By Claim \ref{nwtl-comp-claim}, we need to express, for each node $s=\wt(i)$,
the rank-$m$ types of $\sleft_m(\w,s)$ and $\sright_m(\w,s)$ in
$\tltree$ over $T_\w$, as well as the rank-$m$ type of $\tau^\w(i)$,
in order to express a quantifier-rank $m$ formula, as it will be a
Boolean combination of such formulas. Given $s$, we need to define
$\wt(i_p)$ -- the outer position in whose scope $s$ occurs -- and then
from that point evaluate two FO formulas, defining rank-$m$ types of
words over the alphabet of rank-$m$ types of finite trees. By Kamp's
theorem \cite{Kamp}, each such FO formula is equivalent to an LTL formula whose
propositional letters are rank-$m$ types of trees.

Assume we have an LTL formula $\gamma$ expressing the rank-$m$ type
$\tau_0$ of $\sright_m(\w,s)$. By Kamp's theorem and the separation
property for LTL, it is written using only propositional letters,
Boolean connectives, $\dm$ and $\U$ (that is, no $\dmminus$ and
$\S$). We now inductively take conjunction of each subformula of
$\gamma$ with $\neg(\Yr\top)$ (i.e., a $\tltree$ formula which is true
in left successors), replace LTL connectives $\dm$ and $\U$ by $\Xd$
and $\Ud$, and replace each propositional letter $\tau$ by
$\beta_\tau$, to obtain a $\tltree$ formula $\gamma'$. Then
$(T_\w,\wt(i_p)) \models \gamma'$ iff $\sright_m(\w,s)$ has type
$\tau_0$. Thus, for a formula 
\begin{eqnarray*}
\gamma'' & = & \big(\alpha_{outer}\wedge \gamma'\big) \vee
\neg\alpha_{outer}\Sd (\alpha_{outer}\wedge \gamma')
\end{eqnarray*}
is true in $(T_\w,\wt(i))$ iff the rank-$m$ type of $\sright_m(\w,s)$
is $\tau_0$. 

The proof for $\sleft_m(\w,s)$ is similar. Since this word is finite,
by Kamp's theorem and the separation property, there is an LTL formula
$\gamma$ that uses $\dmminus$, $\S$, propositional letters and Boolean
connectives such that $\gamma$ evaluated in the last position of the
word expresses its rank-$m$ type. Since there is exactly one path from
each node to the root, to translate $\gamma$ into a $\tltree$ formula
$\gamma'$ we just need to replace propositional letters by the corresponding
formulas $\beta_\tau$, and $\dmminus$ by $\Yd$. Then, as for the case 
of $\sright_m(\w,s)$, we have that $\gamma'$ evaluated in $\wt(i_p)$
expresses the type of $\sleft_m(\w,s)$. Then finally the same formula
as in the case of $\sright_m(\w,s)$ evaluated in $s$ expresses that
type.

Finally we need a $\tltree$ formula that expresses $\tau_m^\w(i)$, the
rank-$m$ type of $t^\w(i)$, when evaluated in $(T_\w,\wt(i))$. 
We can split this into two cases. If $\alpha_{outer}$ is 
true in $\wt(i)$, then, as explained earlier,
the rank-$m$ type of $t^\w(i)$ is a Boolean combination of
propositional 
letters, and thus definable. 

 So we now consider the case when $\alpha_{outer}$ is 
not true in $\wt(i)$. Then $\tau_m^\w(i)$ 
is given by a Boolean combination of formulas that specify (1)
the label of $i_p$, and (2) the rank-$m$ type of the subtree of
$t^\w(i_p)$ rooted at the right child of $\wt(i_p)$ with $s$ as a 
distinguished node. This type can be expressed by a formula $\gamma$
in $\tltree$ over $t^\w_0(i_p)$ by \cite{marx-pods04}. Hence if in
$\gamma$ we recursively take the conjunction of each subformula with
$\neg\alpha_{outer}$, we obtain a formula $\gamma'$ of $\tltree$ that
expresses the type of $(t^\w_0(i_p),s)$ when evaluated in
$(T_\w,s)$. Thus, $\tau^\w(i)$ is expressible by a Boolean combination
of formulas $\gamma'$ and $\neg\alpha_{outer}\Sd (\alpha_{outer}
\wedge a)$ where $a$ is a propositional letter.

This completes the proof of translation of \FO\ into $\tltree$ over
nested $\omega$-words, and thus the proof of Lemma
\ref{nwtl-lemma-two} and 
Theorem \ref{nwtl-thm}.}

Recall that $\FO^k$ stands for a fragment of \FO\ that consists of
formulas which use at most $k$ variables in total.
First, from our
translation from $\nwtl$ to \FO\ we get:
\begin{corollary}
Over nested words,
every \FO\ formula with at most one free variable is equivalent to an
$\FO^3$ formula.
\end{corollary}

\OMIT{

Furthermore, for \FO\ {\em sentences}, we can eliminate the since operator.

\begin{corollary}
\label{nwtl-cor}
For every \FO\ sentence $\Phi$ over finite or infinite nested words,
there is a formula $\phi$ of $\nwtl$ that does not use the since
operator $\Sp$ such that $\w\models \Phi$ iff $(\w,1)\models\phi$.
\end{corollary}

{\em Proof}. 
In the proof of Theorem \ref{nwtl-thm}, we show that every \FO\
sentence over a nested word $\w$ can be translated into an
\FO\ sentence over the tree $T_\w$, and then, by the separation property
of $\tltree$ \cite{marx-pods04} is equivalent to a $\tltree$ formula
that does not use $\Sd$ and $\Yd$. Then, given that in the translation
of $\tltree$ into $\nwtls$ we only use $\Ss$ in the rule
$\crc{(\phi\Sd\psi)} = \crc\phi\Ss\crc\psi$, we see that the
equivalent $\nwtls$ formula does not use $\Ss$. Thus, given that in the
proof of Lemma \ref{nwtl-lemma-one}, no since operator is used in the
translations of $\Us$ into $\Uss$ and $\Uss$ into $\Up$, the
corollary follows for the finite case.

For the infinite case, we note that in the proof of Theorem
\ref{nwtl-thm}, for the case of \FO\ sentences we only need to specify
the type of $\sright_m(\w,s)$ where $s$ is the root. Thus, one can see
that in this case the use of $\Sd$ in $\tltree$ formulas is not
required, and hence the resulting formulas are translated into
$\nwtls$ formulas without $\Ss$.
\fth
}

It is well known that LTL over $\omega$-words has the separation
property, and in particular, every LTL formula is equivalent to an LTL
formula without the past connectives when evaluated in the first
position of an $\omega$-word. In the case of nested words, however,
the situation is quite different from LTL. The following proposition
shows that past connectives are necessary even when one evaluates
formulae in the first position of a nested word.  We let $\nwtlf$ be
the future fragment of $\nwtl$ (i.e. the fragment that does not use
$\Sp$ and the operators $\dmminus$ and $\dmmminus$).

\begin{proposition}
\label{nwtlf-prop}
There are \FO\ sentences over nested words that cannot be expressed in
$\nwtlf$.
\end{proposition}

{\em Proof}. 
We shall look at finite nested words; the proof for the infinite case
applies verbatim. To evaluate a formula $\phi$ of $\nwtlf$ in position
$i$ of a nested word $\w$ of length $n$ one only needs to look at
$\w[i,n]$. That is, if $\w$ and $\w'$ of length $n$ and $n'$
respectively are such that $\w[i,n]\cong\w[i',n']$, then
$(\w,i)\models\phi$ iff $(\w',i')\models\phi$ for every formula $\phi$
of $\nwtlf$. 

Furthermore, for every collection of $\nwtlf$ formulas
$\Psi=\{\psi_1,\ldots,\psi_l\}$, one can find a number $k=k(\Psi)$
such that $$\w[i,n] \efeq_k \w[i',n'] \ \ \ \text{implies}\ \ \ \
(\w,i)\models\psi_p \Leftrightarrow (\w',i')\models\psi_p, \ \
\text{for all }p\leq l.$$ In particular, if $b^r$ stands for the word
of length $r$ in which all positions are labeled $b$ and the matching
relation is empty, there are numbers $k_1 >k_2$
depending only on $\Psi$, such that 
\[b^{k_1} \models \psi_p 
\Leftrightarrow b^{k_2}\models\psi_p, \ \ \text{for all }p\leq l.\]

Now consider the following $\nwtl$ formula: $$\alpha \ \ = \ \
\dmm\top \ \wedge\ \dmm\dmminus a,$$ saying that the first position is
a call, and the predecessor of its matching return is labeled $a$.  We
claim that this is not expressible in $\nwtlf$.

Assume to the contrary that there is a formula $\beta$ of $\nwtlf$
equivalent to $\alpha$. Let $\Psi$ be the collection of all
subformulas of $\beta$, including $\beta$ itself, and let $k_1$ and
$k_2$ be constructed as above. We now consider two nested words $\w_1$
and $\w_2$ of length $k_1+2$ whose
underlying words are $bab^{k_1}$ of length $n=k_1+2$, such that the matching relation
$\M_1$ of $\w_1$ has one edge $\mu_1(1,3)$, and the matching relation
$\M_2$ of $\w_2$ has one edge $\mu_2(1,n+1-k_2)$. In other words, the only
return position of $\w_1$ is $r_1=3$, and  the only
return position of $\w_2$ is $r_2=n+1-k_2$, and thus
$\w_1[r_1,n]=b^{k_1}$ and $\w_2[r_2,n]=b^{k_2}$. 
Further notice that for every $i > 1$ we have $\w_1[i,n]\cong
\w_2[i,n]$. 

Observe that $(\w_1,1)\models\alpha$ and $(\w_2,1)\models\neg\alpha$. 

We now prove by induction on formulas in $\Psi$ that for each such 
formula $\gamma$ we have $(\w_1,1)\models\gamma$ iff
$(\w_2,1)\models\gamma$, thus proving that $\beta$ and $\alpha$ cannot
be equivalent. 

\begin{enumerate}[$\bullet$]
\item The base case of propositional letters is immediate.
\item The Boolean combinations are straightforward too.
\item Let $\gamma=\dm\psi$. Then 
$$\begin{array}{cl}
& (\w_1,1) \models \gamma \\
\LRA & (\w_1,2)\models\psi\\
\LRA & (\w_2,2)\models\psi\\
\LRA & (\w_2,1)\models\gamma,
\end{array}
$$
since $\w_1[2,n]\cong \w_2[2,n]$.
\item Let $\gamma=\dmm\psi$. Then 
$$\begin{array}{cl}
& (\w_1,1) \models \gamma \\
\LRA & (\w_1,3)\models\psi\\
\LRA & b^{k_1} \models \psi\\
\LRA & b^{k_2} \models \psi\\
\LRA & (\w_2,n+1-k_2)\models\psi\\
\LRA & (\w_2,1)\models\gamma,
\end{array}
$$
since $\psi\in\Psi$.
\item Let $\gamma = \phi\Up\psi$. Assume
  $(\w_1,1)\models\gamma$. Consider three cases.
\begin{enumerate}[(1)]
\item[]Case 1: $(\w_1,1)\models\psi$. By the hypothesis
  $(\w_2,1)\models\psi$ and we are done. 

\item[]Case 2: The witness for $\phi\Up\psi$ occurs beyond the only
  return. Then $(\w_1,1)\models\phi$ and
  $(\w_1,r_1)\models\phi\Up\psi$. Since $\phi\Up\psi\in\Psi$ we have
  $(\w_2,r_2)\models \phi\Up\psi$, and by the hypothesis,
  $(\w_2,1)\models\phi$, so $(\w_2,1)\models\phi\Up\psi$.

\item[]Case 3: The witness for $\phi\Up\psi$ occurs inside the
  call. Since for every position $i> 1$ we have $(\w_1,i)\models\phi$
  iff $(\w_2,i)\models\phi$ and likewise for $\psi$, the same summary
  path witnesses $\phi\Up\psi$ in $\w_2$. 
\end{enumerate}
Thus, $(\w_2,1)\models\gamma$. 

Now assume $(\w_2,1)\models\gamma$. In the proof of
$(\w_1,1)\models\gamma$ is the same as above in Cases 1 and 2. For
Case 3, assume that in the path which is a witness for $\phi\Up\psi$
the position in which $\psi$ is true is the 2nd or the 3rd position in
the word. Then the same path witnesses $(\w_1,1)\models\gamma$, as in
the proof of Case 3 above. Next assume it is a position with index $j$
higher than $3$ (which is still labeled $b$) where $\psi$ first
occurs. Then $\phi$ must be true in all positions $i$ with $3 \leq i
\leq j$ in $\w_2$. Hence $\phi$ is true in all such positions in
$\w_1$ as well, and thus the summary path in $\w_1$ that skips the
first call (i.e. jumps from $1$ to $3$) witnesses $\phi\Up\psi$. Hence, in all the cases
$(\w_2,1)\models\gamma$ implies $(\w_1,1)\models\gamma$, which
completes the inductive proof, and thus shows the inexpressibility of
$\alpha$ in $\nwtlf$. 
\qed
\end{enumerate}

\noindent
Note also that adding all other until/since pairs 
to $\nwtl$ does not
change its expressiveness. That is, if we let $\nwtlp$ be 
$\nwtl+\{\Ul, \Sl, \Uc, \Sc, \Ur, \Sr\}$, then:
\begin{corollary}
$\nwtlp=\FO$.
\end{corollary}

Later, when we provide our upper bounds for model-checking,
we shall pride the upper bounds
with respect to $\nwtlp$  rather than just $\nwtl$.

{\em Remark}\ In the conference version, we had a corollary stating
that the since operator can be eliminated for formulae evaluated in
the first position of a nested word. It relied on the proof of Theorem
\ref{nwtl-thm} and the {\em separation property} for $\tltree$ claimed
in \cite{marx-pods04}. The latter, as was discovered recently, is
incorrect. The proof of  Theorem
\ref{nwtl-thm} relies only on the expressive completeness of $\tltree$
which is correct \cite{Schl92,marx-tods} and thus is not affected.

\subsection{The {\em within} operator}
\label{sec-within}
We now go back to the three until/since operators originally proposed
for temporal logics on nested words, based on the the linear, call,
and abstract paths. In other words, our basic logic, denoted by 
$\ltlv$, is 
$$
\begin{array}{l} 
   \phi,\phi' \  :=  \ \top\ \mid\  a \ \mid\ \call\ 
\mid\ \retr\ \mid \ \neg \phi \ \mid \ \phi \vee \phi' \
  \mid \\ 
\ \ \ \  \dm \phi \ \mid \  \dmm \phi \ \mid \ \dmminus \phi \ \mid\
  \dmmminus \phi\ \mid \\
\ \ \ \  \phi \Ul \phi'  \ \mid \   \phi\Sl\phi'\ \mid\
 \phi \Uc \phi'  \ \mid \   \phi\Sc\phi'\ \mid\
 \phi \Ur \phi'  \ \mid \   \phi\Sr\phi'
\end{array}
$$

We now extend this logic with the {\em within} $\WW$ operator proposed
in \cite{AEM04}. Recall that $(\w,i)\models\WW\phi$ iff $i$ is a call,
and $(\w[i,j],i)\models\phi$, where $j=r(i)$ if $i$ is a matched call,
$j=|\w|$ if $i$ is an unmatched call and $\w$ is finite, and $j =
\infty$ otherwise.
  We denote this extended logic by $\ltlv+\WW$.

\begin{theorem}
\label{within-thm}
$\ltlv+\WW=\FO$ over both finite and infinite nested words.
\end{theorem}

\proof
The translation from $\ltlvp+\WW$ into $\FO$ is similar to the
translation used in the proof of Theorem \ref{expcompl-one}. To prove
the other direction, we show how to translate $\nwtls$ into
$\ltlvp+\WW$. Recall that by Lemma \ref{nwtl-lemma-two}, we know that
$\nwtls = \FO$ over both finite and infinite nested words.  More
precisely, for every formula $\varphi$ in $\nwtls$, we show how to
construct a formula $\alpha_\varphi$ in $\ltlvp + \WW$ such that for
every nested word $\w$ (finite or infinite) and position $i$ in it, we
have that $(\w, i) \models \varphi$ if and only if $(\w, i) \models
\alpha_\varphi$.

Since $\ltlvp$ includes the same past modalities as $\nwtls$,
$\alpha_\varphi$ is trivial to define for the atomic formulas, Boolean
combinations and next and previous modalities:
\begin{eqnarray*}
\alpha_{\true} & := & \true,\\
\alpha_{\call} & := & \call,\\
\alpha_{\retr} & := & \retr,\\
\alpha_a & := & a,\\
\alpha_{\neg \phi} & := & \neg \alpha_\phi,\\
\alpha_{\phi \vee \psi} & := & \alpha_\phi \vee \alpha_\psi,\\
\alpha_{\dm \phi} & := &  \dm \alpha_\phi,\\
\alpha_{\dmm \phi} & := & \dmm \alpha_\phi,\\
\alpha_{\text{\mbox{\small $\circleddash$}} \phi} & := & \dmminus \alpha_\phi,\\
\alpha_{\text{\mbox{\small $\circleddash$}}_\M \phi} & := & \dmmminus \alpha_\phi.
\end{eqnarray*}
Thus, we only need to show how to define $\alpha_{\phi \Us \psi}$ and
$\alpha_{\phi \Ss \psi}$. Formula $\alpha_{\phi \Us \psi}$ is defined as: 
\begin{multline*}
\alpha_{\phi \Us \psi} \ \ := \  \ \bigg[\mret \wedge
\alpha_\psi\bigg] \ \vee \  \bigg[\neg \mret \wedge 
\bigg(\bigg(\beta_\varphi \Ur (\neg \mret \wedge \alpha_\psi)\bigg) 
\ \vee \\ 
\bigg(\beta_\varphi \Ur (\beta_\varphi \wedge \dm (\mret \wedge
\alpha_\psi))\bigg) \ \vee \ 
\bigg(\beta_\varphi \Ur \WW \eventually (\alpha_\psi \wedge \dmminus
(\gamma \Sc (\alpha_\varphi \wedge \neg \dmminus
\top)))\bigg)\bigg)\bigg]   
\end{multline*}
where $\mret$ is defined as $\retr\wedge\prev_{\M}\top$, to capture
matching return positions, $\eventually \theta$ is defined as $\top \U
\theta$ and formulas $\beta_\varphi$, $\gamma$ are defined as:
\begin{eqnarray*}
\beta_\varphi & := & \alpha_\varphi \vee \mret,\\
\gamma & := & \beta_\varphi \Sr (\alpha_\varphi
\wedge \neg \retr \wedge \dmminus ((\alpha_\varphi \wedge \neg \retr)
\Sl \call)).
\end{eqnarray*}
Moreover, formula $\alpha_{\phi \Ss \psi}$ is defined as: 
\begin{eqnarray*}
\alpha_{\phi \Ss \psi} & := & \alpha_\psi \vee (\alpha_\varphi \wedge
\dmminus (\gamma \Sc (\beta_\varphi \Sr (\alpha_\psi \wedge \neg \mret)))).
\end{eqnarray*}
This concludes the proof of the theorem. \qed

\subsection{\texorpdfstring{$\caret$}{CaRet} and other within operators}

\newcommand{\first}{{\tt first}}

The logic $\caret$, as defined in \cite{AEM04}, did not have all the
operators of $\ltlv$. In fact it did not have the previous operators
$\dmminus$ and $\dmmminus$, and it only had linear and abstract until
operators, and the call since operator.
That is, $\caret$ was defined as 
$$
\begin{array}{rcl} 
  \phi,\phi' & := &  \ \top\ \mid\ a \ \mid \ \call\ \mid\ 
 \retr\ \mid\ \neg \phi \ \mid \ \phi \vee \phi' \
  \mid \\ 
&& \dm \phi \ \mid \  \dmm \phi \ \mid \ \dmminus_c \phi \ \mid\\
&& \phi \Ul \phi'  \ \mid \   \phi\Ur\phi'\ \mid\  \phi\Sc\phi'\,,
\end{array}
$$
and we assume that $a$ ranges over $\Sigma\cup\{\pret\}$, 
where $\pret$
  is true in pending returns. Notice that $\pret$ is not expressible
  with the remaining operators. Recall that the operator $\dmminus_c$
  is the previous operator corresponding to call paths; formally,
  $(\w, i) \models \dmminus_c \phi$ iff $\C(i)$ is defined and $(\w,
  \C(i)) \models \phi$.

A natural question is whether there is an expressively-complete
extension of this logic. It turns out that the past modality
$\dmminus$, together with two {\em within} operators
based on $\C$ and $\R$ (the innermost call and its return) functions
provide such an extension.  
We
define two new formulas $\C \phi$ and $\R\phi$ with the semantics as
follows: 

\begin{enumerate}[$\bullet$]

\item $(\w,i) \models \C \phi$
iff $(\w[j,i],j) \models \phi$, where $j = \C(i)$ if $\C(i)$ is
defined, and $j = 1$ otherwise.

\item $(\w,i) \models \R \phi$ if $(\w[i,j],i) \models \phi$, 
where $j = \R(i)$ if $\R(i)$ is defined, and $j=|\w|$ (if $\w$ is
finite) or $\infty$ (if $\w$ is infinite) otherwise. 
\end{enumerate}

The logic obtained by adding $\C$ and $\R$ to $\caret$ is denoted by
$\caret+\{\C,\R\}$. 

\begin{theorem}
\label{expcompl-one}
$\caret+\{\C,\R\} \ =\ \FO$ over both finite and infinite nested words.
\end{theorem}

As a corollary (to the proof) we obtain the following: 

\begin{corollary}
For every FO formula $\phi(x)$ over finite or infinite nested words, there is a
formula $\psi$ of $\caret+\{\C,\R\}$ that does not use the $\Ur$
operator, such that $\w \models \phi(i)$ iff $(\w,i) \models \psi$.  
\end{corollary}

The proof of this result is somewhat involved, and
relies on different techniques. The operators used in $\caret$ do not
correspond naturally to tree translations of nested words, and the
lack of all until/since pairs makes a translation from $\nwtl$
hard. We thus use a composition argument {\em directly} on nested
words. The theorem is proved for finite nested words, but 
the same techniques can be used to prove the infinite case. 

\renewcommand{\min}{{\rm min}}
\newcommand{\maxx}{{\rm max}}

We extend the vocabulary with two constants $\min$ and $\maxx$, and
assume that $\min$ is always interpreted as the first element of the
nested word and $\maxx$ as the last element. 

Let $\w$ be a finite nested word of length $n$ and   
and $i$ an element in $\w$. Let
 $c_1,\dots,c_m$, where $m \geq 0$, be all elements in $\w$ such that,
 for each $j \in [1,m]$, $c_j < i$ and there is an
 element $r_j$ such
  that $\M(c_j,r_j)$ and $i \leq r_j$. Assume without loss of
 generality that $c_1 < c_2 < \dots < c_m$.  

Fix $k \geq 0$.  Let $\Gamma$ be the set of all rank-$k$ types of
nested words with distinguished constants $\min$ and $\maxx$
(including the rank-$k$ type of the empty nested word). We define the
word $\Omega_k(\w,i) = a_0 a_1 \cdots a_m$ over alphabet $\Gamma
\times \Gamma$ as follows:

\begin{enumerate}[$\bullet$] 

\item The element $a_0$ is labeled
with the tuple whose first component is
 the rank-$k$ type of
 $(\w[1,c_1-1],\min,\maxx)$ and whose second component is the rank-$k$  
type of 
$(\w[r_1+1,n],\min,\maxx)$ if $m \neq 0$ (notice that if $c_1 = 1$ then 
$\w[1,c_1-1]$ is the empty nested word, and the same is true of
$\w[r_{i+1},n]$ if $r_m = n$); otherwise, it 
is labeled with the tuple whose first component is
 the rank-$k$  type of
 $(\w[1,i-1],\min,\maxx)$ and whose second component is the rank-$k$  
type of 
$(\w[i,n],\min,\maxx)$ (notice that if $i = 1$ then 
$\w[1,i-1]$ is the empty nested word);

\item 
for each $0 < j < m$, the element $a_j$ is labeled with the tuple whose
 first component is 
 the rank-$k$ type of $(\w[c_j,c_{j+1}-1],\min,\maxx)$ and 
whose second component is 
 the rank-$k$ type of $(\w[r_{j+1}+1,r_j],\min,\maxx)$; and 

\item if $m
 \neq 0$ then the element $a_m$
 is labeled with the 
the tuple whose first component is the 
rank-$k$ type of $(\w[c_m,i-1],\min,\maxx)$ and 
whose second component is the rank-$k$ type of 
$(\w[i,r_m],\min,\maxx)$.    
\end{enumerate} 

The following is our composition argument: 

\begin{lemma}[Composition Method] \label{lemma:comp} Let $\w_1$ and
$\w_2$ be two nested words, and let $i$ and $i'$ be two elements in
$\w_1$ and $\w_2$, respectively.  
Then $\Omega_k(\w_1,i) \equiv_{k+2}
\Omega_k(\w_2,i')$ implies $(\w_1,i,\min,\maxx) \equiv_k
(\w_2,i',\min,\maxx)$.  \end{lemma}

\proof
First we need to introduce some terminology.  Let $\w$ be a finite nested
word of length $n$ and $i$ be a position in $\w$. Assume elements
$c_1,\dots,c_m,r_1,\dots,r_m$ are defined as above.  With each element $s$
of $\w$ we associate an element $[s]$ of $\Omega_k(\w,i)$ as
follows:

\begin{enumerate}[$\bullet$]

\item If $m\neq 0$ and $s$ belongs to $\w[1,c_1-1]$ or
$\w[r_1+1,n]$, then $[s]$ is the first element of
$\Omega_k(\w,i)$. In such case we say that $\w[0,c_1-1]$ and
$\w[r_1+1,n]$ are the left and right intervals represented by
$[s]$, respectively. 

If $m = 0$ and $s$ is an arbitrary
element of $\w$, then $[s]$ is also the first (and unique) element of
$\Omega_k(\w,i)$. In such case we say that $\w[0,i-1]$ and
$\w[i,n]$ are the left and right intervals represented by $[s]$,
respectively.

\item 
If $m \neq 0$ and $s$ belongs to $\w[c_m,i-1]$ or $\w[i,r_m]$, then
$[s]$ is the last element of $\Omega_k(\w,i)$. 
In such case we say that
$\w[c_m,i-1]$ and $\w[i,r_m]$ are the left and right intervals 
represented by $[s]$, 
respectively.  

\item 
If $m \neq 0$ and $s$ belongs to $\w[c_\l,c_{\l+1}-1]$ or
$\w[r_{\l+1}+1,r_\l]$, for some $1 \leq \l < m$, then
$[s]$ is the $(\l+1)$-th element of $\Omega_k(\w,i)$. 
In such case we say that
$\w[c_\l,c_{\l+1}-1]$ and 
$\w[r_{\l+1}+1,r_\l]$ are the left and right intervals represented by $[s]$, 
respectively.  

\end{enumerate} We denote by $[s]^L$ and $[s]^R$ the left and right
intervals represented by $[s]$, respectively.

We now prove the lemma.  For each round $j$ ($0 \leq j \leq k$) of the
$k$-round game on $(\w_1,i,\min,\maxx)$ and $(\w_2,i',\min,\maxx)$,
\dupl's response $b_j$ in $\w_2$ to an element $a_j$ in $\w_1$, played
by \spoiler\, is defined as follows (the strategy for the case when
\spoiler\ picks a point in $\w_2$ is completely symmetric). Assume
that \spoiler\ plays element $[a_j]$ in $\Omega_k(\w_1,i)$ in round
$j$ of the $(k+2)$-round game on $\Omega_k(\w_1,i)$ and
$\Omega_k(\w_2,i')$. Then given that $\Omega_k(\w_1,i) \equiv_{k+2}
\Omega_k(\w_2,i')$, \dupl\ uses her winning strategy to choose a
response $[q_j]$ in $\Omega_k(\w_2,i')$ to $[a_j]$. Thus, by
definition of $\Omega_k$, we have that the right and left intervals
represented by $[a_j]$ have the same rank-$k$ type as the right and
left intervals represented by $[q_j]$, respectively. Hence, if $a_j$
belongs to the left interval represented by $[a_j]$, then the \dupl\
can find response $b_j$ to $a_j$ according to the winning strategy for
the $k$-round game on $[a_j]^L$ and $[q_j]^L$, and if $a_j$ belongs to
the right interval represented by $[a_j]$, then the \dupl\ can find
response $b_j$ to $a_j$ according to the winning strategy for the
$k$-round game on $[a_j]^R$ and $[q_j]^R$.

Assume that for round $0 \leq j < k$ the elements played by following
this strategy are (1) $([p_1],\dots,[p_j])$ in $\Omega_k(\w_1,i)$, (2)
$([q_1],\dots,[q_j])$ in $\Omega_k(\w_2,i')$, (3) $(a_1,\dots,a_j)$ in
$\w_1$, and (4) $(b_1,\dots,b_j)$ in $\w_2$. We note that by
definition of the strategy, for every $i \in [1,j]$, we have that $a_i
= p_i$ or $b_i = q_i$. Since we assume that the $[p_j]$'s and
$[q_j]$'s are played according to a winning strategy for \dupl\ in the
$(k+2)$-round game on $\Omega_k(\w_1,i)$ and $\Omega_k(\w_2,i')$, it
is the case that: \begin{eqnarray*}
(\Omega_k(\w_1,i),[p_1],\dots,[p_j]) & \equiv_{k-j+2} &\\
(\Omega_k(\w_2,i'),[q_1],\dots,[q_j]).  \end{eqnarray*} By the way the
strategy is defined, for each $\l \in [1,j]$, if $\bar a_\l^L$ and
$\bar a_\l^R$ are the subtuples of $(a_1, \ldots, a_j)$ containing the
elements from $(a_1, \ldots, a_j)$ that belong to $[a_\l]^L$ and
$[a_\l]^R$, respectively, then the corresponding subtuples $\bar
b_\l^L$ and $\bar b_\l^R$ of $(b_1,\dots,b_j)$ contain the elements
from $(b_1, \ldots, b_j)$ that belong to $[b_\l]^L$ and $[b_\l]^R$,
respectively. Further, by definition of the strategy, we also have
that $([a_\l]^L,\bar a^L_\l,\min,\maxx) \equiv_{k - j} ([b_\l]^L,\bar
b^L_\l,\min,\maxx)$ and $([a_\l]^R,\bar a^R_\l,\min,\maxx) \equiv_{k -
j} ([b_\l]^R,\bar b^R_\l,\min,\maxx)$.


We now show how to define \dupl's response in the round $j+1$.  Let us
assume without loss of generality that for round $j+1$ of the game on
$(\w_1,i,\min,\maxx)$ and $(\w_2,i',\min,\maxx)$, \spoiler\ picks an
element $a_{j+1}$ in $\w_1$ that belongs to the left interval
represented by $[a_{j+1}]$ (all the other cases can be treated in a
similar way). \dupl\ response $b_{j+1}$ in $\w_2$ is defined as
follows. First, there must be an element $[s]$ in $\Omega_k(\w_2,i')$
such that \begin{eqnarray*}
(\Omega_k(\w_1,i),[p_1],\dots,[p_j],[p_{j+1}]) & \equiv_{k-j+1} & \\
(\Omega_k(\w_2,i'),[q_1],\dots,[q_j],[s]), \end{eqnarray*} where
$p_{j+1} = a_{j+1}$. The latter, together with the way that the
strategy is defined, implies that there is an element $b$ in $[s]^L$
such that $([a_{j+1}]^L,\bar a',a_{j+1},\min,\maxx) \equiv_{k - j-1}
([s]^L,\bar b',b,\min,\maxx)$, where $\bar a'$ is the subtuple of
$(a_1, \ldots, a_j)$ containing all the elements in $(a_1,\dots,a_j)$
that belong to $[a_{j+1}]^L$ and $\bar b'$ is the corresponding
subtuple of $(b_1,\dots,b_j)$. We then set $b_{j+1} = b$.

We show by induction that, for each $j \leq k$, if $(a_1,\dots,a_j)$
and $(b_1,\dots,b_j)$ are the first $j$ moves played by \spoiler\ and
\dupl\ on $(\w_1,i,\min,\maxx)$ and $(\w_2,i',\min,\maxx)$,
respectively, according to the strategy defined above, then
$((a_1,\dots,a_j),(b_1,\dots,b_j))$ defines a partial isomorphism
between $(\w_1,i,\min,\maxx)$ and $(\w_2,i',\min,\maxx)$. This is
sufficient to show
 that $(\w_1,i,\min,\maxx) \equiv_k (\w_2,i',\min,\maxx)$.
  
Assume $j = 0$. Since $\Omega_k(\w_1,i) \equiv_{k+2}
\Omega_k(\w_2,i')$, it must be th case that the labels of the last
elements of $\Omega_k(\w_1,i)$ and $\Omega_k(\w_2,i')$ coincide.
Thus, $([i]^R,i) \equiv_0 ([i']^R,i')$, and we conclude that $i$ and
$i'$ have the same label, and $i$ is a call (resp. return) iff $i'$ is
a call (resp. return). Further, if $i = \min$ then $\Omega_k(\w_1,i)$
has only one element and that element is 
   labeled $(\tau_\e,\tau)$, for some $\tau \neq \tau_\e$. Since
   $\Omega_k(\w_1,i) \equiv_{k+2} \Omega_k(\w_2,i')$,
   $\Omega_k(\w_2,i)$ also has a single element and that element is
   labeled
   $(\tau_\e,\tau)$. It follows that $i' = \min$. The converse can be
   proved analogously. In the same way it is possible to show that $i
   = \maxx$ iff $i' = \maxx$.

Assume that the property holds for $j$. Also, assume without loss of
generality that for the round $j+1$ of the game on
$(\w_1,i,\min,\maxx)$ and $(\w_2,i',\min,\maxx)$, \spoiler\ picks an
element $a_{j+1}$ in $\w_1$ that belongs to the right interval
represented by $[a_{j+1}]$ (all the other cases can be treated in a
similar way). We prove that $b_{j+1}$ as defined above preserves the
partial isomorphism.  First we show that $a_{j+1} = i$ iff $b_{j+1} =
i'$. In this case $[a_{j+1}]$ is the last element of
$\Omega_k(\w_1,i)$, and $\Omega_k(\w_1,i) \equiv_{k+2}
\Omega_k(\w_2,i')$ implies that $[b_{j+1}]$ is the last element of
$\Omega_k(\w_2,i')$. Since $a_{j+1} = i$ is the first element of
$[a_{j+1}]^R$, $b_{j+1}$ has to be the first element of $[b_{j+1}]^R$,
which is $i'$.  

In the same way it is possible to prove that $a_{j+1} = \min$ iff
$b_{j+1} = \min$, and that $a_{j+1} = \maxx$ iff $b_{j+1} = \maxx$.

Further, it is also clear that the label of $a_{j+1}$ in $\w_1$ is $a$
iff the label of $b_{j+1}$ in $\w_2$ is $a$, for each $a \in
\Sigma$. Next we consider the remaining cases.

\begin{enumerate}[$\bullet$]

\item $a_{j+1} \in \call$. Then $([a_{j+1}]^R,\bar
a',a_{j+1},\min,\maxx) \equiv_{k - j-1}\!([b_{j+1}]^R,\bar
b',b_{j+1},\min,\maxx)$, where $\bar a'$ is the subtuple of $(a_1,
\ldots, a_j)$ containing all the elements in $(a_1,\dots,a_j)$ that
belong to $[a_{j+1}]^L$ and $\bar b'$ is the corresponding subtuple of
$(b_1,\dots,b_j)$. This immediately implies that $b_{j+1} \in
\call$. The converse is proved analogously.

\item $a_{j+1} \in \rett$. This is similar to the previous
  case. 

\item Suppose first that $a_{j+1} < a_\l$ holds for some $\l \in
[1,j]$. Since $a_{j+1}$ belongs to $[a_{j+1}]^R$, we have that $a_\l$
belongs to $[a_\l]^R$ and, thus, we only need to consider the cases
$[a_\l] = [a_{j+1}]$ and $[a_{j+1}] < [a_\l]$. If $[a_\l] =
[a_{j+1}]$, then $([a_\l]^R,a_\l,a_{j+1}) \equiv_{0}
([b_\l]^R,b_\l,b_{j+1})$ and, therefore, $b_{j+1} < b_l$ also
holds. If $[a_{j+1}] < [a_\l]$, then $[b_{j+1}] < [b_\l]$ and, thus,
$b_{j+1} < b_\l$ holds since $b_\l$ and $b_{j+1}$ belong to $[b_\l]^R$
and $[b_{j+1}]^R$, respectively. 

Suppose, on the other hand, that $a_\l < a_{j+1}$ holds for some $\l
\in [1,j]$. We need to consider three cases: $[a_\l] = [a_{j+1}]$,
$[a_\l] < [a_{j+1}]$ and $[a_{j+1}] < [a_\l]$. If $[a_\l] =
[a_{j+1}]$, then $([a_\l]^R,a_\l,a_{j+1}) \equiv_{0}
([b_\l]^R,b_\l,b_{j+1})$ and, therefore, $b_\l < b_{j+1}$ also
holds. If $[a_{j+1}] > [a_\l]$, then $a_\l$ belongs to $[a_\l]^L$ and
$[b_{j+1}] < [b_\l]$ and, thus, $b_\l < b_{j+1}$ holds since $b_\l$
belongs to $[b_\l]^L$ while $b_{j+1}$ belongs to
$[b_{j+1}]^R$. Finally, if $[a_\l] > [a_{j+1}]$, then $[b_\l] >
[b_{j+1}]$ and, thus, $b_\l < b_{j+1}$ holds since $b_{j+1}$ belongs
to $[b_{j+1}]^R$ and every element in $[b_{j+1}]^R$ is bigger than
every element in either $[b_\l]^R$ or $[b_\l]^L$.

The converse is proved analogously. 

\item Suppose first that $\M(a_{j+1},a_\l)$ holds for some $\l \in
[1,j]$.  Since $a_{j+1}$ belongs to the right interval represented by
$[a_{j+1}]$, it is the case that $[a_\l]$ also belongs to
$[a_{j+1}]^R$. Thus, given that $([a_\l]^R,a_\l,a_{j+1}) \equiv_{0}
([b_\l]^R,b_\l,b_{j+1})$, we conclude that $\M(b_{j+1},b_\l)$ holds.

Second, $\M(a_\l,a_{j+1})$ holds for some $\l \in [1,j]$. It is not
hard to see that $[a_\l] = [a_{j+1}]$. 
We need to consider two cases: If $a_\l$ belongs to
$[a_{j+1}]^R$, then $([a_\l]^R,a_\l,a_{j+1}) \equiv_{0}
([b_\l]^R,b_\l,b_{j+1})$, and thus, $\M(b_\l,b_{j+1})$ holds. If
$a_\l$ belongs to $[a_{j+1}]^L$, then 
$a_\l$ is the first element of $[a_{j+1}]^L$ and
$a_{j+1}$ is the last element of $[a_{j+1}]^R$. Thus, since 
$([a_{j+1}]^L,a_\l,\min) \equiv_0 ([b_{j+1}]^L,b_\l,\min)$, we conclude that $b_\l$ is
the first element of $[b_{j+1}]^L$. Further, since 
$([a_{j+1}]^R,a_{j+1},\maxx) \equiv_0 ([b_{j+1}]^L,b_{j+1},\maxx)$, 
we conclude that
$b_{j+1}$ is the last element of $[b_{j+1}]^R$. Therefore,
$\M(b_\l,b_{j+1})$ holds.

The converse is proved analogously. 
\end{enumerate} 
This concludes the proof of the lemma.\qed

We now present the proof of Theorem \ref{expcompl-one}. 

\aProof{Theorem \ref{expcompl-one}}{We first show that every
$\caret+\{\R,\C\}$ formula $\varphi$ is equivalent to an $\FO$ formula
$\alpha_\varphi(x)$ over nested words, that is, for every nested word
$\w$ it is the case that $(\w,i) \models \varphi$ iff $\w \models
\alpha_\varphi(i)$.  The translation is standard, and can be done by
recursively defining $\alpha_\varphi(x)$ from $\varphi$ as shown
below.  We use the notation $\theta(x)^{[y,z]}$ for the {\em
relativization} of $\theta(x)$ to elements in the interval $[y,z]$,
that is, $\theta(x)^{[y,z]}$ is obtained from $\theta(x)$ by replacing
each subformula of the form $\exists u \beta$ with $\exists u (y \leq
u \wedge u \leq z \wedge \beta)$ and each subformula of the form
$\forall u \beta$ with $\forall u (y \leq u \wedge u \leq z
\rightarrow \beta)$.  Here is the translation:

{\small
\begin{eqnarray*}
\alpha_a(x) &\!\!\!:=\!\!\!& P_a(x),\\
\alpha_\call(x) &\!\!\!:=\!\!\!& \call(x),\\
\alpha_\rett(x) &\!\!\!:=\!\!\!& \rett(x),\\
\alpha_\intt(x) &\!\!\!:=\!\!\!& \neg \call(x) \wedge \neg \rett(x),\\
\alpha_{\pret}(x) &\!\!\!:=\!\!\!& \rett(x) \wedge \neg \exists y \M(y,x),\\
\alpha_{\neg \varphi}(x) &\!\!\!:=\!\!\!& \neg \alpha_\varphi(x),\\
\alpha_{\varphi \vee \psi}(x) &\!\!\!:=\!\!\!& \alpha_\varphi(x) \vee
\alpha_\psi(x), \\
\alpha_{\dm \varphi}(x) &\!\!\!:=\!\!\!& \exists y\, (x < y \,\wedge \,\neg \exists z (x
< z \wedge z < y) \,\wedge\, \alpha_\varphi(y)),\\
\alpha_{\dmminus \varphi}(x) &\!\!\!:=\!\!\!& \exists y\, (y < x \,\wedge \,\neg \exists z (y
< z \wedge z < x) \,\wedge\, \alpha_\varphi(y)),\\
\alpha_{\dmm \varphi}(x) &\!\!\!:=\!\!\!& \exists y\, (\M(x,y) \,\wedge\,
\alpha_\varphi(y)),
\\
\alpha_{{\text{\small $\mathbf{\circleddash}$}}_c \varphi}(x) &\!\!\!:=\!\!\!& \exists y \exists z \, (y < x \wedge x
< z \,\wedge \, \M(y,z) \,\wedge\, 
\alpha_\varphi(y) \,\wedge\\ 
&& \ \ \forall u \forall v (u < x \wedge
x < v \wedge \M(u,v) \,\rightarrow\, u = y \vee u < y)),
\end{eqnarray*}
\begin{eqnarray*}
\alpha_{\varphi \Ul \psi}(x) &\!\!\!:=\!\!\!& \exists y\, ((x < y \vee x = y)
\,\wedge\, \alpha_\psi(y) \,\wedge\\ && \ \ \ \ \forall z (z < y \wedge (z = x
\vee x < z) \,\rightarrow\, \alpha_\varphi(z))),\\ 
\alpha_{\varphi \Ur \psi}(x) &\!\!\!:=\!\!\!& \exists y\, ((x < y \vee x = y)
\,\wedge\, \alpha_\psi(y) \,\wedge\, \forall u \forall v\,(u < y
\wedge y < v \wedge \M(u,v) \,\rightarrow\, u < x) \,\wedge\,  \\
&& \ \ \ \   
\forall z (z < y \wedge (z = x
\vee x < z) \wedge \forall u \forall v\,(u < z
\wedge z < v \wedge \M(u,v) \,\rightarrow\, u < x) \,\rightarrow\,
\alpha_\varphi(z))),
\end{eqnarray*}
\begin{eqnarray*}
\alpha_{\varphi \Sc \psi}(x) &\!\!\!:=\!\!\!& \alpha_\psi(x) \,\vee\, \exists y\,
(y < x \,\wedge\, \alpha_\call(y) \,\wedge\, 
\forall z\,(\M(y,z) \rightarrow x < z) 
\,\wedge\, \alpha_\psi(y) \,\wedge\, \\ 
&& \ \ \ \ \ \forall z\,(((z = x) \vee (\alpha_\call(z) \wedge z < x
\wedge y < z \wedge \forall u\,(\M(z,u) \rightarrow x < u)))
\,\rightarrow\, \alpha_\varphi(z))) ,\\  
\alpha_{\C \varphi}(x) &\!\!\!:=\!\!\!&
(\neg \exists y \exists z \,(\M(y,z) \,\wedge\, y <
x \,\wedge\, x < z) \,\wedge\, \forall z \,(\neg \exists u (u < z)
\,\rightarrow\, \alpha_\varphi(z)^{[z,x]}\,\big) \,\,\vee\, \\
&& \ \ \ (\exists y \exists z \,(\M(y,z) \,\wedge\, y <
x \,\wedge\, x < z \,\wedge\, \\
&& \ \ \ \ \ \ \ \ \ \ \ \ \ \ \ \ \ \
\forall u \forall v (u < x \wedge
x < v \wedge \M(u,v) \,\rightarrow\, u = y \vee u < y) \,\wedge\,
\alpha_\varphi(y)^{[y,x]})), \\ 
\alpha_{\R \varphi}(x) &\!\!\!:=\!\!\!&
(\neg \exists y \exists z \,(\M(y,z) \,\wedge\, y <
x \,\wedge\, x < z) \,\wedge\, \forall z \,(\neg \exists u (z < u)
\,\rightarrow\, \alpha_\varphi(x)^{[x,z]}\,\big) \,\,\vee\, \\
&& \ \ \ (\exists y \exists z \,(\M(y,z) \,\wedge\, y <
x \,\wedge\, x < z \,\wedge\, \\
&& \ \ \ \ \ \ \ \ \ \ \ \ \ \ \ \ \ \
\forall u \forall v (u < x \wedge
x < v \wedge \M(u,v) \,\rightarrow\, u = y \vee u < y) \,\wedge\,
\alpha_\varphi(x)^{[x,z]})). 
\end{eqnarray*}}

We now show the other direction, that is, $ \FO \subseteq
\caret+\{\R,\C\}$.  We start by proving the result for $\FO$ sentences
(that is, we prove that for every $\FO$ sentence $\varphi$ there is an
$\caret+\{\R,\C\}$ formula $\psi$, such that $\w \models \varphi$ iff
$(\w,1) \models \psi$), and then extend it to the case of $\FO$
formulas with one free variable.  Let $\phi$ be an FO sentence. We use
induction on the quantifier rank to prove that $\phi$ is equivalent to
an $\caret+\{\R,\C\}$ formula.

For $k = 0$ the property trivially holds, as $\phi$ is a Boolean
combination of formulas of the form $P_a(\min)$, $P_a(\maxx)$, $\min <
\maxx$, $\M(\min,\maxx)$, etc. All of them can be easily expressed in
$\caret+\{\R,\C\}$.

We now prove for $k + 1$ assuming that the property holds for $k$.
Since every $\FO$ sentence of quantifier rank $k+1$ is a Boolean
combination of $\FO$ sentences of the form $\exists x \phi(x)$, where
$\phi(x)$ is a formula of quantifier rank $k$, we just have to show
how to express in $\caret+\{\R,\C\}$ a sentence of this form.

Let $\Gamma$ be the set of all rank-$k$ types of nested words over
alphabet $\Sigma \cup \{\min,\maxx\}$. We distinguish by
$\tau_\varepsilon$ the rank-$k$ type of the empty nested word. By
induction hypothesis, for each $\tau \in \Gamma$ there is an
$\caret+\{\R,\C\}$ formula $\xi_\tau$ such that $(\w,1) \models
\xi_\tau$ iff the rank-$k$ type of $(\w,\min,\maxx)$ is $\tau$.  


Let $\Lambda$ be the set of all rank-$(k+2)$ types of words over
alphabet $\Gamma \times \Gamma$. We first construct, for each $\lambda
\in \Lambda$, an $\caret+\{\R,\C\}$  formula $\alpha_\lambda$ over
alphabet $\Sigma$  such that,
$$(\w,i) \models \alpha_\lambda \ \ \ \Longleftrightarrow \ \ \ 
\text{the rank-$(k+2)$ type of $\Omega_k(\w,i)$ is $\lambda$,}$$
 for each nested 
word $\w$ and position $i$ of $\w$. 
    
Fix $\lambda \in \Lambda$.  From Kamp's theorem \cite{Kamp}, there is
an LTL formula $\beta_\lambda$ over alphabet $\Gamma \times \Gamma$
such that a word $u$ satisfies $\beta_\lambda$ evaluated on its last
element iff the rank-$(k+2)$ type of $u$ is $\lambda$. By the
separation property of LTL, we can assume that $\beta_\lambda$ only mentions
past modalities $\dmminus$ and $\S$. Moreover, given that $\varphi\,
\S\, \psi\ \equiv\ \psi \vee (\varphi \wedge \dmminus (\varphi\, \S\,
\psi))$, we can also assume that $\beta_\lambda$ is a Boolean
combination of formulas of the form either $\theta$ or $\dmminus
\theta'$, where $\theta$ does not mention any temporal modality and
$\theta'$ is an arbitrary past LTL formula. Thus, since
$\caret+\{\R,\C\}$ is closed under Boolean combinations, to show how
to define $\alpha_\lambda$ from $\beta_\lambda$, we only need to
consider two cases: (1) $\beta_\lambda$ is an LTL formula over $\Gamma
\times \Gamma$ without temporal modalities, and (2) $\beta_\lambda$ is
of the form $\dmminus \theta$, where $\theta$ is an arbitrary past LTL
formula over $\Gamma \times \Gamma$. Next we consider these two cases.

\begin{enumerate}[$\bullet$] 

\item Assume that $\beta_\lambda$ is an LTL formula without temporal
  modalities. Then $\alpha_\lambda$ is defined to be
  $\beta_\lambda^\circ$, where $(\ )^\circ$ is defined recursively as
  follows. Given $(\tau, \tau') \in \Gamma \times \Gamma$, $(\tau,
  \tau')^\circ$ is defined as follows, where we assume that $\tau_a$
is the rank-$k$ type of any nested word with a single element labeled
$a$ ($a \in \Sigma$): 
\begin{enumerate}[(1)]
\item If $\tau, \tau' \neq \tau_\e$, then $(\tau,
  \tau')^\circ$ is defined as the
  disjunction of the following formulas: \begin{enumerate}[(a)] 

\item
  $(\neg \rett \wedge
    \dmminus_c \top \wedge \dmminus (\neg \call 
\wedge \C \xi_\tau) \wedge \R \xi_{\tau'})$;

\item $\bigvee_{\{a \mid
\tau = \tau_a\}} 
(\neg \rett \wedge
    \dmminus_c \top \wedge \dmminus (\call 
\wedge a) \wedge \R \xi_{\tau'})$; 

\item
    $(\neg \rett \wedge \neg \dmminus_c \top \wedge \dmminus \C
    \xi_\tau \wedge \R \xi_{\tau'})$; 

\item $(\pret \wedge \dmminus
    \C \xi_\tau \wedge \R \xi_{\tau'})$; 

\item $\bigvee_{\{a \mid
    \tau' = \tau_a\}} (\rett \wedge \neg \pret \wedge a \wedge
    \dmminus (\neg \call \wedge \C \xi_{\tau}))$; 

\item
    $\bigvee_{\{(a,b) \mid \tau = \tau_a, \tau'=\tau_b\}} (\rett
    \wedge \neg \pret \wedge b \wedge \dmminus (\call \wedge a))$.
    
\end{enumerate} 

\item if $\tau' = \tau_\e$ then
    $(\tau,\tau')^\circ$ is simply $\neg \top$; and 

\item if $\tau =
    \tau_\e$ and $\tau' \neq \tau_\e$, then $(\tau,\tau')^\circ$ is
    defined as $\neg \dmminus \top \wedge \R \xi_{\tau'}$.
    \end{enumerate}

Furthermore, if $\psi$ and $\varphi$ are
LTL formulas without temporal modalities, then \begin{eqnarray*} (\neg
\varphi)^\circ &:=& \neg \varphi^\circ,\\ (\varphi \vee \psi)^\circ
&:=& \varphi^\circ \vee \psi^\circ.  \end{eqnarray*}

\item Assume that $\beta_\lambda$ is a formula of the form $\dmminus
\theta$, where $\theta$ is an arbitrary past LTL formula. Then
$\alpha_\lambda$ is defined to be $\beta_\lambda^\star$, where $(\
)^\star$ is defined recursively as follows. Given $(\tau, \tau') \in
\Gamma \times \Gamma$, $(\tau, \tau')^\star$ is defined as 
follows: 

\begin{enumerate}[(1)] 

\item If $\tau,\tau' \neq \tau_\e$, then
$(\tau,\tau')^\star$ is defined as the disjunction of the following formulas:

\begin{enumerate}[(a)] 

\item $\dmminus ((\neg \call \vee (\call \wedge \neg \dmm \top))
\wedge \C \xi_{\tau}) \wedge \dmm \dm ((\neg \rett \vee \pret) \wedge
\R \xi_{\tau'}) $; 

\item $\bigvee_{\{a \mid \tau=\tau_a\}}
\dmminus (\call \wedge \dmm \top \wedge a) \wedge \dmm \dm (\neg \rett
\wedge \R \xi_{\tau'})$; 

\item $\bigvee_{\{(a,b) \mid \tau=\tau_a, \tau' = \tau_b\}}
\dmminus (\call \wedge \dmm \top \wedge a) \wedge \dmm \dm (\rett
\wedge b)$;

\item $\bigvee_{\{a \mid \tau' = \tau_a\}} \dmminus (\neg \call \wedge
\C \xi_{\tau}) \wedge \dmm \dm (\rett \wedge \neg \pret \wedge a)$.

\end{enumerate}

\item if $\tau = \tau_e$ and $\tau' \neq \tau_e$, then
$(\tau,\tau')^\star$ is defined as $\neg \dmminus \top \wedge \dmm \dm
\R \xi_{\tau'}$;

\item if $\tau \neq \tau_\e$ and $\tau' = \tau_e$, then
$(\tau,\tau')^\star$ is defined as $\dmminus \C \xi_{\tau} \wedge \neg
\dmm \dm \top$; and

\item if $\tau,\tau' = \tau_\e$ then $(\tau,\tau')^\star$ is
defined as $\neg \dmminus \top \wedge \neg \dmm \dm \top$. 

\end{enumerate} 

Furthermore, if $\psi$
and $\varphi$ are past LTL formulas, then \begin{eqnarray*} (\neg
\varphi)^\star &:=& \neg \varphi^\star,\\ (\varphi \vee \psi)^\star
&:=& \varphi^\star \vee \psi^\star,\\ (\dmminus \varphi)^\star &:=&
\dmminus_c \,\varphi^\star,\\ (\varphi\, \S\, \psi)^\star &:=&
\varphi^\star\, \S^c\, \psi^\star.  \end{eqnarray*} \end{enumerate}

Now, let $\exists x \phi(x)$ be an $\FO$ sentence such that the
quantifier rank of $\phi(x)$ is $k$. Then, from our composition method
$\phi(x)$ can be expressed in $\caret+\{\R,\C\}$ as the formula
$\bigvee_{\lambda \in \Lambda'} \alpha_\lambda$, where $\Lambda'
\subseteq \Lambda$ is the set of all rank-$(k+2)$ types of words over
alphabet $\Gamma \times \Gamma$ that belong to $\{\Omega_k(\w,i) \
\mid \ \w \models \phi(i)\}$. Thus, $\exists x \phi(x)$ can be expressed as the
following $\caret+\{\R,\C\}$ formula: $\top \,\Ul \,(\bigvee_{\lambda
\in \Lambda'} \alpha_\lambda)$. This concludes the proof of the
theorem.

Finally, from the composition method and the previous proof we see
that the equivalence $\FO = \caret+\{\R,\C\}$ also holds for unary
queries over nested words.}

\section{Model-Checking and Satisfiability}
\label{mc-sec}

\noindent In this section we show that both satisfiability and model-checking
are decidable in single-exponential-time for $\nwtl$, and
in polynomial time in the size of the model.
Here we assume the model of the procedural
program is given as a Recursive State Machine (RSM) \cite{RSM}.
(Runs of an RSM can naturally be viewed as nested words
 when matching function calls (or ``box entries'') and returns 
 (or ``box exits'') along the run are paired together.) 
In fact we prove this bound
for $\nwtlp$, an \FO-complete extension of $\nwtl$ with all of $\Ul,
\Sl, \Uc, \Sc, \Ur, \Sr$. We use automata-theoretic techniques: 
translating formulae into equivalent automata on nested words. We then
show that the logic based on adding the
{\em within} operator to $\nwtlp$, (and even just adding 
{\em within} to $\caret$) 
 requires doubly-exponential time for
model-checking, but is exponentially more succinct.

\subsection{Nested word automata}

A {\em nondeterministic \buchi\ nested word automaton\/} (BNWA) $A$ 
over an alphabet $\Sigma$
is a structure $(Q, Q_0, Q_f, P, P_0, P_f, \cd, \id, \rd )$ consisting of
a finite set of states $Q$,
a set of initial states $Q_0\subseteq Q$,
a set of \buchi\ accepting states   $Q_f\subseteq Q$,
a set of hierarchical symbols $P$,
a set of initial hierarchical symbols $P_0\subseteq P$,
a set of final hierarchical symbols $P_f\subseteq P$,
a call-transition relation $\cd \subseteq Q \times \Sigma \times Q \times P $,
an  internal-transition relation $\id \subseteq Q \times \Sigma \times Q  $, 
and a return-transition relation $\rd \subseteq Q \times P \times \Sigma \times Q  $.
The automaton $A$ starts in an initial state and reads the nested word from left to
right. The state is propagated along the linear edges as in the case of
a standard word automaton.
However, at a call, the nested word automaton propagates state
along the linear edge as well as a hierarchical symbol 
along the nesting edge 
(if there is no matching return,
then the latter is required to be in  $P_f$ for acceptance). 
At a matched return, the new state is determined based on the
state propagated along the linear edge as well as the symbol along the
incoming
nesting edge (edges incident upon unmatched returns are assumed to be labeled with
initial hierarchical symbols).

Formally, a {\em run\/} $r$ of the automaton $A$ over a nested word
$\w=(a_1 a_2 \ldots , \M,\call,\retr)$ is a sequence $q_0, q_1, \ldots$ of
states along the linear edges,
and a sequence $p_i$, for every call position $i$, of hierarchical symbols 
along nesting edges,
such that $q_0\in Q_0$ and for each position $i$, if $i$ is a call 
then $(q_{i-1},a_i,q_i,p_i)\in\cd$; 
if $i$ is internal, then $(q_{i-1},a_i,q_i)\in\id$; if $i$ is a return such
that $\M(j, i)$, then $(q_{i-1}, p_j, a_i,q_i)\in\rd$;
and if $i$ is an unmatched return then $(q_{i-1},p,a_i,q_i)\in\rd$ for some $p\in P_0$.
The run $r$ is accepting if 
(1) for all pending calls $i$, $p_i\in P_f$, and 
(2) the final state $q_\ell\in Q_f$ if $\w$ is a finite word of length $\ell$, 
and for infinitely many positions $i$, $q_i\in Q_f$, if $\w$ is a nested $\omega$-word.  
The automaton $A$
accepts the nested word $\w$ if it has an accepting run over $\w$.

Nested word automata have the same expressiveness as 
the monadic second order logic over
nested words, and the language emptiness problem for them can be
decided in polynomial-time~\cite{nested}.

\subsection{Tableau construction}

We now show how to build a BNWA
accepting the satisfying models of a formula of $\nwtlp$.
This 
leads to decision procedures for satisfiability and
model checking.

Given a formula $\varphi$, we wish to construct a \buchi\ nested word
automaton $A_\varphi$ whose states correspond to sets of subformulas
of $\varphi$.  Intuitively, given a nested word $\w$, a run $r$, which
is a linear sequence $q_0q_1\ldots$ of states and symbols $p_i$
labeling nesting edges from call positions, should be such that each state $q_i$ is
precisely the set of formulas that hold at position $i+1$. The label
$p_i$ is used to remember abstract-next formulas that hold at
position $i$ and the abstract-previous formulas that hold at matching return.
For clarity of presentation, we first focus on formulas with next
operators $\dm$ and $\dmm$, and until over summary-down paths.  

Given a formula $\varphi$, the closure of $\varphi$, denoted by
$\Cl(\varphi)$, is the smallest set that satisfies the following
properties: 
\begin{enumerate}[$\bullet$]
\item
$\Cl(\varphi)$ contains $\varphi$, $\call$, $\retr$,
$\int$, and $\next\retr$; 
\item
if either $\neg \psi$, or $\next\psi$ or
$\dmm\psi$ is in $\Cl(\varphi)$ then $\psi \in \Cl(\varphi)$; 
\item 
if $\psi \vee \psi' \in \Cl(\varphi)$, then $\psi,\psi' \in \Cl(\varphi)$;
\item
if $\psi \dpuntil \psi'\in\Cl(\varphi)$, then $\psi$, $\psi'$,
$\next(\psi\dpuntil \psi')$, and 
$\dmm(\psi\dpuntil\psi')$ are in
$\Cl(\varphi)$; and
\item
if $\psi \in \Cl(\varphi)$ and $\psi$ is not of
the form $\neg \theta$ (for any $\theta$), then $\neg \psi \in
\Cl(\varphi)$. 
\end{enumerate}
It is straightforward to see that the size of
$\Cl(\varphi)$ is only linear in the size of $\varphi$.  Henceforth,
we identify $\neg \neg \psi$ with the formula $\psi$.

An \emph{atom} of $\varphi$ is a set $\Phi \subseteq \Cl(\varphi)$
that satisfies the following properties:
\begin{enumerate}[$\bullet$]
 \item For every $\psi \in \Cl(\varphi)$, 
           $\psi\in \Phi$ iff $\neg \psi \not \in \Phi$ .
 \item For every formula $\psi \vee \psi' \in \Cl(\varphi)$,
           $\psi \vee \psi' \in \Phi$ iff ($\psi \in \Phi$ or $\psi' \in \Phi$).
 \item 
       For every formula $\psi\dpuntil \psi'\in \Cl(\varphi)$,
           $\psi \dpuntil \psi' \in \Phi$ iff
           either $\psi' \in \Phi$ or ($\psi \in \Phi$ and $\next\retr\not\in\Phi$ and
                      $\next(\psi \dpuntil \psi') \in \Phi$) or
 ($\psi\in\Phi$ and $\dmm(\psi \dpuntil \psi') \in \Phi$).
 \item $\Phi$ contains exactly one of the elements in the set $\{\call, \retr, \int\}$.
\item If $\dmm\psi\in\Phi$ for some $\psi$, then $\call\in\Phi$.
\end{enumerate}
These clauses capture local consistency requirements.
In particular, a summary-down until formula $\psi\dpuntil\psi'$ holds at a position if either
the second argument $\psi'$ holds now, or $\psi$ holds now and satisfaction of $\psi\dpuntil\psi'$
is propagated along a call, internal, or nesting edge.

A hierarchical-atom of $\varphi$ is a set $\Phi\subseteq \Cl(\varphi)$ such that
if $\psi\in \Phi$ then $\dmm\psi\in\Cl(\varphi)$.
A hierarchical-atom contains possible abstract-next obligations to be propagated across
nesting edges.

Given a formula $\varphi$, we build a nested word automaton $A_\varphi$ as follows.
The alphabet $\Sigma$ is $2^\AP$, where $AP$ is the set of atomic
propositions. 
\begin{enumerate}[(1)]
\item
Atoms of $\varphi$ are states of $A_\varphi$;
\item
An atom $\Phi$ is an initial state iff $\varphi\in\Phi$;
\item
Hierarchical-atoms of $\varphi$ are hierarchical symbols of $A_\varphi$;
\item
All hierarchical symbols are initial;
\item
For atoms $\Phi,\Psi$ and a symbol $a\subseteq \AP$,
$(\Phi,a,\Psi)$ is an internal transition of $A_\varphi$ iff
(a) $\int\in\Phi$; and
(b) for $p\in\AP$, $p\in a$ iff $p\in\Phi$; and
(c) for each $\next\psi\in\Cl(\varphi)$,
$\psi\in\Psi$ iff $\next\psi\in\Phi$.
\item
For atoms $\Phi,\Psi_l$, a hierarchical-atom $\Psi_h$, and a symbol $a\subseteq \AP$,
$(\Phi,a,\Psi_l,\Psi_h)$ is a call transition of $A_\varphi$ iff
(a) $\call\in\Phi$; and
(b) for $p\in\AP$, $p\in a$ iff $p\in\Phi$; and
(c) for each $\next\psi\in\Cl(\varphi)$,
$\psi\in\Psi_l$ iff $\next\psi\in\Phi$; and
(d) for each $\dmm\psi\in\Cl(\varphi)$,
$\psi\in\Psi_h$ iff $\dmm\psi\in\Phi$.
\item
For atoms $\Phi_l,\Psi$, hierarchical-atom $\Psi_h$, and a symbol $a\subseteq \AP$,
$(\Phi_l,\Phi_h,a,\Psi)$ is a return transition of $A_\varphi$ iff
(a) $\retr\in\Phi_l$; and
(b) for $p\in\AP$, $p\in a$ iff $p\in\Phi_l$; and
(c) for each $\next\psi\in\Cl(\varphi)$,
$\psi\in\Psi$ iff $\next\psi\in\Phi_l$; and
(d) for each $\dmm\psi\in\Cl(\varphi)$,
$\psi\in\Phi_h$ iff $\psi\in\Phi_l$.

\end{enumerate}
The transition relation ensures that the current symbol is consistent
with the atomic propositions in the current state, and next operators
requirements are correctly propagated.

The sole final  hierarchical symbol is the empty hierarchical-atom.
This ensures that, in an accepting run,
at a pending call, no requirements are propagated along the nesting edge.
For each until-formula $\psi$ in the closure, let $F_\psi$ be the set
of atoms that either do not contain $\psi$ or contain the second argument of $\psi$.
Then a nested word $\w$ over the alphabet $2^\AP$ satisfies  $\varphi$ iff
there is a run $r$ of $A_\varphi$ over $\w$ such that 
all pending call edges are labeled with the sole final hierarchical symbol, and
for each until-formula $\psi\in\Cl(\varphi)$, for infinitely many positions $i$,
$q_i\in F_\psi$. 
This multi-\buchi\ accepting condition  can
be translated to \buchi\ acceptance as usual by adding a counter.

Now we proceed to show how to handle various forms of until operators.
In each case, we specify the changes needed to the definition of the closure
and local consistency requirements for atoms. 
\begin{enumerate}[\hbox to6 pt{\hfill}]
\item\noindent{\hskip-10 pt\bf Global paths:}\
If $\psi \Ul \psi'\in\Cl(\varphi)$, then $\psi$, $\psi'$,
$\next(\psi \Ul \psi')$ are in $\Cl(\varphi)$.
Local consistency of $\Phi$ requires that
for every formula $\psi\Ul \psi'\in \Cl(\varphi)$,
           $\psi \Ul \psi' \in \Phi$ iff
           either $\psi' \in \Phi$ or ($\psi \in \Phi$ and
                      $\next(\psi \Ul \psi') \in \Phi$).
\item\noindent{\hskip-10 pt\bf Summary-up paths:}\
If $\psi \upuntil \psi'\in\Cl(\varphi)$, then $\psi$, $\psi'$,
$\next(\psi \upuntil \psi')$, and $\dmm(\psi \upuntil \psi')$
 are in $\Cl(\varphi)$.
Local consistency of $\Phi$ requires that
for every formula $\psi\upuntil \psi'\in \Cl(\varphi)$,
           $\psi \upuntil \psi' \in \Phi$ iff
           either $\psi' \in \Phi$, or 
           ($\psi \in \Phi$ and $\call\in\Phi$ and $\dmm(\psi \upuntil \psi') \in \Phi$), or
($\psi \in \Phi$ and $\call\not\in\Phi$ and $\next(\psi \upuntil \psi') \in \Phi$).
\item\noindent{\hskip-10 pt\bf Abstract paths:}\
If $\psi \Ur \psi'\in\Cl(\varphi)$, then $\psi$, $\psi'$,
$\next(\psi \Ur \psi')$, $\next\dmmminus\top$, and $\dmm(\psi \Ur \psi')$
 are in $\Cl(\varphi)$.
Local consistency of $\Phi$ requires that
for every formula $\psi\Ur \psi'\in \Cl(\varphi)$,
           $\psi \Ur \psi' \in \Phi$ iff
           either $\psi' \in \Phi$, or 
           ($\psi \in \Phi$ and $\call\in\Phi$ and $\dmm(\psi \Ur \psi') \in \Phi$), or
($\psi \in \Phi$ and $\call\not\in\Phi$ and $\next\retr\not\in\Phi$ and $\next(\psi \Ur \psi') \in \Phi$), 
or ($\psi\in\Phi$ and $\next\retr\in\Phi$ and $\next\dmmminus\top\not\in\Phi$ and
$\next(\psi \Ur \psi') \in \Phi$).
The last case accounts for propagation of the eventuality across unmatched returns.
\item\noindent{\hskip-10 pt\bf Call paths:}\
Recall that positions along a call path are related by the innermost call operator:
a call path jumps from a call position $i$ to a position $j$ such that $i=\C(j)$.
Thus, a call path can be simulated by a summary-down path consisting of
call edges, summary edges and internal edges, where the formula is asserted
only before following the call edge. This effect is captured by using an auxiliary
operator as follows.
If $\psi \Uc \psi'\in\Cl(\varphi)$, then $\psi$, $\psi'$, $\psi\aUc \psi'$,
$\next(\psi\aUc \psi')$, and 
$\dmm(\psi\aUc \psi')$ are in
$\Cl(\varphi)$.
Local consistency of $\Phi$ requires that
for every formula $\psi\Uc\psi'\in \Cl(\varphi)$,
           $\psi \Uc \psi' \in \Phi$ iff
           either $\psi' \in \Phi$ or ($\psi \in \Phi$ and $\call\in \Phi$ and
  $\next(\psi \aUc \psi') \in \Phi$);
and 
           $\psi \aUc \psi' \in \Phi$ iff
           either $\psi\Uc \psi'\in\Phi$, or $\dmm(\psi \aUc \psi') \in \Phi$, or
 ($\next\retr\not\in\Phi$ and $\next(\psi \aUc \psi') \in \Phi$).
\item\noindent{\hskip-10 pt\bf Summary paths:}\
 The summary-until is handled using the fact that
$\psi\puntil\psi'$ is equivalent to
$\psi\upuntil(\psi\dpuntil\psi')$.
\end{enumerate}\smallskip

\noindent Note that the definition of $A_\varphi$ stays unchanged, as
the correct propagation of requirements is handled by next and
abstract-next formulas ensured by local consistency. The eventual
satisfaction of until formulas is handled the same way as before: for
each until-formula $\psi$ in the closure, let $F_\psi$ be the set of
atoms that either do not contain $\psi$ or contain the second argument
of $\psi$, and it is required that each such $F_\psi$ is visited
infinitely often.

The past-time formulas (previous, abstract-previous, and various forms of since
operators) are handled in a symmetric manner.
Thus, we have shown:

\begin{theorem}\label{exp-buchi}
For a formula $\varphi$ of $\nwtlp$,
one can effectively construct a nondeterministic \buchi\ nested word automaton $A_\varphi$
of size $2^{O(|\varphi|)}$ accepting 
the models of $\varphi$.
\end{theorem}

Since the automaton $A_\varphi$ is exponential in the size of $\varphi$, we
can check satisfiability of $\varphi$ in exponential-time by testing emptiness
of $A_\varphi$. \EXPTIME-hardness follows from the corresponding hardness result for
$\caret$.

\begin{corollary}
The satisfiability problem for $\nwtlp$ is \EXPTIME-complete.
\end{corollary}

When programs are modeled by nested word automata $A$ (or
equivalently, pushdown automata, or recursive state machines), and
specifications are given by formulas $\varphi$ of  $\nwtlp$, 
we can use the classical automata-theoretic approach:
negate the specification, build the NWA $A_{\neg\varphi}$ accepting
models that violate $\varphi$, take the product with the program
$A$, and test for emptiness of $L(A)\cap L(A_{\neg\varphi})$.  Note
that the program typically will be given more compactly, say, as a
Boolean program~\cite{BallRajamani}, and thus, the NWA $A$ may itself
be exponential in the size of the input.

\begin{corollary}
Model checking $\nwtlp$ specifications with respect to Boolean programs
is \EXPTIME-complete.  If the program model is given as a recursive
state machine or nested word automaton, the running time is
polynomial in the model and exponential in the $\nwtlp$ formula,
and remains \EXPTIME-complete.
\end{corollary}

\subsection{Checking the {\em within} operator}

We now show that adding {\em within} operators makes model-checking
doubly exponential. 
Given a formula $\varphi$ of $\nwtl$ or $\nwtlp$, let $p_\varphi$ be a special
proposition that does not appear in $\varphi$. Let $W_\varphi$ be the
language of nested words $\w$ such that for each position $i$,
$(\w,i)\models p_\varphi$ iff $(\w,i)\models \WW \varphi$.  We 
construct a doubly-exponential automaton $B$
that captures  $W_\varphi$.  
First, using the tableau construction for $\nwtlp$,
we construct an exponential-size automaton $A$ 
that captures nested words that satisfy $\varphi$.
Intuitively, every time a proposition $p_\varphi$ is encountered,
we want to start a new copy of $A$, and a state of $B$ keeps track of
states of multiple copies of $A$. 
At a call, $B$ guesses whether the call has a matching return or not.
In the latter case, 
as in case of determinization
construction for nested word automata~\cite{nested}, 
we need to 
maintain pairs of states of $A$ so that the join at return positions
can be done correctly.
A state of $B$, then, is either a set of states of $A$ or a set of pairs
of states of $A$. We explain the latter case.
The intended meaning is that a
pair $(q,q')$ belongs to the state of $B$, while reading position $i$
of a nested word $\w$, if the subword from $i$ to the first unmatched return
can take the automaton $A$ from state $q$ to state $q'$. 
When reading an internal symbol $a$, a summary $(q,q')$ in the current state
can be updated to $(u,q')$, provided $A$ has an internal transition from $q$ to
$u$ on symbol $a$. Let $B$ read a call symbol $a$.
Consider a summary $(q,q')$ in the current state, and a call-transition 
$(q,a,q_l,q_h)$ of $A$. Then $B$ guesses the return transition $(u_l,q_h,b,u)$ that will
be used by $A$ at the matching return, and sends the summary
$(q_l,u_l)$ along the call edge and the triple $(b,u,q')$ along the nesting edge.
While processing a return symbol $b$, the current state of $B$ must contain
summaries only of the form $(q,q)$ where the two states match, and
for each summary $(b,u,q')$ retrieved from the state along the nesting edge,
the new state contains $(u,q')$.
Finally, $B$ must enforce that $\WW\varphi$ holds when $p_\varphi$ is read.
Only a call symbol $a$ can contain 
the proposition 
$p_\varphi$, and when reading
such a symbol, $B$ guesses a call transition $(q_0,a,q_l,q_h)$, where $q_0$ is the initial
state of $A$, and a return transition $(u_l,q_h,b,q_f)$, 
where $q_f$ is an accepting state of $A$,
and sends the summary $(q_l,u_l)$ along the call edge and 
the symbol $b$ along the nesting edge.

\begin{lemma}
For every formula $\varphi$ of $\nwtlp$, there is a nested word automaton
that accepts the language $W_\varphi$ and has size doubly-exponential
in $|\varphi|$.
\end{lemma}

Consider a formula $\varphi$ of $\nwtlp+\WW$.
For every within-subformula $\WW\varrho$ of
$\varphi$,   
let $\varrho'$ be obtained from $\varrho$ by
substituting each top-level subformula $\WW\psi$ in $\varrho$ by the
proposition $p_\psi$.  
Each of these primed
formulas is a formula of $\nwtlp$.  Then, if we take the product of the
nested word automata accepting $W_{\varrho'}$ corresponding to all the
within-subformulas $\varrho$, together with the nested word automaton
$A_{\varphi'}$, the resulting language 
captures the set of
models of $\varphi$.  Intuitively, the automaton for $W_{\varrho'}$ is
ensuring that the truth of the proposition $p_\varrho$ reflects the truth
of the subformula $\WW\varrho$.  If $\varrho$ itself has a within-subformula
$\WW\psi$, then the automaton for $\varrho$ treats it as an
atomic proposition $p_\psi$, and the automaton checking $p_\psi$,
running in parallel, makes sure that the truth of $p_\psi$ correctly
reflects the truth of $\WW\psi$.

For the lower bound, the decision problem for LTL games can be reduced 
to the satisfiability problem for formulas with linear untils and within
operators~\cite{Madhu04}, and this shows that
for $\caret$  extended with the within operator,
the satisfiability problem is 2\EXPTIME-hard.
We thus obtain:
\begin{theorem}
For the logic $\nwtlp$ extended with the within operator $\WW$
the satisfiability problem and the model checking problem with respect to
Boolean programs, are both 2\EXPTIME-complete.
\end{theorem}

\paragraph{Remark: checking $\w\models\varphi$ for finite nested words} For finite
nested words, one evaluates the 
complexity of checking whether the given word satisfies a formula,
 in terms of the length $|\w|$ of the word and the size of the formula. 
A straightforward recursion on
subformulas shows that for $\nwtl$ formulas the complexity of
this check is $O(|\w|\cdot|\phi|)$, and for both logics with
{\em within} operators, $\caret+\{\C,\R\}$ and $\ltlv+\WW$, it is 
$O(|\w|^2\cdot|\phi|)$.

\subsection{On {\em within} and succinctness} We saw that adding
within operators to $\nwtlp$ increases the complexity of
model-checking by one exponent. Thus there is no
polynomial-time translation from $\nwtlp + \WW$ to $\nwtlp$. We now
prove a stronger result that gives a space bound as well: while
$\nwtlp+\WW$ has the same power as $\nwtlp$, its formulae can be
exponentially more succinct than formulas of $\nwtlp$. That is, there
is a sequence $\phi_n$, $n\in\nn$, of $\nwtlp+\WW$ formulas such that
$\phi_n$ is of size $O(n)$, and the smallest formula of $\nwtlp$
equivalent to $\phi_n$ is of size $2^{\Omega(n)}$.  For this result,
we require nested $\omega$-words to be over the alphabet $2^{AP}$.

\begin{theorem}\label{succinct-theo}
$\nwtlp + \WW$ is exponentially more succinct than $\nwtlp$.
\end{theorem}

{\em Proof}. 
The proof is based upon succinctness results in \cite{EVW02,LMS}, by
adapting their examples to nested words. 

From the FO completeness of $\nwtlp$, we have that $\nwtlp + \WW$ can be
translated into $\nwtlp$. We show that at least an exponential blow-up
is necessary for such translation. More precisely, we construct a
sequence $\{\phi_n\}_{n \geq 1}$ of $\nwtlp + \WW$ formulas of size
$O(n)$, such that the shortest $\nwtlp$ formula that is equivalent to
$\phi_n$ is of size $2^{\Omega(n)}$. Our proof is a modification of
similar proofs given in \cite{EVW02,LMS}.  Assume $\Sigma =
\{a_0,\dots,a_n\}$, and let $\phi_n$ be the following $\nwtlp + \WW$
formula (here, $\always^\ppath \theta$ and $\peventually^\ppath \theta$ are
abbreviations for $\neg (\top \Up \neg \theta)$ and $\top \Sp \theta$,
respectively):
\begin{eqnarray*}
\always^\ppath \bigg(\call \to \WW \always^\ppath \bigg((\bigwedge_{i
=1}^n (a_i \, \leftrightarrow \, \peventually^\ppath(a_i \wedge \neg
\dmminus \top))) \ \rightarrow \ (a_0 \, \leftrightarrow \,
\peventually^\ppath (a_0 \wedge \neg \dmminus \top))\bigg)\bigg).
\end{eqnarray*}
It is not hard to see that $\w \models \phi_n$ iff for all positions
$i,j$ in $\w$ such that $\M(i,j)$ holds, if position $\ell$ in
$\w[i,j]$ coincides with $i$ on $a_1,\dots,a_n$, then $\ell$ also
coincides with $i$ on $a_0$.

It is shown in Theorem \ref{exp-buchi} that for each $\nwtlp$
formula $\alpha$, the language 
\begin{eqnarray*}
L_\alpha & = & \{\,\w \,\mid\, \w \text{ is a nested $\omega$-word
such that $\w \models \alpha$}\,\} 
\end{eqnarray*}
is recognized by a nondeterministic nested word automaton of size
$2^{O(|\alpha|)}$. Thus, to prove the theorem, it is enough to show
that every such automaton for $L_{\varphi_n}$ is of size
$2^{2^{\Omega(n)}}$.  Let $A$ be a nondeterministic nested word
automaton for $L_{\varphi_n}$.  Assume that $b_0,\dots,b_{2^n-1}$ is
an enumeration of the symbols in $2^{\Sigma \setminus
\{a_0\}}$. For every $K \subseteq
\{0,\dots,2^n-1\}$ let $\w_K$ be the word $c_0 \cdots c_{2^n-1}$ over
alphabet $2^\Sigma$, where for each $i \leq 2^n-1$:
\begin{equation*}
c_i = 
\begin{cases}
b_i & i \in K\\
b_i \cup \{a_0\} & \text{otherwise}
\end{cases}
\end{equation*}
It is not hard to see that for each $K \subseteq \{0,\dots,2^n-1\}$,
the nested $\omega$-word $(\w_K^\omega,\M)$, where $\M = \{(j, 3 \cdot
2^n - j + 1) \mid 1 \leq j \leq 2^{n}\}$, is such that
$(\w_K^\omega,\M) \models \phi_n$. Let $(q^1_K,p^1_K,q^2_K)$ and
$(q^1_{K'},p^1_{K'},q^2_{K'})$ be pairs of states such that (1) there
exists an accepting run of $A$ on $(\w_K^\omega,\M)$ such that $A$ is
in state $q^1_K$ and has hierarchical symbol $p^1_K$ in call position
$2^n$, and $A$ is in state $q^2_K$ in internal position $2 \cdot 2^n$;
(2) there exists an accepting run of $A$ on $(\w_{K'}^\omega,\M)$ such
that $A$ is in state $q^1_{K'}$ and has hierarchical symbol $p^1_{K'}$
in call position $2^n$, and $A$ is in state $q^2_{K'}$ in internal
position $2 \cdot 2^n$. Next we show that $(q^1_K,p^1_K,q^2_K) \neq
(q^1_{K'},p^1_{K'},q^2_{K'})$ if $K \neq K'$. On the contrary, assume
that $(q^1_K,p^1_K,q^2_K) = (q^1_{K'},p^1_{K'},q^2_{K'})$. Then $A$ accepts
$(\w_K \w_{K'} \w_K^\omega,\M)$, which leads to a contradiction since $(\w_K
\w_{K'} \w_K^\omega,\M) \not\models \varphi_n$. Given that the number
of different $K$'s is $2^{2^n}$, the latter implies that the number of
different triples of states and hierarchical symbols of $A$ is at
least $2^{2^n}$. Thus, if $m$ is equal to the number of states of $A$
plus the number of hierarchical symbols of $A$, then $m^3 \geq
2^{2^n}$ and, hence, $m \geq 2^{2^{n-2}}$. Therefore, the size of $A$
is $2^{2^{\Omega(n)}}$. This concludes the proof of the theorem.
\qed

\section{Finite-Variable Fragments}
\label{fv-sec}

\noindent We have already seen that \FO\ formulas in one free variable over
nested words can be written using just three distinct variables, as in
the case of the usual, unnested, words. For finite nested words this
is a consequence of a tree representation of nested words and the
three-variable property for \FO\ over finite trees \cite{marx-tods},
and for infinite nested words this is a consequence Theorem
\ref{nwtl-thm}. 

In this section we prove two results. First, we give a model-theoretic
proof that \FO\ formulas with zero, one, or two free variables over
nested words (finite or infinite) are equivalent to $\FO^3$ formulas.
Given the $\FO=\FO^3$ collapse, we ask whether there is
a temporal logic expressively complete for $\FO^2$, the two-variable
fragment. We adapt techniques from \cite{EVW02}
to find a temporal logic that has the same expressiveness as $\FO^2$
over nested words (in a vocabulary that has successor relations
corresponding to the ``next'' temporal operators).

\subsection{The three-variable property}

\newcommand{\sland}{\;\land\;}
\newcommand{\abs}[1]{ \vert #1 \vert }

We give a model-theoretic, rather than a syntactic,
argument, that uses 
Ehrenfeucht-Fra\"{\i}ss\'e games and shows that over nested words, 
formulas with at most two free variables are equivalent to
$\FO^3$ formulas. Note that for finite nested words, the translation
into trees, already used in the proof of Theorem \ref{nwtl-thm}, can
be done using at most three variables. This means that the result of
\cite{marx-tods} establishing the 3-variable property for finite
ordered unranked trees gives us the 3-variable property for finite
nested words. We prove that $\FO=\FO^3$ over arbitrary
nested words.

\begin{theorem}\label{3var th}
Over finite or infinite nested words, every \FO\ formula with at most
$2$ free variables is equivalent to an $\FO^3$ formula.
\end{theorem}

{\em Proof}.
As we mentioned already, in the finite case this is a direct
consequence of \cite{marx-tods} so we concentrate on the infinite
case.  
It is more convenient for us to prove the result for ordered
unranked forests in which a subtree rooted at every node is
finite. The way to translate a nested $\omega$-word into such a
forest is as follows: when a 
matched call $i$ with $\M(i,j)$ is encountered,
it defines a subtree with $i$ as its root, and $j+1$ as the next
sibling (note that this is different from the translation into binary
trees we used before). 
If $i$ is an internal position, or a pending call or a pending
return position, then it has no descendants and its next sibling is $i+1$.
Matched returns do not have next sibling, nor do they have any descendants.
The nodes in the forest are labeled with $\call$, $\retr$,
and the propositions in $\Sigma$, as in the original nested word.

It is routine to define, in \FO, relations $\desc$ and $\sib$ for 
descendant and younger sibling in such a forest. Furthermore, from
these relations, we can define the usual $\leq$ and $\M$ in nested
words using at most $3$ variables as follows. For $x \leq y$, the
definition is given by
$$(y \desc x) \vee \exists z\Big(x \desc z \wedge \exists x\big(z
\lsib z \wedge y \desc x\big)\Big)$$
and for $\mu(x,y)$, by
$$(y \desc x) \wedge \forall z\big((z \desc x) \to \exists x(x=z
\wedge x \leq y)\big).$$ 
Thus, it suffices to prove the three-variable property for such
ordered forests, which will be referred to as $\A$, $\B$, etc.
We shall use pebble games. 
Let 
$\G^v_m(\A, a_1,b_1,\B,b_1,b_2)$ be the $m$-move, $v$-pebble game on structures $\A$
and $\B$ where initially pebbles $x_i$ are placed on $a_i$ in $\A$ and
$b_i$ in $\B$.  \dupl\ has a winning strategy for $\G^v_m(\A,
a_1,b_1,\B,b_1,b_2)$ iff $\A,a_1,a_2$ and $\B,b_1,b_2$ agree on all
formulas with at most $v$ variables and quantifier-depth $m$.  
We know from \cite{IK}
that to prove Theorem \ref{3var th}, it suffices to show
the following,

\begin{claim}
For all $k$, if \dupl\ has a winning strategy for the game
$\G^3_{3k+2}(\A,a_1,a_2;$ $\B,b_1,b_2)$, then she also has a winning
strategy for the game $\G^k_{k}(\A,a_1,a_2;\B,b_1,b_2)$.
\end{claim}

We will show how \dupl\ can win the $k$-pebble game by maintaining a
set of 3-pebble sub-games on which she will copy \spoiler's
moves and decide on good responses using her winning strategy for
these smaller 3-pebble games.   The choice of these sub-games will
partition the universe $\abs{\A}\cup\abs{\B}$ so that each play by
\spoiler\ in the $k$-pebble game will be answered in one $3$-pebble
game.  This is similar to the proof that linear orderings have the
3-variable property \cite{IK}.

The subgames, $\G^3_{m}(\A,a_1,a_2;\B,b_1,b_2)$, that \dupl\
maintains will all be {\em vertical} in which $a_2 \desc a_1$ and
$b_2 \desc b_1$ hold, or {\em horizontal} in which $a_1 \lsib a_2$ and
$b_1 \lsib b_2$ hold.

The following lemma gives the beginning strategy of \dupl\ in which she
replaces an arbitrary game configuration with a set of configurations
each of which is vertical or horizontal.

\begin{lemma}\label{Rajeev le}
If \dupl\ wins $\G^3_{m+4}(\A,a_1,a_2;\B,b_1,b_2)$. Then there are
points $a_1',a_2'$ from $\A$ and $b_1',b_2'$ from $\B$ 
such that \dupl\ wins the horizontal game
$\G^3_{m+2}(\A,a'_1,a'_2;\B,b'_1,b'_2)$ and the vertical games
$\G^3_{m+2}(\A,a'_i,a_i;\B,b'_i,b'_i)$ for $i= 1,2$.
\end{lemma}

\proof
For this proof since $\A$ and $\B$ are fixed, we will 
describe a game only by listing the chosen points, 
e.g., $(a_1,a_2;b_1,b_2)$.  We simulate two moves of the
  game, $\G^3_{m+4}(a_1,a_2;b_1,b_2)$, in which we choose
  \spoiler's moves and then \dupl\ answers according to her winning
  strategy.  Let $u+v$ denote the least common ancestor of $u$ and
  $v$. First, we have \spoiler\ place pebble $x_3$ on $a_1'$, the
  unique child of $a_1+a_2$ that is an ancestor of $a_1$.  (Note that
  if $a_1' = a_1$ then this move can be skipped and similarly for the
  second move if $a_2' = a_2$.)  \dupl\ answers by placing $x_3$ on
  some point $b_1'$.  Second, \spoiler\ should move pebble $x_1$ from $a_1$
  to $a_2'$, the unique child of $a_1+a_2$ that is an ancestor of
  $a_2$.  \dupl\ moves $x_1$ to some point $b_2'$.

Since \dupl\ has
moved according to her winning strategy, we have that she still has a
winning strategy for the three games in the statement of the lemma.
Furthermore, since $a_1'$ and $a_2'$ are siblings and we have two
remaining moves, $b_1'$ and $b_2'$ must be siblings as well.
\qed

Using Lemma \ref{Rajeev le} we initially partition the universe
according to four subgames:
\begin{enumerate}[$\bullet$]
\item  $(a_r,a_p;b_r,b_p)$ with domain everything not below $a_p$ or
  $b_p$. Here $a_p=a_1+a_2$, i.e., the parent of $a_1'$,
  $b_p=b_1+b_2$, i.e., the parent of $b_1'$ and $a_r$ and $b_r$ are
  the roots of $\A$ and $\B$, (the roots are not necessary but then
  the subgames are all on horizontal or vertical pairs), or
\item  $(a_1', a_1;b_1',b_1)$ with domain everything below $a_1'$ or $b_1'$,
\item  $(a_2', a_2;b_2',b_2)$, with domain everything below $a_2'$ or
  $b_2'$, 
\item  $(a_1', a_2';b_1',b_2')$, with the remaining domain.
\end{enumerate}

We now have to explain, inductively, how all moves of \spoiler\ in the
$k$-pebble game are answered by \dupl\ and how, in the process, the
universe is further partitioned.  We inductively assume that \dupl\
has a winning strategy for each of the 3-pebble, m-move sub-games.
There are two cases:

{\bf Vertical:}  \spoiler\ places a new pebble on a point $a$ that is 
in the domain of a vertical game: $(a_1,a_2;b_1,b_2)$.  We thus know that $a_1$ is a proper
ancestor of $a$.  The interesting case is where
neither of $a$ and $a_2$ is above the other so, without loss of generality,
assume that $a<a_2$.  We place $x_3$ on $a_2'$, the child of $a+a_2$
  that is above $a_2$.  Let \dupl\
  move according to her winning strategy, placing $x_3$ on some point
  $b_2'$.  We split the
  original game into $(a_1,a_2';b_1,b_2')$ and  
$(a_2',a_2;b_2',b_2)$ 
so \dupl\ has a winning strategy for these 3-pebble, $m-1$
  move sub-games.  Next, in the  $(a_1,a_2';b_1,b_2')$ game we place
  $x_3$ on $a_p$, the parent of $a_2'$ and we let \dupl\ answer
  according to her winning strategy, placing $x_3$ on some point,
  $b_p$.  We then split off the game $(a_1,a_p;b_1,b_p)$.

Returning to the game $(a_1,a_2';b_1,b_2')$,  we have \spoiler\ place $x_3$ on $a'$,
the sibling of  $a_2'$ above $a$, and let \dupl\ answer according to her
winning strategy, placing $x_3$ on some point, $b'$.  

Finally, we let \spoiler\ move $x_1$ to $a$, and let \dupl\ reply with $x_1$
on some point $b$.  

The sub-games are thus:  $(a_1,a_p;b_1,b_p)$, $(a',a_2'; b', b_2')$,
$(a',a; b', b)$, and $(a_2',a_2;b_2',b_2)$ and \dupl\ has winning
strategies for the $\G^3_{m-3}$ game on all of them.  

{\bf Horizontal:}
In this case, we have the configuration, $(a_1,a_2;b_1,b_2)$,
consisting of a pair of siblings.  The only interesting case occurs
when \spoiler\ puts a new pebble on some vertex, $a$, s.t. $a_1 < a <
a_2$.  In this case, we have \spoiler\ place pebble $x_3$ on $a'$, the
sibling of $a_1$ above $a$.  \dupl\ will place pebble $x_3$ on some
vertex, $b'$, which must be a sibling of $b_1$ and $b_2$.  

Next, in the game below $a'$ and $b'$, we let \spoiler\ place pebble $x_2$
on $a$ and we let \dupl\ answer according to her winning strategy in
this game, placing $x_2$ on some vertex, $b$.  The domain of the
original configuration is thus split into domains for three sub-games:  
$(a_1,a';b_1,b')$,
$(a',a_2;b',b_2)$, and
$(a',a;b',b)$.  On each of these, \dupl\ has a winning strategy
for the 3-pebble, $m-2$ move game.  

We now complete the proof that \dupl\ wins $\G^k_k(a_1,a_2;b_1,b_2)$.
Whenever \spoiler\ places a new pebble on some point, say $a$, in the
original game, \dupl\ will answer as described above, i.e., in one of
the little games we will have \dupl\ wins $\G^3_{3r} (a,a';b,b')$ where
there are $r$ moves remaining in the big game.

\dupl\ then answers in the big game by placing the corresponding
pebble on $b$.  To see that the resulting moves are a win for \dupl,
we must just consider  any two pebbled points, $a_i,a_j\in \A$, and
$b_i,b_j\in \B$.  If they came from the same sub-game, then they agree
on relations $\desc,\lsib$ because \dupl\ wins the sub-game.  Otherwise,
$a_i,b_i$ came from one sub-game, $G_i$, and 
$a_j,b_j$ came from another sub-game, $G_j$.  By our choice of the
domains and transitivity of $\desc,\lsib$, it thus follows that $a_i,a_j$
stand in the same relation with respect to $\desc,\lsib$ as $b_i,b_j$ do.

\subsection{The two-variable fragment}
In this section, we construct a temporal logic that captures the two-variable
fragment of \FO\ over nested words. 
Note that for finite unranked trees, a navigational
logic capturing $\FO^2$ is known \cite{marx-twovars,marx-tods}: it
corresponds to a fragment of XPath. However, translating the basic
predicates over trees into 
the vocabulary of nested words requires $3$ variables, and thus we
cannot apply existing results even in the finite case. 

Our temporal logic will be based on several next and eventually
operators. 
Since $\FO^2$ over a linear ordering cannot define the
successor relation but temporal logics have next operators, we
explicitly introduce successors into the vocabulary of
$\FO$.  
These successor relations in effect partition the linear edges into three 
disjoint types; {\em interior} edges, {\em call} edges, and {\em return} edges,
and the nesting edges
(except those from a position to its linear successor) into two disjoint types; 
{\em call-return\/} summaries, and {\em call-interior-return\/} summaries.

\begin{enumerate}[$\bullet$]
\item $S^i(i,j)$ holds iff $j=i+1$ and either $\M(i,j)$ or 
 $i$ is not a call and $j$ is not a   return.
\item $S^c(i,j)$ holds iff $i$ is a call and $j=i+1$ is not a return;
\item $S^r(i,j)$ holds iff $i$ is not a call and $j=i+1$ is a return.
\item $S^{cr}(i,j)$ holds iff $\M(i,j)$ and there is a path from $i$ to $j$ using only
call and return edges.
\item $S^{cir}(i,j)$ holds iff $\M(i,j)$ and neither $j=i+1$ nor $S^{cr}(i,j)$.
\end{enumerate} 

Let $T$ denote the set $\{c,i,r,cr,cir\}$ of all edge types.
In addition to the built-in predicates $S^t$ for $t\in T$, we
add the {\em transitive closure} of all unions of subsets 
of these relations.
That is, for each non-empty set $\Gamma\subseteq T$ of 
edge types,  let $S^\Gamma$ stand for the union $\cup_{t\in\Gamma} S^t$,
and let $\leq^\Gamma$ be the reflexive-transitive closure of $S^\Gamma$.
Now when we refer to $\FO^2$ over nested words, we mean $\FO^2$ in the
 vocabulary of the unary predicates plus all the $\leq^\Gamma$'s,
the five successor relations, and the built-in unary $\call$ and $\retr$
predicates. 

We define a temporal logic $\ucaret$  that has
future and past versions of next operators parameterized by edge types,
and eventually operators parameterized
by a set of edge types. 
For example, $\Diamond^{\{c\}}$ means eventually along
a path containing only call edges.
Its formulas are given by:
$$
\begin{array}{rcl} 
  \phi & := &  \top\ \mid\ a \ \mid\ \call\ \mid\ \retr\ \mid \ \neg \phi \ 
\mid \ \phi \vee \phi' \
  \mid \\ 
&& \dm^t \phi \ \mid  \ \dmminus^t \phi \ \mid\  \eventually^\Gamma \phi \ 
\mid \   \peventually^\Gamma \phi
\end{array}
$$
where $a$ ranges over $\Sigma$, $t$ ranges over $T$,  and $\Gamma$ ranges over non-empty subsets of 
$T$. 
The semantics is defined in the obvious way.  For example,
$(\w,i) \models \eventually^\Gamma \varphi$ iff 
for some position $j$,  $i \leq^{\Gamma} j$ and 
$(\w,j)\models \varphi$; $(\w,i) \models \dm^t \varphi$ 
iff for some position $j$, $S^t(i,j)$  and $(\w,j) \models \varphi$;
and $(\w,i) \models \call$ iff $\call(i)$ holds in $\w$.

For an $\FO^2$ formula $\phi(x)$ with one free variable $x$, let
$\qdp{\phi}$ be its quantifier depth,
and for a $\ucaret$ formula $\phi'$,  let $\odp{\phi'}$ 
be its operator depth.

\begin{theorem} 
  \label{translation theorem}\hfil
\begin{enumerate}[\em(1)]
\item  {$\ucaret$} is expressively complete for $\FO^2$ over nested words.

\item If formulas are viewed as DAGs (i.e identical subformulas are shared), then
every $\FO^2$ formula $\phi(x)$ can be converted to an equivalent
  {$\ucaret$} formula $\phi'$
of size $2^{\mathcal O(|\phi| (\qdp \phi + 1))}$ and $\odp{\phi'} \leq  
10 \; \qdp \phi$.
  The translation is computable in time polynomial in
  the size of $\phi'$.
\item Model checking of $\ucaret$ can be carried out with the
same worst case complexity as for NWTL.
\end{enumerate}
\end{theorem}

{\em Proof}.
The translation from  $\ucaret$ into $\FO^2$ is standard and can be
done with negligible blow-up in the size of the formula, so we
concentrate on the other direction.
The proof generalizes the proof of an analogous result
 for unary temporal logic over words from \cite{EVW02}. 

Given an $\fotwo$ formula $\phi(x)$ the translation procedure works
a follows. When $\phi(x)$ is atomic, \ie, of the form $a(x)$, it
outputs $a$.  When $\phi(x)$ is of the form $\psi_1 \vee \psi_2$
or $\neg \psi$---we say that $\phi(x)$ is \emph{composite}---it
recursively computes $\psi_1'$ and $\psi_2'$, or $\psi'$ and outputs
$\psi_1' \vee \psi_2'$ or $\neg \psi'$.  The two cases that remain are
when $\phi(x)$ is of the form $\exists x\, \phi^*(x)$ or $\exists
y\, \phi^*(x,y)$. In both cases, we say that $\phi(x)$ is
\emph{existential}. In the first case, $\phi(x)$ is equivalent to
$\exists y\, \phi^*(y)$ and, viewing $x$ as a dummy free variable in
$\phi^*(y)$, this reduces to the second case.
  
In the second case, we can rewrite $\phi^*(x,y)$ in the form
\[  \phi^*(x,y)  =  \beta(\chi_0(x,y),..,\chi_{r-1}(x,y),
    \xi_0(x),..,\xi_{s-1}(x), \zeta_0(y),.., \zeta_{t-1}(y))\]
where $\beta$ is a propositional formula, each formula $\chi_i$ is an
atomic order formula, each formula $\xi_i$ is an atomic or existential
$\fotwo$ formula with $\qdp{\xi_i} < \qdp \phi$, and each formula
$\zeta_i$ is an atomic or existential $\fotwo$ formula with
$\qdp{\zeta_i} < \qdp \phi$.
  
In order to be able to recurse on subformulas of $\phi$ we have to
separate the $\xi_i$'s from the $\zeta_i$'s. We first introduce a case
distinction on which of the subformulas $\xi_i$ hold or not. We obtain
the following equivalent formulation for $\phi$:
\[  \bigvee_{\overline{\gamma} \in \{\true, \false\}^s} (
  \bigwedge_{i<s} (\xi_i \leftrightarrow \gamma_i) \wedge \exists y \,
  \beta( \chi_0, \dots, \chi_{r-1}, \gamma_0, \dots, \gamma_{s-1},
  \zeta_0, \dots, \zeta_{t-1})) \enspace. \]
We proceed by a case distinction on which order relation holds between
$x$ and $y$, where $x \leq y$. 
We consider  mutually exclusive cases, determined by
the following formulas, which we call \emph{order types}.
 
\begin{enumerate}[$\bullet$]
\item $\Psi_0$ is $x=y$.
\item For each $t\in T$,  $\Psi_t$ is  $S^{t}(x,y)$.
\item For each $t\in T$,  $\Phi_t$ is
$\exists z\ (S^{t}(x,z)\wedge z <^t y)$.
\item Let $o=t_1,t_2,\ldots t_k$  be a sequence over $T$ such that
$2\le k\le 5$, all $t_i$'s are distinct, and a call never appears before return
(that is, if $t_i=c$ then $t_j\not=r$ for $j>i)$.
Then $\Psi_o$ stands for
{\small
\begin{eqnarray*}
\hspace{15mm}\exists z_1,z'_1,z_2,z'_2,\ldots z_k\
(\,S^{t_1}(x,z_1) \wedge z_1\le^{T_1} z'_1 \wedge S^{t_2}(z'_1,z_2) \wedge
z_2\le^{T_2}z'_2\wedge \cdots \wedge z_k\le^{T_k} y\,)
\end{eqnarray*}}\noindent
where for $1\le i\le k$, the set $T_i$ equals the set $\{t_1,t_2\ldots t_i\}$,
but with $r$ removed if both $c$ and $r$ belong to this set.
\end{enumerate}

\noindent We claim  that these order types are
mutually exclusive and complete, and are expressible in $\ucaret$ (and hence, in 
$\FO^2$). First, let us show that the order types form a disjoint partition, 
meaning for all pairs $(x,y)$
such that $x \leq y$,  we
have exactly one of these relationships holding true. 
To see this, suppose $x < y$. Then either  $S^t(x,y)$ holds for some type $t$
(and the successor relations $S^t$ are disjoint, for distinct $t$'s), 
or there is a path from $x$ to $y$
that uses at least two edges. 
The key observation is that a path from $x$ to $y$ is a summary path iff
the path does not contain a call edge followed later by a return edge.
Also, there is a unique summary path from $x$ to $y$.
We can now classify the paths by the edge types that this unique summary
path contains, and the order in which they first appear in the path.
For example, $\Phi_{c}(x,y)$ holds when there is a path from $x$ to $y$ using 2 or more
call edges; $\Phi_{c,cir}(x,y)$ holds when there is a path from $x$ to $y$
which begins with a call edge, uses at least one call-interior-return summary edge, 
and uses only these two types of edges; 
$\Phi_{r,i,c}(x,y)$ holds when there is a path from $x$ to $y$ that can be split into
three consecutive parts:
a part containing only return edges, a part containing at least one internal 
and only internal and return edges, 
and a part containing at least one call and only call and internal edges.
Note that some of these order types are empty: for example, two summary edges can never
follow one another, and hence $\Phi_{cr}(x,y)$ can never hold. 
Emptiness of some of the order types is not relevant to the proof.

When we assume that one of these order types is true,
each atomic order formula evaluates to either $\true$ or $\false$, in
particular, each of the $\chi_i$'s evaluates to either $\true$ or
$\false$; we will denote this truth value by $\chi_i^\tau$. 
For example, when $\Psi_{cr}(x,y)$ holds then
(1) $S^t(x,y)$ is true for $t=cr$ and false for $t\not=cr$, and
(2) $\le^\Gamma$ is true if $\Gamma$ contains $cr$ or if $\Gamma$ contains both $c$ and $r$,
and false otherwise.

We can
finally rewrite $\phi$ as follows, where $\Upsilon$ stands for the
set of all order types:
\begin{eqnarray*}
  \bigvee_{\overline{\gamma} \in \{\true, \false\}^s} (
  \bigwedge_{i<s} (\xi_i \leftrightarrow \gamma_i) \wedge
  \bigvee_{\tau \in \Upsilon} \exists y (\tau \wedge
  \beta(\chi_0^\tau, \dots, \chi_{r-1}^\tau, \overline{\gamma},
  \overline{\zeta}))) \enspace.
\end{eqnarray*}
If $\tau$ is an order type, $\psi(x)$ an
$\fotwo$ formula, and $\psi'$ an equivalent $\utl$ formula, there is
a way to obtain a $\utl$ formula $\tau\langle\psi\rangle$
equivalent to $\exists y (\tau \wedge \psi(y))$, as follows.
Assume that $x \leq y$.
\begin{enumerate}[$\bullet$]
\item
For the order type $\Psi_0$, $\tau\langle\psi'\rangle$ is $\psi'$ itself.
\item For each $t\in T$,  
for the order type $\Psi_t$, $\tau\langle\psi'\rangle$ is $\next^t\psi'$.
\item For each $t\in T$,  
for the order type $\Phi_t$, $\tau\langle\psi'\rangle$ is $\next^t\next^t\eventually^{\{t\}}\psi'$.
\item 
For order type $\Psi_o$, where $o=t_1,t_2,\ldots t_k$  is a sequence over $T$, 
$\tau\langle\psi'\rangle$ is 
$\next^{t_1}\eventually^{T_1}\next^{t_2}\cdots$ $\eventually^{T_k}\psi'$,
where for $1\le i\le k\le 5$, the set $T_i$ equals the set $\{t_1,t_2\ldots t_i\}$,
but with $r$ removed if both $c$ and $r$ belong to this set.
\end{enumerate}
The case corresponding to past operators is analogous.
Our procedure will therefore recursively compute $\xi_i'$ for $i<s$
and $\zeta_i'(x)$ for $i<t$ and output

\begin{equation}
\bigvee_{\overline{\gamma} \in \{\true,
  \false\}^s} \!\!\!\! ( \bigwedge_{i<s} (\xi_i' \leftrightarrow
  \gamma_i) \wedge \ \bigvee_{\tau \in \Upsilon} 
\tau\left\langle\beta(\chi_0^\tau, ..,
      \chi_{r-1}^\tau, \overline{\gamma}, \zeta_0'(x), \dots,
      \zeta_{t-1}'(x)) \right\rangle ) \enspace.
\label{eq:translation}
\end{equation}

Now we verify that $|\phi'|$ and $\odp{\phi'}$ are bounded as stated
in the theorem. 
Note that the size $|\phi'|$ is measured by viewing the $\ucaret$
formula as a DAG, i.e., sharing identical subformulas.
That $\odp{\phi'} \leq 10\, \qdp{\phi}$ is easily
seen from the operator depth in the translation table above. 
The proof that $|\phi'| \leq 2^{c |\phi| (\qdp{\phi } + 1)}$ for
some constant $c$ is inductive on the quantifier depth of $\phi$. The
base case is trivial, and the only interesting case in the inductive
step is when $\phi$ is of the form $\exists y \phi^*(x,y)$ as above.
In this case, we have to estimate the length of (\ref{eq:translation}).
There are $2^s \leq 2^{|\phi|}$ possibilities for $\overline{\gamma}$
in (\ref{eq:translation}), and each disjunct in (\ref{eq:translation}) has
length at most $d \, |\phi| \, \max_{i<s, j<t}(|\xi_i'|, |\zeta_j'|)$
for some constant $d$. By induction hypothesis, the latter is bounded
by $d \, |\phi| \, 2^{c |\phi| \qdp{\phi}}$, which implies the claim,
provided $c$ is chosen large enough.

It is straightforward to verify that our translation to $\phi'$ can
be computed in time polynomial in $|\phi'|$.

Model checking of $\ucaret$ can be achieved with the same complexity
as for NWTL using a variant of the tableaux construction in Section
\ref{mc-sec}.
\qed

\section{Conclusion}

\noindent We have provided several new temporal logics over nested words
and shown that they are first-order expressively complete. 
We have furthermore shown that first-order logic over nested
words has the three-variable property, and we have also
provided a temporal logic over nested words that is complete
for two-variable first-order logic.
We have
shown, via an automata-theoretic approach based on 
nested word automata, 
that satisfiability for 
the logic $\nwtlp$ is
EXPTIME-complete,
and that model checking runs in time polynomial in the size of the
RSM model and exponential in the size of the formula.
When the within modality is added to $\nwtl$, the complexity 
of model checking 
becomes doubly exponential.
We note that it remains open whether the original temporal logic \caret,
proposed for nested words 
in \cite{AEM04}, is first-order complete, but we conjecture that it is
not.

\section*{Acknowledgments}
\noindent The authors were supported by: Alur -- NSF CPA award 0541149; Arenas
-- \mbox{FONDECYT} grants 1050701, 7060172 and 1070732; Arenas and Barcel\'o
-- grant P04-067-F from the Millennium Nucleus Centre for Web
Research; Immerman -- NSF grants CCF-0541018 and CCF-0830174; Libkin
-- EC grant MEXC-CT-2005-024502 and EPSRC grant E005039.

\appendix
\section{Proof of Lemma \ref{nwtl-lemma-one}}
\noindent
For translating each $\nwtlss$ formula $\phi$ into an equivalent
$\nwtl$ formula $\beta_\phi$, we need to consider only the case of
until/since operators. The formula $\psi \Uss \theta$ is translated
into
{\small
\begin{multline}
\label{trans-eq-app}
{
\beta_\theta \vee \bigg(\beta_\psi \wedge \bigg(\bigg((\beta_\psi \vee
\rett) \wedge (\neg \mcall \to \dm \beta_\psi) \wedge (\mcall \to (\dmm
\dm \beta_\psi \vee \dmm \dm \beta_\theta))\bigg) \Up} \\ 
{
\bigg((\beta_\psi \vee \rett) \wedge (\neg \mcall \to \dm \beta_\theta)
\ \wedge} \\ 
{
\dm (\neg \rett \wedge \gamma)))\bigg)\bigg)\bigg),}
\end{multline}
}
where $\gamma$ is a formula defined as follows:
{\small
\begin{multline*}
{
\bigg((\beta_\psi \vee
\rett) \wedge (\neg \mcall \to \dm \beta_\psi) \wedge (\mcall \to (\dm
\beta_\psi \vee \dmm \dm \beta_\psi)) \wedge (\dm \rett \to
\call) \bigg)\ \Up}\\ 
{
\bigg((\beta_\psi \vee \rett) \wedge (\neg\mcall \to \dm \beta_\theta)
\wedge (\mcall \to (\dm \beta_\theta \vee \dmm  \dm  \beta_\theta))\bigg)}
\end{multline*}
}
The proof that the translation is correct is by induction on the
structure of $\nwtlss$ formulas. Again we need to consider only the
case of until/since operators. Assume that $\psi$, $\theta$ are
equivalent to $\beta_\psi$ and $\beta_\theta$, respectively. We need
to prove that $\psi \Uss \theta$ is equivalent to (\ref{trans-eq-app}).
\\
\\
($\Leftarrow$) We first show that if 
$(\w, i)$ satisfies (\ref{trans-eq-app}), then $(\w, i) \models \psi \Uss
\theta$. Given that $(\w, i)$
satisfies (\ref{trans-eq-app}), either $(\w, i) \models \beta_\theta$ or
$(\w, i)$ satisfies the second disjunct of
(\ref{trans-eq-app}). Since $\beta_\theta$ and $\theta$ are assumed to be
equivalent, in the former case $(\w, i) \models \psi \Uss
\theta$. Thus, assume that the latter case holds. Then $(\w, i)
\models \psi$, since $\psi$ and $\beta_\psi$ are equivalent, and there
exists a summary path $i = i_0 < i_1 < \cdots < i_p$ such that:
{\small
\begin{eqnarray}
\hspace{-8mm}(\w, i_k) & \!\!\models\!\! & (\beta_\psi \vee \rett)
\wedge (\neg \mcall \to \dm \beta_\psi) \wedge (\mcall \to (\dmm \dm
\beta_\psi \vee \dmm  \dm \beta_\theta)), \ \ \ \ 0 \leq k < p,
\label{gamma-eq-one}\\    
\hspace{-8mm}(\w, i_p) & \!\!\models\!\! & (\beta_\psi \vee \rett)
\wedge (\neg \mcall \to \dm \beta_\theta) \wedge (\mcall \to (\dm
\beta_\theta \vee \dmm \dm \beta_\theta \vee \dm (\neg \rett \wedge
\gamma))). \label{gamma-eq-two}  
\end{eqnarray}}\noindent
We consider three cases. 
\begin{enumerate}[(a)]
\item[(I)] Assume that there exists a position $i_k$ ($k \in [0,
p-1]$) such that $i_k$ is a matched call position and $(\w, i_k)
\models \dmm \dm \beta_\theta$, and let $i_q$ ($q \in [0, p-1]$) be
the first such position. Then only one semi-strict path with
endpoints $i = i_0$ and $r(i_q)+1$ can be obtained from the sequence
$i_0 < i_1 < \cdots < i_q < r(i_q)+1$ by removing all positions $i_k$
(with $k \in [1, q]$) such that $i_{k-1}$ is a matched call position and
$r(i_{k-1}) = i_k$; let $i_0 = j_0 < j_1 < \cdots < j_\ell = r(i_q)+1$
be that semi-strict path. Next we show that: 
\begin{eqnarray*}
(\w, j_k) & \models & \psi \ \ \ \ \ 0 \leq k < \ell,\\
(\w, j_\ell) & \models & \theta,
\end{eqnarray*}
from which we conclude that $(\w, i) \models \psi \Uss \theta$. 

Given that $j_\ell = r(i_q)+1$, $(\w, i_q) \models \dmm \dm
\beta_\theta$ and we assume that $\theta$ and $\beta_\theta$ are
equivalent, we have that $(\w, j_\ell) \models \theta$. Next we show
that $(\w, j_k) \models \psi$ for every $k \in [0,\ell-1]$. If $k =
0$, then the property holds since $(\w, i) \models \psi$ and we assume
that $\psi$ and $\beta_\psi$ are equivalent. Assume that $k \in
[1,\ell-1]$. If $j_k$ is not a return position, then $(\w, j_k)
\models \psi$ since $(\w, j_k) \models \beta_\psi \vee \rett$ (recall
that $j_k$ is a position in the summary path $i_0 < i_1 < \cdots <
i_q$ since $k < \ell$). If $j_k$ is a return position, then we have to
consider two cases. If $j_{k-1} = j_k-1$, then we have that $j_k-1$ is
not a call position since $j_k$ is a return position, $j_0 < j_1 <
\cdots < j_\ell$ is a semi-strict path and $k-1 \in
[0,\ell-2]$. Given that $j_{k-1}$ is a position in the summary path
$i_0 < i_1 < \cdots < i_{q-1}$, we conclude that $(\w, j_k-1) \models
\neg \mcall \to \dm \beta_\psi$. Thus, from the fact that $j_{k-1}$ is
not a call position, we conclude that $(\w, j_k) \models
\beta_\psi$. Hence, $(\w, j_k) \models \psi$. Otherwise, $j_{k-1} \neq
j_k-1$, and we conclude that $j_{k-1}$ is a matched call position and
$j_k = r(j_{k-1}) + 1$. Thus, since $i_q$ is the smallest one
satisfying $\dmm\dm\beta_\theta$ and $j_{k-1} < i_q$, 
and we know from  (\ref{gamma-eq-one}) that
$(\w, j_{k-1}) \models \mcall \to (\dmm \dm \beta_\psi \vee \dmm \dm
\beta_\theta)$, we see that 
 $(\w, j_{k-1}) \models
\mcall \to \dmm 
\dm \beta_\psi$ and, since  $j_{k-1}$ is a
matched call, we conclude that $(\w, j_k) \models \beta_\psi$ and,
therefore, $(\w, j_k) \models \psi$.

\item[(II)] Assume that condition (I) does not hold, and also assume
that either $i_p$ is not a matched call position or $i_p$ is a matched
call position and $(\w, i_p) \models \dm \beta_\theta
\vee \dmm \dm \beta_\theta$. Then given that $(\w, i_p) \models \neg
\mcall \to \dm \beta_\theta$, we have that there exists a position
$i_{p+1} > i_p$ 
such that $(\w, i_{p+1}) \models  \beta_\theta$ and $i_{p+1}$ is
either $i_p+1$ or $r(i_p)+1$. Only one
semi-strict path with endpoints $i = i_0$ and $i_{p+1}$ can be
obtained from the sequence $i_0 < i_1 < \cdots < i_p < i_{p+1}$ by
removing all positions $i_k$ (with $k \in [1, p]$) such that $i_{k-1}$
is a matched call position and $r(i_{k-1}) = i_k$; let $i_0 = j_0 < j_1 <
\cdots < j_\ell = i_{p+1}$ be that semi-strict path. Next we
show that:
\begin{eqnarray*}
(\w, j_k) & \models & \psi \ \ \ \ \ 0 \leq k < \ell,\\
(\w, j_\ell) & \models & \theta,
\end{eqnarray*}
from which we conclude that $(\w, i) \models \psi \Uss \theta$. 

Given that $(\w, i_{p+1}) \models \beta_\theta$ and the hypothesis that
$\theta$ and $\beta_\theta$ are equivalent, we have that $(\w, j_\ell)
\models \theta$. Next we show that $(\w, j_k) \models \psi$ for every
$k \in [0,\ell-1]$. If $k = 0$, then the property holds since $(\w, i)
\models \psi$.
Assume that $k \in [1,\ell-1]$. If $j_k$ is not a return
position, then $(\w, j_k) \models \psi$ since $(\w, j_k) \models
\beta_\psi \vee \rett$ (recall that $j_k$ is a position in the summary
path $i_0 < i_1 < \cdots <i_p$). If $j_k$ is a return position, then we have
to consider two cases. If $j_{k-1} = j_k-1$, then we have that $j_k-1$
is not a call position since $j_0 < j_1 < \cdots < j_\ell$ is a
semi-strict path and $k-1 \in [0,\ell-2]$. Thus, given that
$j_{k-1}$ is a position in the summary path $i_0 < i_1 < \cdots <
i_{p-1}$, we have that $(\w, j_k-1)
\models \neg \mcall \to \dm \beta_\psi$, from which we conclude that
$(\w, j_k) \models \beta_\psi$. Hence, $(\w, j_k) \models
\psi$. Otherwise, $j_{k-1} \neq j_k-1$, and we conclude that $j_{k-1}$
is a matched call position and $j_k = r(j_{k-1}) + 1$. Thus, given
that condition (I) does not hold, we have that $(\w, j_{k-1}) \models
\mcall \to \dmm \dm \beta_\psi$ (since $j_{k-1}$ is a position in the
summary path $i_0 < i_1 < \cdots < i_{p-1}$ and 
$(\w,j_{k-1}) \not\models \dmm\dm\beta_\theta$). Thus, given that
$j_{k-1}$ is a matched call, we conclude from (\ref{gamma-eq-one}) that $(\w, j_k)
\models \beta_\psi$ and, therefore, $(\w, j_k) \models \psi$.  

\item[(III)] We now look at the remaining cases, that is,  condition
  (I) does not hold, $i_p$ is a matched call, and 
$(\w,i_p) \models \neg (\dm\beta_\theta \vee \dmm\dm\beta_\theta)$. By
  (\ref{gamma-eq-two}), this implies 
$(\w, i_p) \models \dm (\neg \rett \wedge \gamma)$.  
{From} $(\w, i_p)
\models \dm \gamma$, we see that there exists a summary path
$i_p+1 = i_{p+1} < i_{p+2} \ldots < i_q$ such that:
{\small
\begin{align*}
(\w, i_k) \ \models \ (\beta_\psi \vee \rett) \wedge & (\neg \mcall
\to \dm \beta_\psi) \wedge \\
& (\mcall \to (\dm \beta_\psi \vee \dmm \dm \beta_\psi)) \wedge (\dm \rett
\to \call) \ \ \ \ \ p+1 \leq k < q,\\    
(\w, i_q) \ \models \ (\beta_\psi \vee \rett) \wedge & (\neg\mcall \to
\dm \beta_\theta) \wedge (\mcall \to (\dm \beta_\theta \vee \dmm \dm
\beta_\theta)).
\end{align*}}\noindent
We first show that $i_q < r(i_p)$. Assume to the contrary that $i_q
\geq r(i_p)$. Since the first position on the path is inside the call
$i_p$, there exists $k \in [p+1,q]$ such that $i_k =
r(i_p)$. Given that $i_{p+1}$ is not a return position (since
$(\w,i_p) \models \neg \dm\rett$), we have that $q >
p+1$ and, therefore, $i_k-1$ is also a position in the summary path $i_{p+1} <
i_{p+2} \ldots < i_q$. But given that $i_k = r(i_p)$ is the matching
return of $i_p$ and $i_p+1 \leq i_k-1$, we have that $i_k-1$ is not a
call position. Thus, $(\w, i_k-1) \not\models \dm \rett \to \call$,
which contradicts the fact that $i_{p+1} < i_{p+2} \ldots < i_q$
witnesses formula $\gamma$. Therefore indeed $i_q < r(i_p)$.

Given that $(\w, i_q) \models (\neg\mcall \to \dm \beta_\theta) \wedge
(\mcall \to (\dm \beta_\theta \vee \dmm \dm \beta_\theta))$, we conclude
that there exists a position $i_{q+1} > i_q$ such that $(\w, i_{q+1})
\models
\beta_\theta$ and $i_{q+1}$ is either $i_q+1$ or $r(i_q)+1$. Only one
semi-strict path with endpoints $i = i_0$ and $i_{q+1}$ can be
obtained from the sequence $i_0 < i_1 < \cdots < i_q < i_{q+1}$ by
removing all positions $i_k$ (with $k \in [1, q]$) such that $i_{k-1}$
is a matched call position and $r(i_{k-1}) = i_k$; let $i_0 = j_0 < j_1 <
\cdots < j_\ell = i_{q+1}$ be that semi-strict path. Next we
show that:
\begin{eqnarray*}
(\w, j_k) & \models & \psi \ \ \ \ \ 0 \leq k < \ell,\\
(\w, j_\ell) & \models & \theta,
\end{eqnarray*}
from which we conclude that $(\w, i) \models \psi \Uss \theta$. 

Given that $(\w, i_{q+1}) \models \beta_\theta$,
we conclude that that $(\w, j_\ell)
\models \theta$. Next we show that $(\w, j_k) \models \psi$ for every
$k \in [0,\ell-1]$. If $k = 0$, then the property holds since $(\w, i)
\models \psi$ and we assume that $\psi$ and $\beta_\psi$ are
equivalent.  Assume that $k \in [1,\ell-1]$. If $j_k$ is not a return
position, then $(\w, j_k) \models \psi$ since $(\w, j_k) \models
\beta_\psi \vee \rett$ (recall that $j_k$ is a position in the
sequence $i_0 < i_1 < \cdots < i_q$). If $j_k$ is a return position,
then we need 
to consider two cases. If $j_{k-1} = j_k-1$, then we have that $j_k-1$
is not a call position since $j_0 < j_1 < \cdots < j_\ell$ is a
semi-strict path and $k-1 \in [0,\ell-2]$. Thus, given that
$j_{k-1}$ is a position in the sequence $i_0 < i_1 < \cdots <
i_{q-1}$ and $j_{k-1} \neq i_p$ (since $i_p$ is a call position), we
have that $(\w, j_k-1) 
\models \neg \mcall \to \dm \beta_\psi$, from which we conclude that
$(\w, j_k) \models \beta_\psi$. Hence, $(\w, j_k) \models
\psi$. If $j_{k-1} \neq j_k-1$, then we have that $j_{k-1}$ is
a matched call position and $j_k = r(j_{k-1}) + 1$. Moreover, in this
case we also have that $j_k < i_p$. Indeed, to see this, assume to the
contrary that $i_p
\leq j_k$. Then given that $i_q < r(i_p)$, we know that $i_{q+1} \leq
r(i_p)$. Thus, given that $i_p$ is a call position, $k <
\ell$ and $j_\ell = i_{q+1}$, we conclude that $i_{p+1} \leq j_k \leq
i_q$. Therefore, given that $j_k$ is a return position and $i_{p+1} <
\cdots < i_q$ is a summary path, there exists
$s \in [p+1,q]$ such that $i_s$ is a call position with matching
return $j_k$. But since $j_{k-1}$ and $j_k$ are both positions in
the summary path $i_{p+1} < \cdots < i_q$ and $j_k = r(j_{k-1})+1$, we
conclude that this path 
contains three positions $a$, $b$ and $c$ such that $a < b < c$ and
$c$ is the matching return of call position $a$, which contradicts the
definition of summary path. So we proved  $j_k < i_p$. Now we have that $(\w,
j_{k-1}) \models \mcall \to
(\dmm \dm \beta_\psi \vee \dmm \dm \beta_\theta)$, from which we
conclude that $(\w, j_k) \models \beta_\psi$ since condition (I) does
not hold and $j_k = r(j_{k-1})+1$. Hence, $(\w, j_k) \models \psi$.
\end{enumerate}

\noindent
($\Rightarrow$) We now show that if $(\w, i) \models \psi \Uss
\theta$, then  $(\w, i)$ satisfies (\ref{trans-eq-app}). Given that $(\w,
i) \models \psi \Uss \theta$, there exists a semi-strict path
$i = i_0 < i_1 < \cdots < i_q$ such that: 
\begin{eqnarray}
\label{m-eq1}
(\w, i_k) & \models & \psi \ \ \ \ \ 0 \leq k < q,\\
\label{m-eq2}
(\w, i_q) & \models & \theta.
\end{eqnarray}
Notice that if $q = 0$, then $(\w, i) \models \theta$ and, therefore,
$(\w, i)$ satisfies the first disjunct of (\ref{trans-eq-app}) since
$\theta$ and $\beta_\theta$ are assumed to be equivalent. Thus, we
suppose that $q > 0$, and we consider two cases.
\begin{enumerate}[(a)]
\item[(I)] Assume that there exists $k \in [0,q-1]$ such that $i_k$
is a matched call position, $i_{k+1} = i_k+1$ and $i_{k+1}$ is not a
return position, and let $i_p$ be the first such position. Then only
one summary path with endpoints $i = i_0$ and $i_p$ can be obtained
from the semi-strict path $i_0 < i_1 < \cdots < i_p$ by adding
positions $r(i_k)$ for every $k \in [0,p-1]$ such that $i_k$ is a
matched call position and $i_{k+1} = r(i_k)+1$; let $i_0 = j_0 < j_1 <
\cdots < j_\ell = i_p$ be that summary path. Next we show that:
\begin{center}
{\small
\begin{tabular}{l}
$(\w, j_k) \ \models \ (\beta_\psi \vee \rett) \wedge (\neg \mcall
\to \dm \beta_\psi) \wedge (\mcall \to \dmm \dm \beta_\psi)$ \ \ \ \ \
$0 \leq k < \ell$,\\ 
$(\w, j_{\ell}) \ \models \  (\beta_\psi \vee \rett) \wedge (\neg \mcall \to
\dm \beta_\theta) \wedge (\mcall \to (\dm \beta_\theta \vee \dmm \dm
\beta_\theta \vee \dm (\neg \rett \wedge \gamma)))$,
\end{tabular}}
\end{center}
from which we conclude that $(\w, i)$ satisfies (\ref{trans-eq-app}).

We start by showing that the first condition above holds.  Let $k \in
[0,\ell-1]$. If $j_k$ is a return position, then we have that $(\w,
j_k) \models (\beta_\psi \vee \rett)$. Otherwise, by definition of
$j_0 < \ldots < j_\ell$, we have that $j_k$ is a position in the
semi-strict path $i_0 < i_1 <
\cdots < i_{p-1}$. Thus, from (\ref{m-eq1}) we
conclude that $(\w, j_k) \models \psi$ and, hence, $(\w, j_k) \models
(\beta_\psi \vee \rett)$ since $\psi$ and $\beta_\psi$ are assumed to
be equivalent. It only remains to show that $(\w, j_k) \models (\neg
\mcall \to \dm \beta_\psi) \wedge (\mcall \to \dmm \dm \beta_\psi)$. If
$j_k$ is a matched call position, then by definition of $i_p$ we have
that $j_k$ and $r(j_k)+1$ are both positions in the semi-strict
path $i_0 < i_1 < \cdots < i_{p-1}$. Thus, from (\ref{m-eq1}) we
conclude that $(\w, r(j_k)+1) \models \psi$ and, therefore, $(\w,j_k)
\models (\mcall \to
\dmm \dm \beta_\psi)$. If $j_k$ is not a matched call position, then we have
that $j_k+1$ is a position in the semi-strict path $i_0 < i_1
< \cdots < i_p$. Thus, from (\ref{m-eq1}) we conclude that $(\w,
j_k+1) \models \psi$ and, therefore, $(\w,j_k)
\models (\neg \mcall \to \dm \beta_\psi)$.  

We now show that the second condition above also holds. Given that
$j_\ell = i_p$ is a matched call position, we have to prove that $(\w,
j_{\ell})
\models \beta_\psi \wedge (\dm \beta_\theta \vee \dmm \dm
\beta_\theta \vee \dm (\neg \rett \wedge \gamma))$. Given that $(\w,
i_p) \models \psi$ and we assume that $\psi$ and $\beta_\psi$ are
equivalent, we have that $(\w, j_{\ell}) \models \beta_\psi$. If $q =
p+1$, then given that $(\w, i_q) \models \theta$ and we assume that
$\theta$ and $\beta_\theta$ are equivalent, we conclude that $(\w,
j_\ell) \models \dm \beta_\theta$. Thus, assume that $q > p+1$. Next we
show that $(\w, j_\ell) \models \dm (\neg \rett \wedge
\gamma)$ in this case. Given that $i_{p+1} = i_p+1$ and $i_{p+1}$ is
not a return position, we have $(\w, i_p+1) \models \neg \rett$, and it only
remains to prove that $(\w, i_p+1) \models \gamma$. Given that $q >
p+1$, only one summary
path with endpoints $i_p+1$ and $i_{q-1}$ can be obtained from the
sequence $i_p+1 = i_{p+1} < i_{p+2} < \cdots < i_{q-1}$ by adding
positions $r(i_k)$ for every $k \in [p+1,q-2]$ such that $i_k$ is a
call position and $i_{k+1} = r(i_k)+1$; let $i_p+1 = s_0 < s_1 <
\cdots < s_m = i_{q-1}$ be that summary path. Next we show that:
{\small
\begin{align*}
(\w, s_k) \ \models \ (\beta_\psi \vee \rett) \wedge & (\neg \mcall
\to \dm \beta_\psi) \wedge \\
& (\mcall \to (\dm \beta_\psi \vee \dmm \dm \beta_\psi)) \wedge (\dm \rett
\to \call) \ \ \ \ \ 0 \leq k < m,\\    
(\w, s_m) \ \models \ (\beta_\psi \vee \rett) \wedge & (\neg\mcall \to
\dm \beta_\theta) \wedge (\mcall \to (\dm \beta_\theta \vee \dmm \dm
\beta_\theta)).
\end{align*}}\noindent
from which we conclude that $(\w, i_p+1) \models \gamma$.

We start by showing that the first condition above holds.  Let $k \in
[0,m-1]$. If $s_k$ is a return position, then we have that $(\w, s_k)
\models (\beta_\psi \vee \rett)$. Otherwise, by definition of $s_0 <
\ldots < s_m$, we have that $s_k$ is a position in the semi-strict
path $i_{p+1} < i_{p+2} < \cdots < i_{q-1}$. Thus, from
(\ref{m-eq1}) we conclude that $(\w, s_k) \models \psi$ and, hence,
$(\w, s_k) \models (\beta_\psi \vee \rett)$ since $\psi$ and
$\beta_\psi$ are assumed to be equivalent. It only remains to show
that:
\begin{eqnarray*}
(\w, s_k) & \models & (\neg \mcall \to \dm \beta_\psi) \wedge (\mcall
\to (\dm \beta_\psi \vee \dmm \dm \beta_\psi)) \wedge (\dm \rett
\to \call).
\end{eqnarray*}
If $s_k$ is a matched call position, then $s_k$ is a
position in semi-strict path $i_{p+1} < i_{p+2} < \cdots <
i_{q-1}$ and either (a) $s_{k+1} = s_k+1$ and $s_k+1$ is a position in
the semi-strict path $i_{p+1} < i_{p+2} < \cdots < i_{q-1}$,
or (b) $s_{k+1} = r(s_k)$ and $r(s_k)+1$ is a position in the
semi-strict path $i_{p+1} < i_{p+2} < \cdots < i_{q-1}$. In
the former case, from (\ref{m-eq1}) we conclude that $(\w, s_k+1)
\models \psi$ and, therefore, $(\w, s_k) \models \mcall \rightarrow
\dm \beta_\psi$. In the latter case, from (\ref{m-eq1}) we conclude
that $(\w, r(s_k)+1) \models \psi$ and, therefore, $(\w, s_k) \models
\mcall \rightarrow \dmm \dm \beta_\psi$. Thus, if $s_k$ is a matched
call position, then $(\w, s_k) \models \mcall \rightarrow (\dm
\beta_\psi \vee \dmm \dm \beta_\psi)$. Assume now that $s_k$ is not a
matched call position. Given that $i_0 < \cdots < i_p < i_{p+1} <
\cdots < i_q$ is a semi-strict path, $i_p$ is a matched call
position, $i_{p+1} = i_p+1$ and $i_p+1$ is not a return position, we
have that $i_q \leq r(i_p)$. Thus, given that $s_k < i_q$, we have
that $i_p < s_k < r(i_p)$, which implies that $s_k$ is either an
internal position or a return position. Therefore, $s_k+1$ is
a position in the semi-strict path $i_{p+1} < i_{p+2} < \cdots
< i_{q-1}$ and, thus, from (\ref{m-eq1}) we conclude that $(\w,
s_k+1) \models \psi$. Hence, $(\w,s_k)
\models \neg \mcall \to \dm \beta_\psi$, and it only remains to prove
that  $(\w, s_k) \models \dm \rett \to \call$. On the contrary,
assume that $(\w, s_k) \models \dm \rett$ and $(\w, s_k)
\not\models \call$. Given that $s_k$ is not a call position and $s_k <
i_{q-1}$, we have that $s_k+1$ is a position in the semi-strict
path $i_{p+1} < \cdots < i_{q-1}$. Thus, given that $(\w, s_k)
\models \dm \rett$, we conclude that there exists a return position in
the sequence $i_p+1 < \cdots < i_{q-1}$. But this leads to a
contradiction since from the fact that $i_q \leq r(i_p)$, we can
conclude that none of the elements $i_{p+1}$, $\ldots$, $i_{q-1}$ is a
return position. 

To conclude this part of the proof, we need to show that the second
condition above holds, that is, $(\w, s_m) \models (\beta_\psi \vee
\rett) \wedge (\neg\mcall \to \dm \beta_\theta) \wedge (\mcall \to
(\dm \beta_\theta \vee \dmm \dm \beta_\theta))$. Given that $s_m =
i_{q-1}$, we have that $(\w, s_m) \models \psi$ and, therefore, $(\w,
s_m) \models \beta_\psi \vee
\rett$. It remains to show that $(\w, s_m) \models (\neg\mcall \to \dm
\beta_\theta) \wedge (\mcall \to (\dm \beta_\theta \vee \dmm \dm
\beta_\theta))$. Given that 
$i_q \leq r(i_p)$, we know that $s_m = i_{q-1}$ is not a return
position. If $s_m$ is an internal position, then $i_q = i_{q-1}+1$
and, thus, $(\w, s_m) \models \neg\mcall \to \dm
\beta_\theta$ since $(\w, i_q) \models \theta$. If $s_m$ is a call
position, then $s_m$ has a matching return and either $i_q =
i_{q-1}+1$ or $i_q = r(i_{q-1})+1$. In the former case, we have that
$(\w, s_m) \models \mcall \to \dm
\beta_\theta$ since $(\w, i_q) \models \theta$. In the latter case,
we have that $(\w, s_m) \models \mcall \to \dmm \dm \beta_\theta$
since $(\w, i_q) \models \theta$. Hence, we conclude that $(\w, s_m)
\models \mcall \to (\dm \beta_\theta \vee \dmm \dm \beta_\theta)$.

\item[(II)] Assume that condition (I) does not hold, that is, assume
that there is no $k \in [0,q-1]$ such that $i_k$ is a matched call
position, $i_{k+1} = i_k+1$ and $i_{k+1}$ is not a return
position. Then only one summary path with endpoints $i = i_0$ and
$i_q$ can be obtained from the semi-strict path $i_0 < i_1 <
\cdots < i_q$ by adding positions $r(i_k)$ for every $k \in [0,q-1]$
such that $i_k$ is a matched call position and $i_{k+1} = r(i_k)+1$;
let $i_0 = j_0 < j_1 < \cdots < j_\ell = i_q$ be that summary
path. Next we show that: 
{\small
\begin{tabular}{l}
\hspace{-2mm}$(\w, j_k) \ \models \ (\beta_\psi \vee \rett) \wedge (\neg \mcall
\to \dm \beta_\psi) \wedge (\mcall \to (\dmm \dm \beta_\psi \vee \dmm
\dm \beta_\theta))$ \ \ \
$0 \leq k < \ell-1$,\\ 
\hspace{-2mm}$(\w, j_{\ell-1}) \ \models \  (\beta_\psi \vee \rett) \wedge (\neg \mcall \to
\dm \beta_\theta) \wedge (\mcall \to (\dm \beta_\theta \vee \dmm \dm
\beta_\theta \vee \dm (\neg \rett \wedge \gamma)))$,
\end{tabular}}

\noindent
from which we conclude that $(\w, i)$ satisfies (\ref{trans-eq-app}).

We start by showing that the first condition above holds.  Let $k \in
[0,\ell-2]$. If $j_k$ is a return position, then we have that $(\w,
j_k) \models (\beta_\psi \vee \rett)$. Otherwise, by definition of
$j_0 < \ldots < j_\ell$, we have that $j_k$ is a position in the
semi-strict path $i_0 < i_1 < \cdots < i_{q-1}$. Thus, from
(\ref{m-eq1}) we conclude that $(\w, j_k) \models \psi$ and, hence,
$(\w, j_k) \models (\beta_\psi \vee \rett)$ since $\psi$ and
$\beta_\psi$ are assumed to be equivalent. It only remains to show
that $(\w, j_k) \models (\neg \mcall \to \dm \beta_\psi) \wedge
(\mcall \to (\dmm \dm \beta_\psi \vee \dmm \dm \beta_\theta))$. If
$j_k$ is a matched call position and $k < \ell-2$, then given that
$i_0 < i_1 < \cdots < i_q$ is a semi-strict path and condition
(I) does not hold, we have that $j_k$ and $r(j_k)+1$ are both
positions in the semi-strict path $i_0 < i_1 < \cdots <
i_{q-1}$. Thus, from (\ref{m-eq1}) we conclude that $(\w, r(j_k)+1)
\models \psi$ and, therefore, $(\w,j_k) \models \mcall \to \dmm \dm
\beta_\psi$. If $j_k$ is a matched call position and $k = \ell-2$,
then given that $i_0 < i_1 < \cdots < i_q$ is a semi-strict
path and condition (I) does not hold, we have that $j_{\ell-1} =
r(j_k)$ and $i_q = j_\ell = r(j_k) + 1$. Thus, given that $(\w, i_q)
\models \theta$, we conclude that $(\w,j_k) \models \mcall \to \dmm
\dm \beta_\theta$. Finally, if $j_k$ is not a matched
call position, then we have that $j_k+1$ is a position in the
semi-strict path $i_0 < i_1 < \cdots < i_{q-1}$ (since $k <
\ell-1$). Thus, from
(\ref{m-eq1}) we conclude that $(\w, j_k+1) \models \psi$ and,
therefore, $(\w,j_k) \models \neg \mcall \to \dm \beta_\psi$.   

To conclude the proof of the lemma, we show that the second condition
above also holds, that is, $(\w, j_{\ell-1}) \models (\beta_\psi \vee
\rett) \wedge (\neg \mcall \to \dm \beta_\theta) \wedge (\mcall \to
(\dm \beta_\theta \vee \dmm \dm \beta_\theta \vee \dm (\neg \rett
\wedge \gamma)))$. If $j_{\ell-1}$ is a return position, we immediately
conclude that $(\w, j_{\ell-1}) \models (\beta_\psi \vee
\rett)$. Thus, assume that $j_{\ell-1}$ is not a return position. But
in this case we conclude that $j_{\ell-1}$ is a position in the
semi-strict path $i_0 < i_1 < \cdots < i_{q-1}$ and, thus,
$(\w, j_{\ell-1}) \models (\beta_\psi \vee \rett)$ since $(\w, j_{\ell-1})
\models \psi$ and we assume that $\psi$ and $\beta_\psi$ are
equivalent. It only remains to show that:
\begin{eqnarray*}
(\w, j_{\ell-1}) & \models & (\neg \mcall \to \dm \beta_\theta) \wedge
(\mcall \to (\dm \beta_\theta \vee \dmm \dm \beta_\theta \vee \dm
(\neg \rett \wedge \gamma))).
\end{eqnarray*}
If $j_{\ell-1}$ is a matched call position, then given that condition
(I) does not hold, we have that $i_q = j_\ell = r(j_{\ell-1}) =
j_{\ell-1}+1$. Thus, given that $(\w, i_q)
\models \theta$ and we assume that $\theta$ and $\beta_\theta$ are
equivalent, we conclude that $(\w, j_{\ell-1}) \models \dm
\beta_\theta$ and, therefore, $(\w, j_{\ell-1}) \models \mcall \to \dm
\beta_\theta$. If $j_{\ell-1}$ is not a matched call position, then we
have that $i_q = j_\ell = j_{\ell-1}+1$. Thus, given that $(\w, i_q)
\models \theta$, we have that $(\w, j_{\ell-1}) \models \dm
\beta_\theta$ and, therefore, $(\w, j_{\ell-1}) \models \neg \mcall \to \dm
\beta_\theta$. This concludes the proof of Lemma \ref{nwtl-lemma-one}.\qed
\end{enumerate}

\end{document}

%% file: nw.pstex_t
\begin{picture}(0,0)%
\epsfig{file=nw.pstex}%
\end{picture}%
\setlength{\unitlength}{1658sp}%
\begingroup\makeatletter\ifx\SetFigFont\undefined%
\gdef\SetFigFont#1#2#3#4#5{%
  \reset@font\fontsize{#1}{#2pt}%
  \fontfamily{#3}\fontseries{#4}\fontshape{#5}%
  \selectfont}%
\fi\endgroup%
\begin{picture}(17120,1904)(2764,-4036)
\put(17101,-4036){\makebox(0,0)[b]{\smash{{\SetFigFont{6}{7.2}{\familydefault}{\mddefault}{\updefault}{\color[rgb]{0,0,0}6}%
}}}}
\put(3601,-4036){\makebox(0,0)[b]{\smash{{\SetFigFont{6}{7.2}{\familydefault}{\mddefault}{\updefault}{\color[rgb]{0,0,0}1}%
}}}}
\put(5401,-4036){\makebox(0,0)[b]{\smash{{\SetFigFont{6}{7.2}{\familydefault}{\mddefault}{\updefault}{\color[rgb]{0,0,0}3}%
}}}}
\put(6301,-4036){\makebox(0,0)[b]{\smash{{\SetFigFont{6}{7.2}{\familydefault}{\mddefault}{\updefault}{\color[rgb]{0,0,0}4}%
}}}}
\put(7201,-4036){\makebox(0,0)[b]{\smash{{\SetFigFont{6}{7.2}{\familydefault}{\mddefault}{\updefault}{\color[rgb]{0,0,0}5}%
}}}}
\put(9001,-4036){\makebox(0,0)[b]{\smash{{\SetFigFont{6}{7.2}{\familydefault}{\mddefault}{\updefault}{\color[rgb]{0,0,0}7}%
}}}}
\put(9901,-4036){\makebox(0,0)[b]{\smash{{\SetFigFont{6}{7.2}{\familydefault}{\mddefault}{\updefault}{\color[rgb]{0,0,0}8}%
}}}}
\put(10801,-4036){\makebox(0,0)[b]{\smash{{\SetFigFont{6}{7.2}{\familydefault}{\mddefault}{\updefault}{\color[rgb]{0,0,0}9}%
}}}}
\put(4501,-4036){\makebox(0,0)[b]{\smash{{\SetFigFont{6}{7.2}{\familydefault}{\mddefault}{\updefault}{\color[rgb]{0,0,0}2}%
}}}}
\put(8101,-4036){\makebox(0,0)[b]{\smash{{\SetFigFont{6}{7.2}{\familydefault}{\mddefault}{\updefault}{\color[rgb]{0,0,0}6}%
}}}}
\put(12601,-4036){\makebox(0,0)[b]{\smash{{\SetFigFont{6}{7.2}{\familydefault}{\mddefault}{\updefault}{\color[rgb]{0,0,0}1}%
}}}}
\put(14401,-4036){\makebox(0,0)[b]{\smash{{\SetFigFont{6}{7.2}{\familydefault}{\mddefault}{\updefault}{\color[rgb]{0,0,0}3}%
}}}}
\put(15301,-4036){\makebox(0,0)[b]{\smash{{\SetFigFont{6}{7.2}{\familydefault}{\mddefault}{\updefault}{\color[rgb]{0,0,0}4}%
}}}}
\put(16201,-4036){\makebox(0,0)[b]{\smash{{\SetFigFont{6}{7.2}{\familydefault}{\mddefault}{\updefault}{\color[rgb]{0,0,0}5}%
}}}}
\put(18001,-4036){\makebox(0,0)[b]{\smash{{\SetFigFont{6}{7.2}{\familydefault}{\mddefault}{\updefault}{\color[rgb]{0,0,0}7}%
}}}}
\put(18901,-4036){\makebox(0,0)[b]{\smash{{\SetFigFont{6}{7.2}{\familydefault}{\mddefault}{\updefault}{\color[rgb]{0,0,0}8}%
}}}}
\put(19801,-4036){\makebox(0,0)[b]{\smash{{\SetFigFont{6}{7.2}{\familydefault}{\mddefault}{\updefault}{\color[rgb]{0,0,0}9}%
}}}}
\put(13501,-4036){\makebox(0,0)[b]{\smash{{\SetFigFont{6}{7.2}{\familydefault}{\mddefault}{\updefault}{\color[rgb]{0,0,0}2}%
}}}}
\end{picture}%